\documentclass[10pt,journal,compsoc]{IEEEtran}

\usepackage[nocompress]{cite}
\usepackage{graphicx}

\usepackage[utf8]{inputenc} %

\usepackage{xcolor}
\usepackage[normalem]{ulem} %
\usepackage{soul}
\usepackage{multirow}

\usepackage[boxruled,linesnumbered]{algorithm2e}
\SetAlFnt{\footnotesize}
\SetAlCapFnt{\small}
\SetAlCapNameFnt{\small}

\usepackage{caption}
\usepackage{subcaption}

\usepackage{tabularx}
\usepackage{makecell}
\usepackage{amsmath}
\usepackage{amssymb}

\usepackage{comment}

\newcommand{\edit}[1]{#1}

\begin{document}

\title{\LARGE \bf
Performant, Multi-objective Scheduling of Highly Interleaved Task Graphs on Heterogeneous System on Chip Devices
}

\author{Joshua~Mack,
        Samet~E.~Arda,
        Umit~Y.~Ogras,
        Ali~Akoglu
\IEEEcompsocitemizethanks{
\IEEEcompsocthanksitem J. Mack and A. Akoglu are with the Electrical and Computer Engineering Dept., University of Arizona, Tucson, AZ 85719 USA. E-mail: \{jmack2545, akoglu\}@email.arizona.edu.%
\IEEEcompsocthanksitem S. E. Arda is with the School of Electrical, Computer and Energy Engineering, Arizona State University, Tempe, AZ 85287 USA. E-mail: sarda1@asu.edu%
\IEEEcompsocthanksitem U. Y. Ogras is with the Electrical and Computer Engineering Dept., University of Wisconsin-Madison, Madison, WI 53706 USA. E-mail: uogras@wisc.edu%
}}

\IEEEtitleabstractindextext{%
\begin{abstract}
Performance-, power-, and energy-aware scheduling techniques play an essential role
in optimally utilizing processing elements (PEs) of heterogeneous systems.
List schedulers, a class of low-complexity static schedulers, 
have commonly been used in static execution scenarios.
However, list schedulers are not suitable for runtime decision making, particularly when multiple concurrent applications are interleaved dynamically. 
For such cases, the static task execution times and expectation of idle PEs assumed by list schedulers lead to inefficient system utilization and poor performance.
To address this problem, we present techniques for optimizing execution of list scheduling algorithms in dynamic runtime scenarios via a family of algorithms inspired by the well-known heterogeneous earliest finish time (HEFT) list scheduler.
Through dynamically arriving, realistic workload scenarios that are simulated in an open-source discrete event heterogeneous SoC simulator, we exhaustively evaluate each of the proposed algorithms across two SoCs modeled after the Xilinx Zynq Ultrascale+ ZCU102 and O-Droid XU3 development boards.
Altogether, depending on the chosen variant in this family of algorithms, we are able to achieve an up to 39\% execution time improvement, up to 7.24x algorithmic speedup, or up to 30\% energy consumption improvement compared to the baseline HEFT implementation.

\end{abstract}

}

\maketitle
\IEEEdisplaynontitleabstractindextext

\ifCLASSOPTIONcompsoc
    \IEEEraisesectionheading{\section{Introduction}\label{sec:introduction}}
\else
    \section{Introduction} \label{sec:introduction}
\fi

\IEEEPARstart{W}{hen} designed with the right mix of accelerators and general-purpose processors, heterogeneous processors offer 
performance and energy-efficiency not attainable with homogeneous processing systems~\cite{khokhar_heterogeneous_1993, gupta_forgotten_2012}.
Battery capacity and power budget are at a premium in both homogeneous and heterogeneous platforms with strict size, weight, and power (SWaP) requirements. The ability to make scheduling decisions that extend the lifetime between charges is as valuable in these devices, if not more-so, than meeting execution deadline requirements alone. 
As such, portable heterogeneous platforms can utilize their full capabilities only if they carefully balance power, energy, and execution time requirements by utilizing task scheduling that is aware of all three.
However, schedulers for heterogeneous systems, which are required to realize this potential, face significant challenges that are not present in homogeneous counterparts. 
First, task performance and power consumption are not uniform across all processing elements (PEs), even in a static context. 
Consequently, the inability to exploit symmetries in the scheduling problem increases the computational complexity of converging to a high-quality scheduling decision. 
Furthermore, the already NP-Complete \cite{ullman1975np} scheduling problem becomes even harder 
when the scheduling problem expands to a runtime scope where applications may interleave at arbitrary times. 

Previous studies showed that list-scheduling algorithms effectively balance the complexity of the scheduler and quality of schedules they generate~\cite{wu2015workflow}.
Among the large body of list-scheduling algorithms, heterogeneous earliest finish time (HEFT)~\cite{Topcuoglu02} is a well-known heuristic that generates competitive static schedules that minimize the total application execution time~\cite{maurya2018benchmarking}.
List-scheduling algorithms, such as HEFT, are typically used as purely static schedulers as they operate on entire directed acyclic graphs (DAGs) with known task dependencies and application structures.
However, many runtime systems, including the Linux OS, employ a ready-queue-based framework that focuses on only the tasks ready to execute. Hence, by definition, there are no dependencies among the tasks waiting for scheduling.
Furthermore, new applications and tasks can be launched before previous ones finish. Existing list schedulers cannot work in these scenarios since they would have to process \edit{DAGs} of new applications and partial DAGs of existing ones.
Consequently, utilization of list schedulers is challenging in runtime environments, where the scheduler only has visibility into ready tasks with minimal or no insight into future interleaved workloads.

This paper addresses the limitations of list schedulers to work in dynamic runtime environments. Overcoming this limitation makes a large body of list-schedulers~\cite{Bittencourt10,Arabnejad14,zhou2017list, xie_energy-efficient_2017,akbar_list-based_2016} practical and opens up the opportunity to use them in operating systems and other runtime environments.
Motivated by HEFT's ability to balance runtime complexity and quality of generated schedules, we use HEFT as our baseline scheduler and use it in dynamic runtime scenarios. 
Furthermore, we develop new dynamic schedulers that can optimize not only for performance but also for energy, which in turn enable balancing the trade-off between performance and energy consumption.
Finally, static scheduling algorithms have typically been evaluated on individual DAG instances against an optimal schedule length ratio. However, in runtime systems, multiple applications can run concurrently. 
Hence, to properly evaluate static scheduling algorithms in dynamic environments, we must consider scenarios with multiple concurrent DAGs and arbitrary application interleaving. 
To address this need, we evaluate the proposed schedulers against the optimal achievable solutions found by dynamically solving Constraint Programming (CP) formulations of each scheduling problem.
With this insight, list scheduling algorithms, like HEFT, can truly be evaluated against the best achievable runtime rather than overly optimistic bounds that assume no communication and maximum parallelism. 
To that end, this paper makes the following contributions:

Taking the original HEFT as a baseline (referred to as HEFT\textsubscript{Base}), we incrementally develop a family of heuristics that we call HEFT\textsubscript{Dyn}, HEFT\textsubscript{RT}, HEFT\textsubscript{EDP}, and HEFT\textsubscript{EDP-LB}.
Each scheduler builds off of the one listed before it. In that sense, HEFT\textsubscript{Dyn} optimizes HEFT\textsubscript{Base} for dynamic workload scenarios through a combination of DAG merging, running task constraints, and dynamic dependencies.
Through this process, we observe that there is room for algorithmic simplification and propose a runtime variant, HEFT\textsubscript{RT}.
We show that HEFT\textsubscript{Dyn} and HEFT\textsubscript{RT} improve average frame execution time over HEFT\textsubscript{Base} by up to 34\% and 45\% respectively. Additionally, we find that HEFT\textsubscript{RT} is algorithmically more efficient than HEFT\textsubscript{Dyn} with a 7.24x average algorithmic speedup.
By reformulating HEFT\textsubscript{RT} to optimize for energy-delay product (EDP) rather than only execution time, we present HEFT\textsubscript{EDP} and show that it reduces energy consumption by an average of 30.4\% and 20.6\% compared to HEFT\textsubscript{RT} on the Odroid-XU3 and ZCU102 platforms, respectively.
Finally, to better balance energy consumption and execution time, we introduce a load balanced-variant, HEFT\textsubscript{EDP-LB}, which effectively reduces energy consumption without drastically sacrificing workload execution time performance relative to HEFT\textsubscript{RT}.

In summary, this family of heuristics enables the use of HEFT in richly interleaving workload scenarios with the ability to optimize for energy, delay, or both. We believe that these strategies remove previous limitations that prevented effective deployment of list scheduling algorithms like HEFT in System on Chip (SoC)-scale resource management.
As technical contributions, this paper:

\begin{itemize}
    \item Presents the first evaluation of the original HEFT algorithm (HEFT\textsubscript{Base}) under dynamically interleaving workload scenarios, exposes the drawbacks of deploying an unmodified list scheduler to a dynamic runtime, and develops an execution-focused heuristic (HEFT\textsubscript{RT}) to address these drawbacks.
    \item Adjusts HEFT\textsubscript{RT} to be energy-aware by developing HEFT\textsubscript{EDP} and HEFT\textsubscript{EDP-LB}, enabling system designers to target various points in the trade space of energy consumption and execution time.
    \item Presents an exhaustive evaluation of the proposed policies with respect to HEFT\textsubscript{Base} along with Minimum Execution Time, Constraint Programming, and Predict Earliest Finish Time~\cite{Arabnejad14} schedulers using dynamically interleaving workload mixtures across simulated SoCs that are validated against Odroid-XU3~\cite{ODROID} and Xilinx Zynq ZCU102~\cite{FPGA} platforms.
    
\end{itemize}

The rest of the paper is organized as follows: 
Section~\ref{sec:background} begins with preliminaries on HEFT. 
Section~\ref{sec:simsetup} describes the simulation environment used to develop each of the presented algorithms.
Section~\ref{sec:algorithmic_contributions} presents an in depth look at barriers to scaling HEFT in interleaving workload scenarios and introduces HEFT\textsubscript{Dyn}, HEFT\textsubscript{RT}, HEFT\textsubscript{EDP}, and HEFT\textsubscript{EDP-LB}. 
Section~\ref{sec:results} presents exhaustive performance analysis across a large variety of workload scenarios.
Section~\ref{sec:peft_rt} presents a brief evaluation of the generalizability of the proposed techniques.
Section~\ref{sec:related_work} explores related work in the area of task scheduling with emphasis on dynamic task scheduling. 
Finally, Section~\ref{sec:conclusions} concludes and provides avenues for future work.
\section{Background} \label{sec:background}
Directed Acyclic Graphs (DAGs) are a standard method of representing applications where each node represents a task along with its computation cost, and each edge represents the communication cost between data-dependent tasks.
HEFT works by performing a single pass over the DAG to rank nodes by relative importance, sorting them in non-increasing order, and scheduling them in this priority order.
In HEFT, for $E$ edges and $P$ PEs, the NP-complete scheduling problem is reduced to polynomial time ($E\times P$) by considering the average computation cost for each node over the available PEs and the average communication cost for each edge as a function of the amount of data transfer between two nodes and average link speed between the PEs. 
Despite this tremendous complexity reduction, HEFT still manages to provide competitive scheduling decisions~\cite{maurya2018benchmarking}. 
This is due to a combination of a good choice of rank metric and the insertion-based scheduling.

\begin{figure}[tb]
    \centering
    \vspace{-3mm}
    \includegraphics[width=0.65\linewidth]{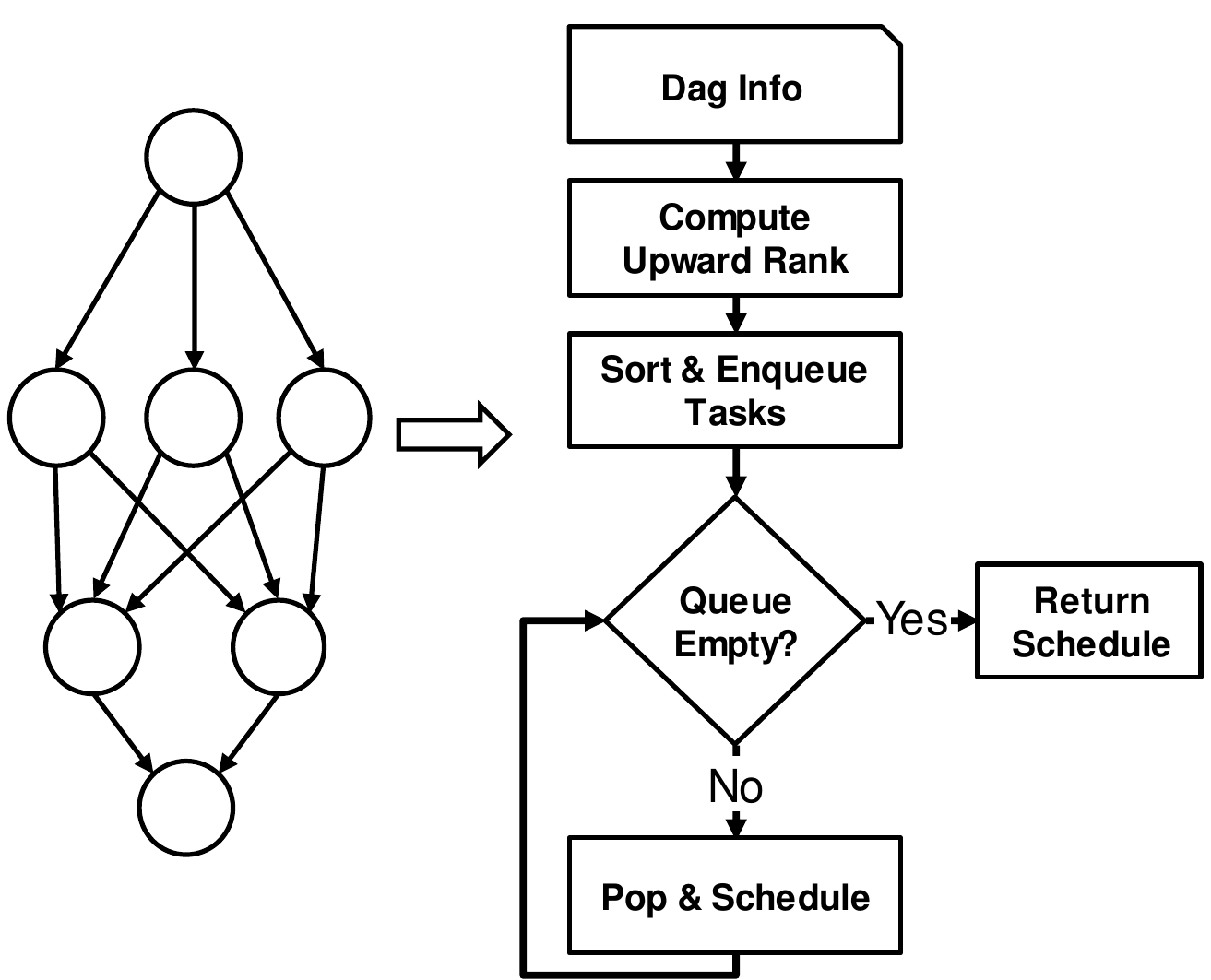}
    \vspace{-1mm}
    \caption{Illustration of the HEFT list scheduling process.}
    \vspace{-3mm}
    \label{fig:list_scheduler_process}
\end{figure}

\begin{figure*}[t]
    \centering
    \includegraphics[trim=1cm 2cm 1cm 2cm, width=0.8\linewidth]{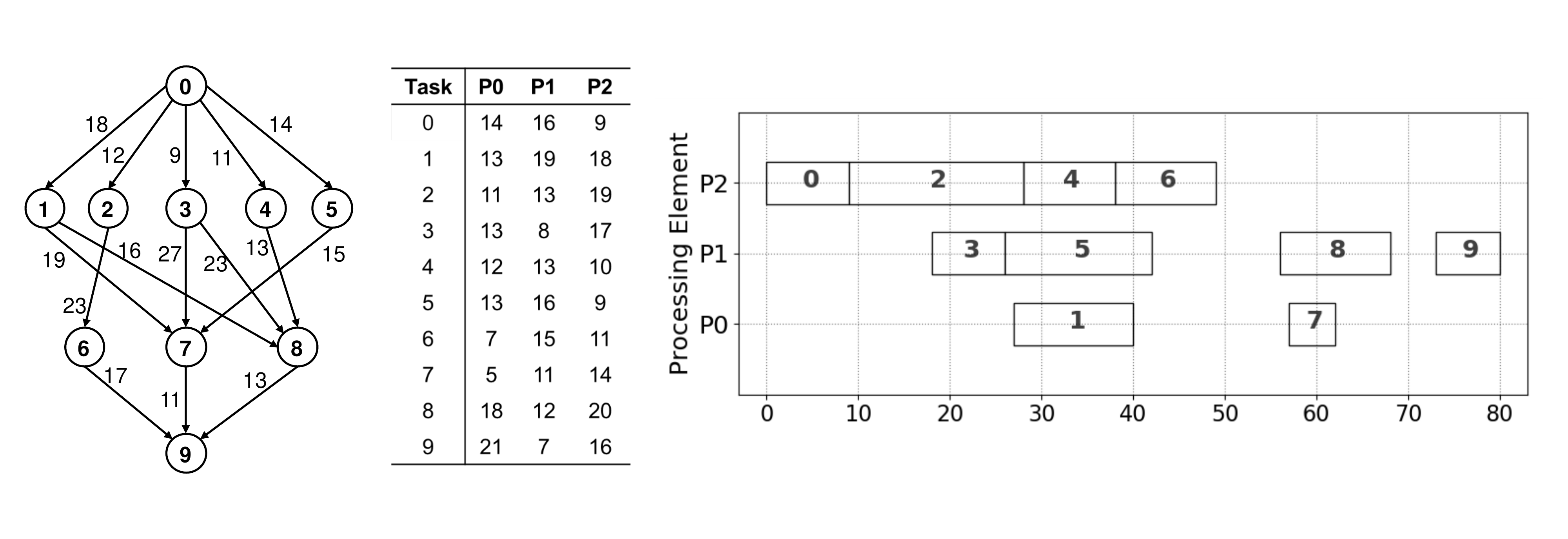}
    \caption{Canonical DAG in~\cite{Topcuoglu02}. Each task is supported on three PEs (P0, P1, P2). Table shows execution time for each task on each PE. Gantt chart shows output schedule generated via the DS3 simulation environment~\cite{arda2020ds3}. In this scenario, all link speeds between distinct PEs are assumed to be 1 in accordance with~\cite{Topcuoglu02}, and the edge weights in the DAG represent the total volume of data that must be transferred over those links.}
    \label{fig:topcuoglu}
\vspace{-12pt}
\end{figure*}

The basic flow of HEFT is shown in Fig.~\ref{fig:list_scheduler_process}.
Other list schedulers perform a similar process with differences in the rank calculation and schedule heuristic.
Given an input DAG, each node in the DAG is assigned a metric of importance.
In standard HEFT, that metric is \textit{upward rank}, which classifies the weight of a node based on its placement in the critical execution path of the DAG.
Once this process is complete, the list of nodes in the DAG is sorted by this metric in non-increasing order, and the resource allocation phase begins.
In this phase, a node is popped and allocated to one of the system resources. 
This process repeats until all the nodes are scheduled.
In HEFT, the allocation process is performed by an insertion-based earliest finish time (EFT) scheduler.
HEFT estimates the earliest possible finish time of a given node 
on each PE by considering both the delay of transferring data from parent tasks to that PE as well as the compute cost of that node on the PE under consideration. 
After that, it assigns the task to the PE that minimizes its earliest finish time.
At the end of this process, a static schedule is generated, with the canonical input/output pair as illustrated with Fig.~\ref{fig:topcuoglu}.
The schedule shown in the figure is the output from our simulation environment and serves as the basis for functional verification of our implementation. 
Further details are presented in Section~\ref{sec:simsetup}.

\section{Experimental Methodology and Setup} \label{sec:simsetup}
\begin{figure}[bt]
	\centering
	\includegraphics[width=.74\linewidth]{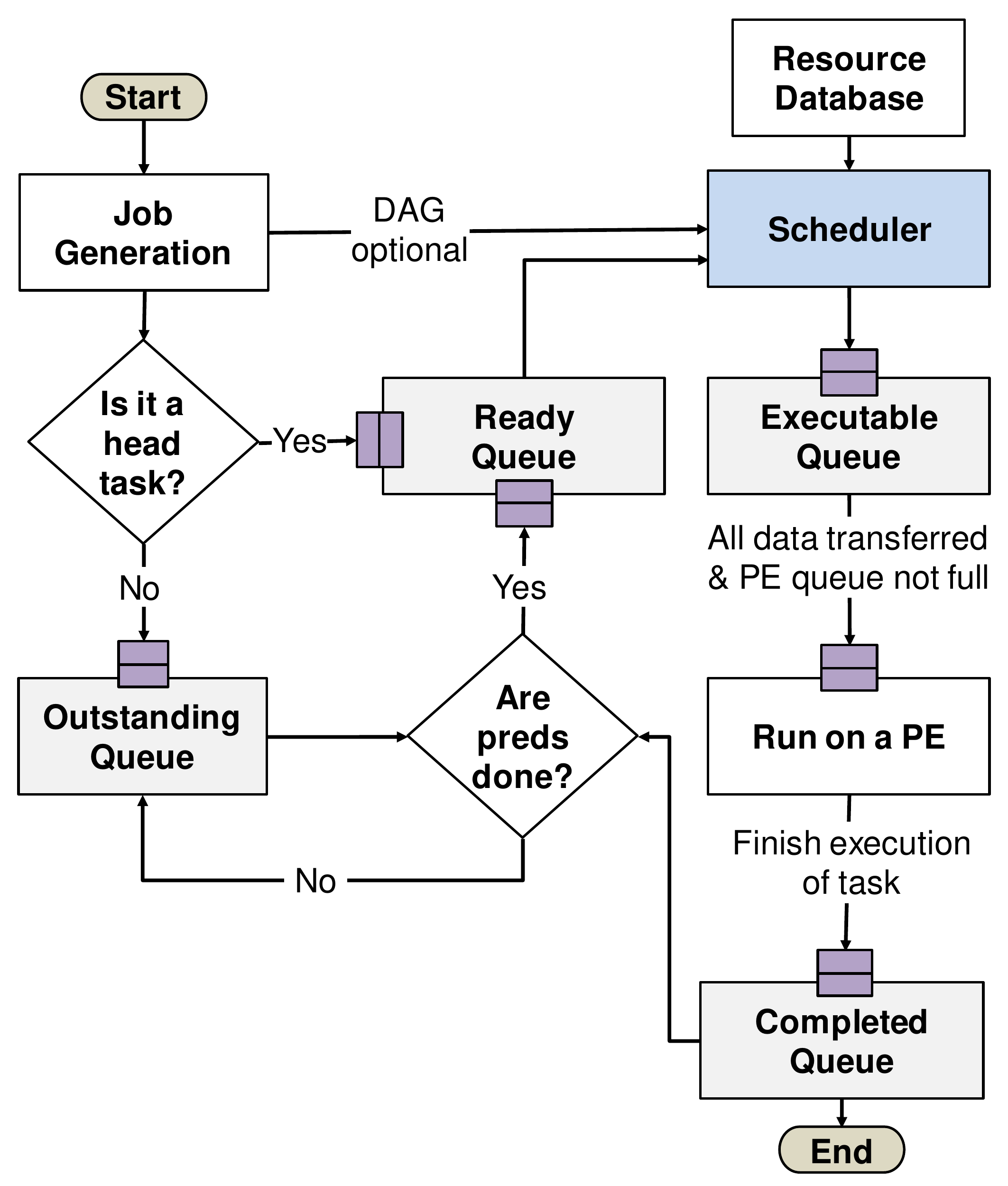}
	\caption{Basic operation of the DS3~\cite{arda2020ds3} simulation framework.}
	\label{fig:framework}
\end{figure}

\subsection{Simulation Framework}
We conduct our evaluations in the open-source DS3 environment~\cite{arda2020ds3} that allows the designer to simulate the behavior of a heterogeneous platform under a variety of realistic SoC workload scenarios, resource management strategies, and hardware configurations.

DS3 enables simulations of benchmark applications, modeled as DAGs, considering a target heterogeneous SoC platform under different scheduling algorithms. 
The list of PEs in the SoC along with the expected computation cost and power consumption of the tasks in the application(s) are stored in the \textit{Resource Database} (see Fig.~\ref{fig:framework}). 
The \textit{Job Generator} injects frames (i.e. instances of applications) into the simulation with an average inter-arrival time determined by a user-defined distribution. 
We refer to this rate of frame injection as the ``frame rate" for the remainder of this study.

In addition, the DS3 enables a loosely defined plug-and-play interface to run a wide range of built-in and custom scheduling algorithms.
The interaction between a particular scheduling algorithm and the framework is depicted in Fig.~\ref{fig:framework}. 
The simulator invokes the scheduler at every scheduling epoch with the list of tasks ready for execution (tasks in \textit{Ready Queue}). Then, it maps ready tasks to PE by utilizing the data in the Resource Database. Finally, the tasks are simulated and placed into the \textit{Completed Queue} upon completion.

For our evaluations, we implement the two hypothetical SoC configurations discussed previously as well as our proposed scheduling algorithms.
Then, we perform trend-based analysis on execution time, throughput, and energy consumption metrics across both SoCs for different workloads composed of a mixture of six WiFi and Radar applications. 

\subsection{Benchmark Applications and SoC Configurations}

\begingroup
\def\arraystretch{1.0}
\begin{table}[tb]
    \centering
    \begin{tabular}{|c|c|c|c|c|}
    \hline
    Application & 
        \begin{tabular}{@{}c@{}} Max\\Width \end{tabular} & 
        Depth & 
        \begin{tabular}{@{}c@{}} Number of\\Nodes \end{tabular} & 
        \begin{tabular}{@{}c@{}} Supported\\Accelerators \end{tabular} \\
    
    \hline
    
    WiFi TX & 
        5 &
        7 &
        27 &
        IFFT \\
    
    \hline
    
    WiFi RX &
        5 &
        10 &
        34 &
        \begin{tabular}{@{}c@{}} FFT,\\Viterbi Decoder \end{tabular} \\

    \hline

    \begin{tabular}{@{}c@{}} Radar\\Correlator \end{tabular} &
        2 &
        6 &
        7 &
        FFT, IFFT \\

    \hline

    \begin{tabular}{@{}c@{}} Temporal\\Mitigation \end{tabular} &
        2 &
        6 &
        10 &
        Matrix Multiply \\

    \hline

    \begin{tabular}{@{}c@{}} Single\\Carrier TX \end{tabular} &
        1 &
        8 &
        8 &
        - \\

    \hline

    \begin{tabular}{@{}c@{}} Single\\Carrier RX\end{tabular} &
        1 &
        8 &
        8 &
        Viterbi Decoder \\
    
    \hline
    \end{tabular}
    \caption{Parallelism, complexity, and accelerator characteristics of each application evaluated in this study.}
    \label{tab:dag_characteristics}
    \vspace{-3pt}
\end{table}
\endgroup

In this study, we utilize six different applications from wireless communications and radar domains: WiFi-transmitter (WiFi-TX), WiFi-receiver (WiFi-RX), range detection (RangeDet), temporal mitigation (TempMit), single-carrier transmitter (SC-TX), and single-carrier receiver (SC-RX) applications. 
As shown in Table~\ref{tab:dag_characteristics}, these applications exhibit different workload characteristics due to variations in DAG structure and support for accelerators. For example, WiFi RX exhibits a large degree of parallelism while the Single Carrier TX consists of a linear chain of DAG nodes.
Execution time and energy profiles of these real-world applications are available in DS3 for two commercially available SoC platforms: the Xilinx Zynq ZCU102 FPGA board~\cite{FPGA} and the Odroid-XU3~\cite{ODROID}.
The ZCU102 FPGA has 4 ARM Cortex-A53 cores coupled with programmable fabric, and the Odroid-XU3 has a Samsung Exynos 5422 SoC with 4 ARM Cortex-A7 and 4 ARM Cortex-A15 cores.
We integrated power consumption and execution time profiling data collected from these boards with state-of-the-art accelerator execution and power measurements from the literature~\cite{fft_acc,viterbi_acc}.
Using this data, DS3 allows us to model two SoC configurations consisting of (1) a big cluster of 4 ARM A53 cores along with two FFT, two Viterbi decoder, and two matrix multiplication accelerators to replicate the ZCU102 with a set of accelerators and (2) the same big.LITTLE cluster of four ARM A15 and four ARM A7 cores to replicate the Odroid.
DS3 is validated against the aforementioned ZCU102 and Odroid-XU3 target platformss with results that are within 3\%, on average, of the reference measurements~\cite{arda2020ds3}.
\edit{Taken together, while we acknowledge that our test applications do not cover the full spectrum of possible workloads, we believe that we cover a reasonable set of the kinds of radar and communications dataflow-based applications that are likely to be deployed on low power heterogeneous SoCs, with Single Carrier RX/TX giving rudimentary communications workflows, WiFi TX/RX giving more advanced communications workflows, Radar Correlator giving a radar-focused workflow, and Temporal Mitigation giving a mixed radar \& comms workflow.}

\subsection{Workload Generation}
We generate workloads by running the benchmark applications at different frame rates.
Initial evaluations are conducted with an upload intensive wireless communication workload composed of 80\% WiFi-TX and 20\% WiFi-RX as each WiFi-RX frame requires nearly 4$\times$ longer to execute.
Later, we expand our analysis to all six applications to demonstrate the generalizability of our conclusions.
Each simulation is run for 100ms of simulation time with the chosen scheduler and a fixed average frame rate.
Each simulation is executed ten times and the results are averaged with the exception of those that utilize the constraint programming scheduler.
Due to their computational complexity, simulations utilizing the constraint programming scheduler are instead only averaged across three iterations.
We evaluate the performance of each scheduler in this study by stressing the scheduler's ability to cope with the volume of incoming jobs.
We establish this by setting a fixed "target" frame rate that defines the expected inter-frame arrival time. 
Then, we monitor if the scheduler can complete jobs at that same rate or if system performance begins to suffer.
We refer to the empirical frame rate observed as the "achieved" frame rate.
Ideally, if a given scheduler is able to cope with the rate of job arrival, the achieved frame rate matches the target frame rate.
However, each scheduler eventually reaches a saturation point where the achieved frame rate diverges from the target frame rate.
Unable to keep up with the target frame rate, each scheduler saturates at some lower point and simultaneously experiences a sharp increase in average job execution time due to this saturation.
We evaluate each scheduler at 18 target frame rates ranging from 0.1 frames/ms to 50 frames/ms and report execution and energy consumption results at their respective achieved frame rates.
Together, this methodology allows us to stress and evaluate each scheduler under various workload levels.

\subsection{Built-in Schedulers}

\noindent The \textit{Minimum Execution Time (MET) scheduler} assigns a ready task to a PE that achieves the minimum expected execution time following a FIFO policy~\cite{heuristic_comparison}.

\noindent 
The \textit{Constraint Programming (CP) scheduler} is based on concepts from IBM ILOG CPOptimizer~\cite{laborie2009ibm}.
An \textit{interval variable} $x$, in constraint programming terminology, is a decision variable, which can represent an activity, operation or a task, as in this study.
The domain of an interval variable $x$ is a subset of $\{\perp\}\cup\{[s,e)|s,e\in \mathbb{Z},s\leq e \}$, where $s$ and $e$ are the start and end of the interval, respectively.
By default interval variables are supposed to be present in the solution of a problem, but they can be specified as being optional.
In this case, $\perp$ is part of the domain of the variable, and thus, the solver determines whether the interval will be present or absent in the solution.
If an interval variable is present: $x = [s,e)$, $(s-e)$ represents the \textit{size} of the interval variable and it is a lower bound on the \textit{length}.
We will apply the following constraints on interval variables:

\begin{itemize}
	\item {\fontfamily{qcr}\selectfont span}$(x,\{x_1,...,x_n\})$ states that if $x$ is present, it starts together with the first present interval in $\{x_1, ..., x_n\}$ and ends together with the last one. Interval $x$ is absent if and only if all the $x_i$ are absent.
	\item {\fontfamily{qcr}\selectfont alternative}$(x,\{x_1,...,x_n\})$ models an exclusive alternative between $\{x_1, ..., x_n\}$: if interval $x$ is present then exactly one of intervals $\{x_1, ..., x_n\}$ is present and $x$ starts and ends together with this chosen one. Interval $x$ is absent if and only if all the $x_i$ are absent.
	\item {\fontfamily{qcr}\selectfont no\_overlap}$(\{x_1,...,x_n\})$ states that permutation $\{x_1,...,x_n\}$ defines a chain of non-overlapping intervals, any interval in the chain being constrained to end before the start of the next interval in the permutation.
    \item {\fontfamily{qcr}\selectfont start\_of}$(x, av)$ returns the start of $x$ when $x$ is present and returns a value $av$ if $x$ is absent (by default if argument $av$ is omitted it assumes $av = 0$).
    \item {\fontfamily{qcr}\selectfont end\_of}$(x, av)$ returns the end of $x$ when $x$ is present and returns a value $av$ if $x$ is absent. 
    \item {\fontfamily{qcr}\selectfont length\_of}$(x, av)$ returns the length (end - start) of $x$ when $x$ is present and returns a value $av$ if $x$ is absent. 
    \item {\fontfamily{qcr}\selectfont end\_before\_start$(x_i,x_j, delay)$} constrains at least the given delay to elapse between the end of $x_i$ and the start of $x_j$. It imposes the inequality {\fontfamily{qcr}\selectfont end\_of}$(x_i)$ + $delay$ $\leq$ {\fontfamily{qcr}\selectfont start\_of}$(x_j)$.
\end{itemize}

Our CP model for dynamic task scheduling minimizes the execution of DAGs in a workload for a given architecture with heterogeneous processing elements.
Consider a set of DAGs $D = \{d_1, d_2, ..., d_n\}$ under consideration for scheduling. 
Then, $d_i$ is an interval variable -- a variable that represents an interval of time via a start and end -- that defines the $i^{th}$ DAG in the system where $i \in n$.
Let $T$ denote the set of tasks $T_i = \{t_{i,1}, t_{i,2}, ..., t_{i, m}\}$ in a DAG $d_i$.
Then, $t_{i,j}$ is an interval variable for a task in the DAG $d_i$ where $i \in n$, $j \in m$, with $n$ the number of DAGs and $m$ the number of tasks in DAG $d_i$.
Finally, $P = \{p_1, p_2, ..., p_z\}$ represents the number of PEs in an architecture. 
Then, $t_{i,j,k}$ is an optional interval variable for task $t_{i,j}$ if it can be executed on PE $p_k$ where $i\in n$, $j\in m$, $k\in z$.
The constraints of the model are listed below:

\begin{equation}
    \forall i \in n \quad \text{{\fontfamily{qcr}\selectfont span}}(d_i,T_i)
    \label{dag_span}
\end{equation}
\begin{equation}
    \label{alternative}
    \begin{split}
    \text{{\fontfamily{qcr}\selectfont alternative}}&(t_{i,j}, \{t_{i,j,1}, t_{i,j,2}, ..., t_{i,j,z}\}) \\
    & \forall i \in n, j \in m
    \end{split}
\end{equation}
\begin{equation} 
    \label{no_overlap}
    \forall i \in n \quad \text{{\fontfamily{qcr}\selectfont no\_overlap}}(T_i)
\end{equation}
\begin{equation} 
    \label{precedence}
    \begin{split}
    \text{{\fontfamily{qcr}\selectfont end\_before\_start}}&(t_{i,j}, succ(t_{i,j}), comm\_cost) \\
    & \forall i \in n, j \in m
    \end{split}
\end{equation}

Constraint \eqref{dag_span} ensures that the interval variable $d_i$ encompasses all tasks $T_i$ and it is bounded by the start of the first task and the end of the last task in the DAG. 
In addition, constraint \eqref{alternative} ensures that the solution only includes one of the optional internal variables $\{t_{i,j,1},t_{i,j,2},...t_{i,j,z}\}$ to represent task variable $t_{i,j}$. 
In other words, task $t_{i,j}$ is mapped \textit{only} to one of the PEs in the system. 
Furthermore, constraint \eqref{no_overlap} ensures that tasks assigned on each PE are not scheduled in overlapping time windows. 
Finally, constraint \eqref{precedence} accounts for the overhead associated with the data transfer ($comm\_cost$) between the tasks and ensures that task $t_{i,j}$ finishes before any of the successor tasks $succ(t_{i,j})$ start their execution.

The objective function to minimize the execution time of DAGs is stated as:
\begin{equation}
  \text{{\fontfamily{qcr}\selectfont minimize}}\Big(
  \text{{\fontfamily{qcr}\selectfont sum}}\big( \text{{\fontfamily{qcr}\selectfont length\_of}}(\{d_1, d_2, ..., d_n\}) \big) \Big)
  \label{dag_objective}
\end{equation}

\begin{figure}
    \centering
    \includegraphics[width=\linewidth]{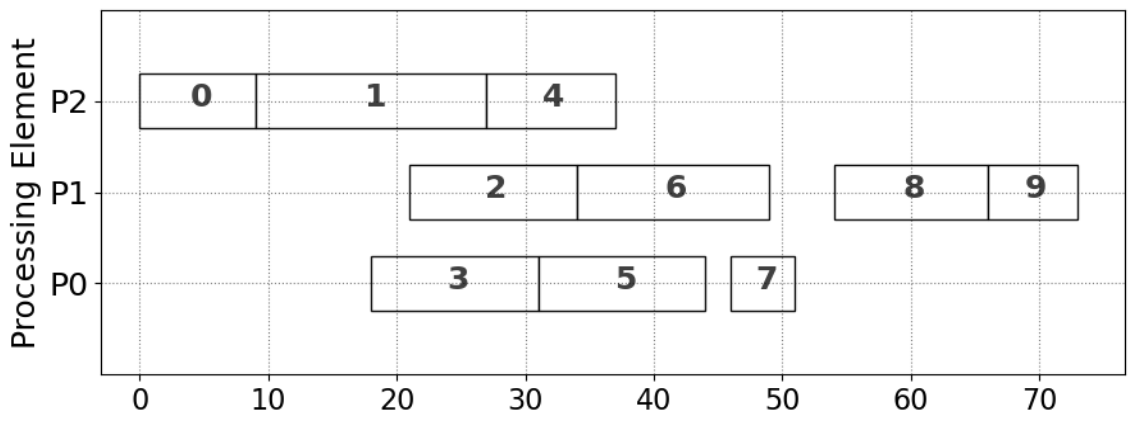}
    \caption{Gantt chart showing optimal schedule for a single instance of the canonical DAG from~\cite{Topcuoglu02}.}
    \label{fig:cp_topcuoglu_soln}
\end{figure}

If we apply this formulation to a single instance of the Canonical DAG in Fig.~\ref{fig:topcuoglu}, $D$ will contain only one DAG, $D={d_1}$, while $T_i$ will consist of ten task interval variables $T_i=\{t_{1,0}, t_{1,1}, t_{1,2}, t_{1,3}, t_{1,4}, t_{1,5}, t_{1,6}, t_{1,7}, t_{1,8}, t_{1,9}\}$. 
Then, the formulation will include, for example, three optional interval variables $\{t_{1,0,0}, t_{1,0,1}, t_{1,0,2}\}$ with a size of 14, 16, and 9, respectively, for the task interval variable $t_{1,0}$.
The simulation framework dynamically calls the CP solver at the injection of each frame to find an optimal schedule, whenever the problem size allows, as a function of the current system state. 
Then, the obtained schedule is stored in a lookup table, and tasks are assigned to the PEs accordingly as the simulation proceeds. 
For the single instance Canonical DAG example above, the CP solver finds the optimal solution as given in Fig.~\ref{fig:cp_topcuoglu_soln}.
Since the CP solver takes hours for large inputs (~100 tasks), we employed a time limit of ten minutes per scheduling decision. 
If the model fails to find an optimal schedule within the time limit, we use the best solution found. 

\edit{
With our experimental setup in place, in Section~\ref{sec:algorithmic_contributions}, we begin by introducing key challenges related to applying list scheduling heuristics in dynamic runtime environments. 
We then derive a number of empirically-motivated algorithmic changes and optimizations to address these challenges until we stabilize to a promising suite of modified algorithms.
In Section~\ref{sec:results}, we proceed to exhaustively evaluate each of these algorithms across a wide variety of workload scenarios and quantify their relative performance against a number of baseline scheduling techniques.
}

\section{Algorithmic Contributions} \label{sec:algorithmic_contributions}
\begin{figure*}[t]
	\centering
	\begin{subfigure}[b]{0.75\linewidth}
	    \centering
	    \includegraphics[width=\linewidth,trim={1cm, 8.9cm, 1cm, 0.5cm},clip]
	    {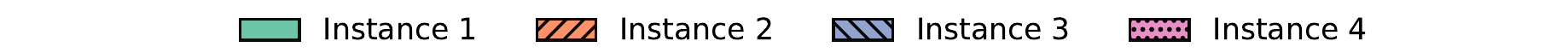}
	\end{subfigure}
	\begin{subfigure}[b]{0.49\linewidth}
		\centering
		\includegraphics[width=\linewidth, height=4.4cm] {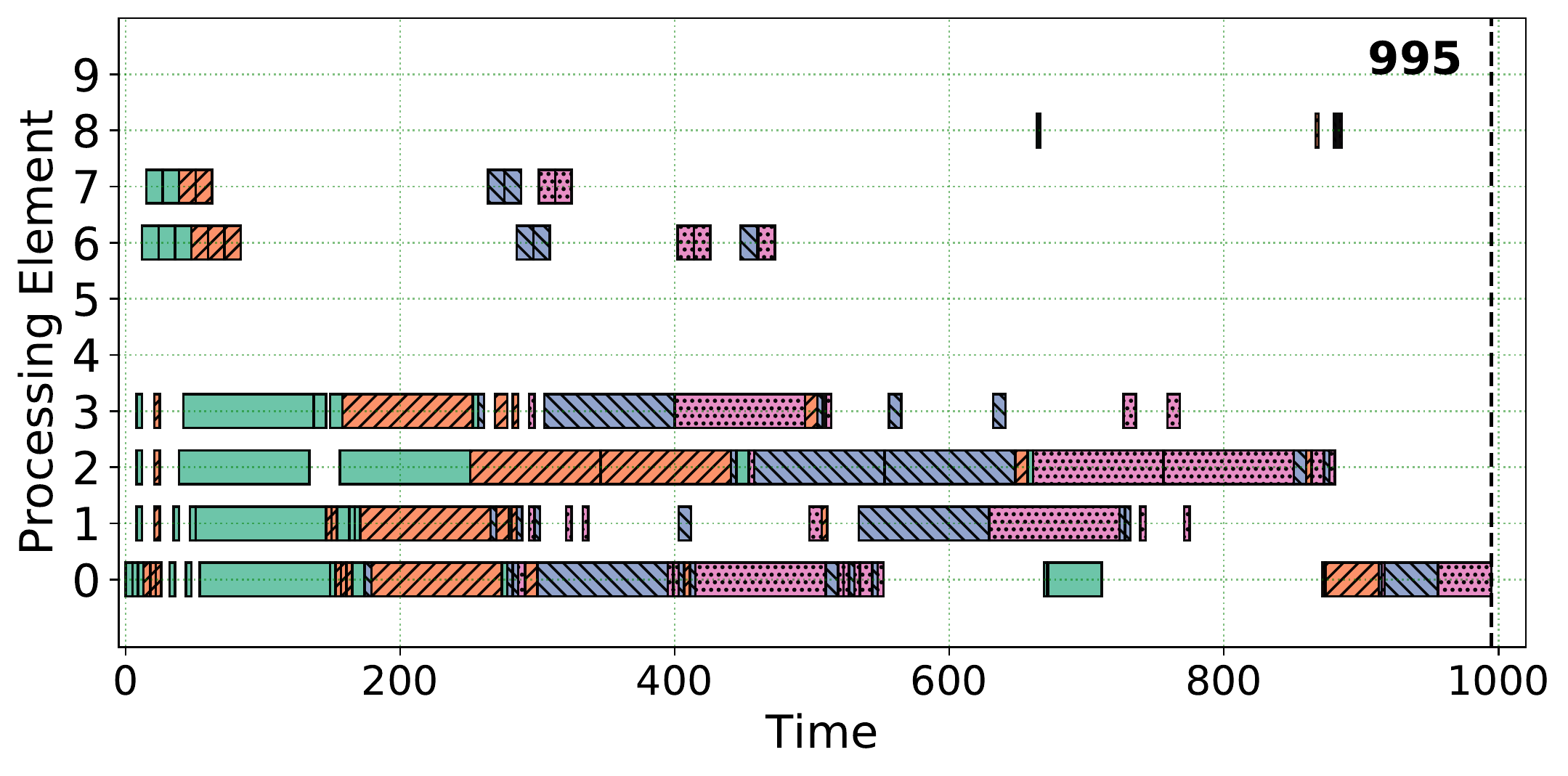}
		\subcaption{} \label{fig:opt0}
	\end{subfigure}
	\hfill
	\begin{subfigure}[b]{0.49\linewidth}
		\centering
		\includegraphics[width=\linewidth, height=4.4cm] {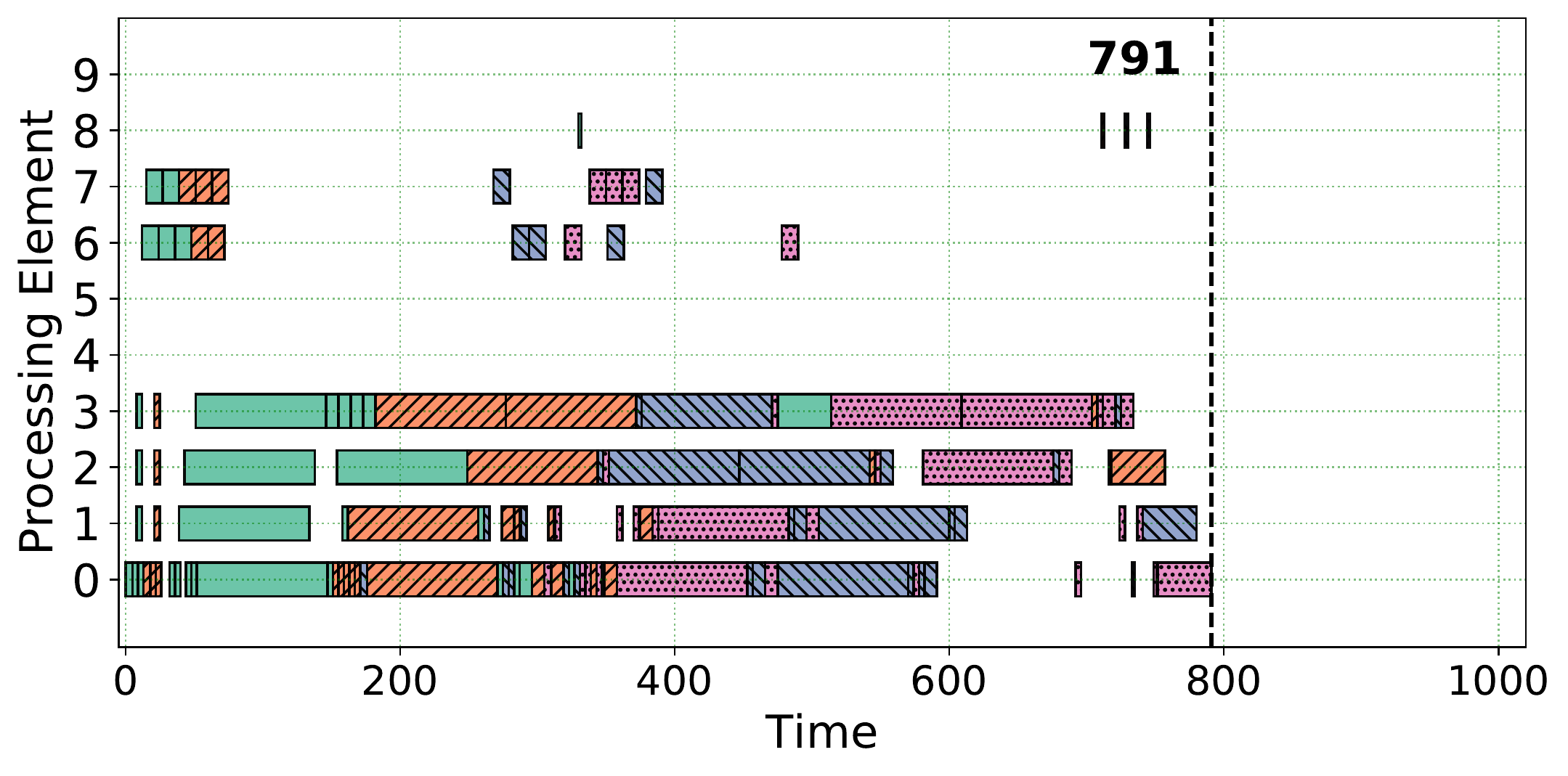}
		\subcaption{} \label{fig:opt1}
	\end{subfigure}
	\begin{subfigure}[b]{0.49\linewidth}
	    \centering
	    \includegraphics[width=\linewidth, height=4.4cm] {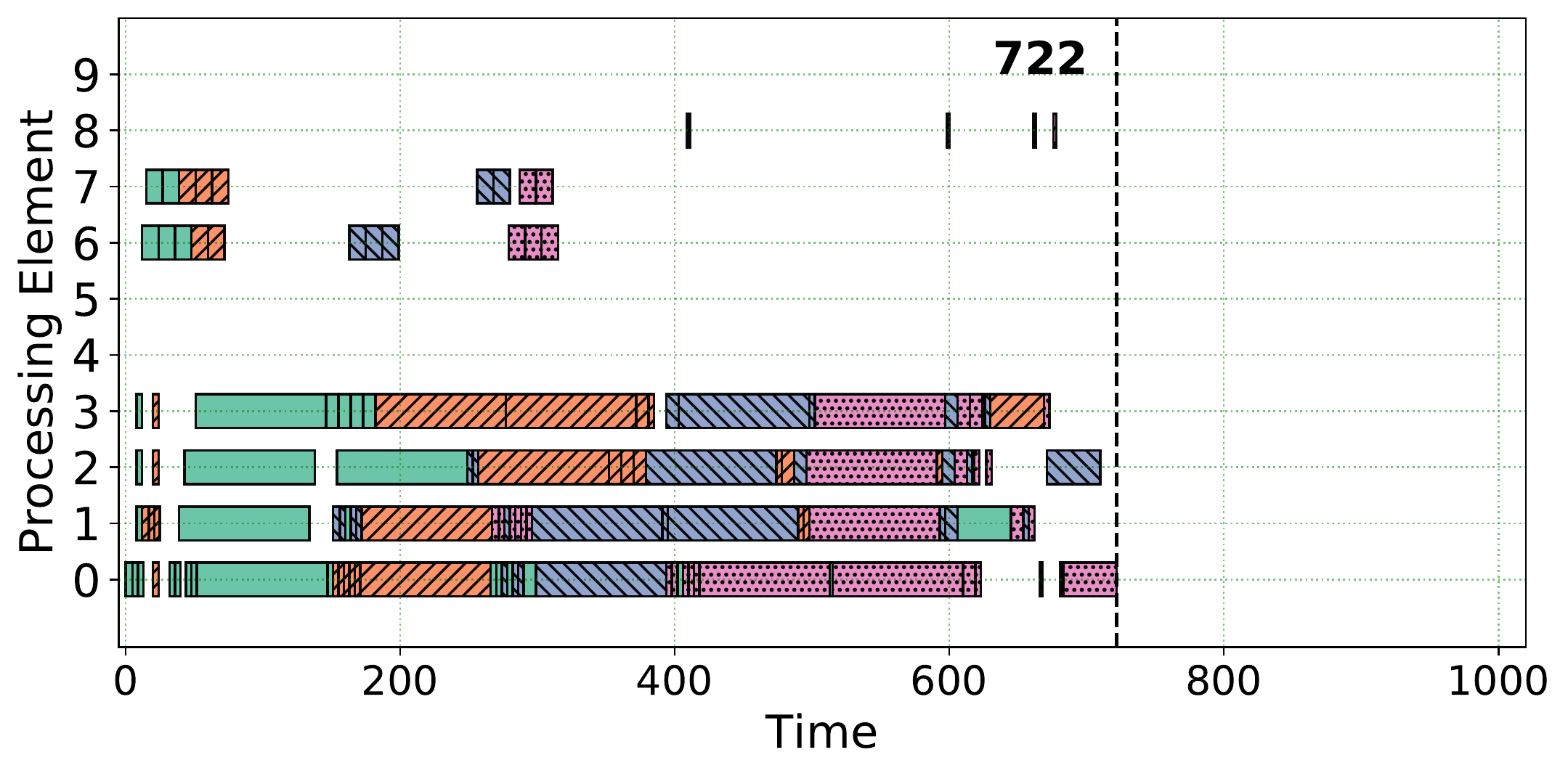}
	    \subcaption{} \label{fig:opt2}
	\end{subfigure}
	\hfill
	\begin{subfigure}[b]{0.49\linewidth}
	    \centering
	    \includegraphics[width=\linewidth, height=4.4cm] {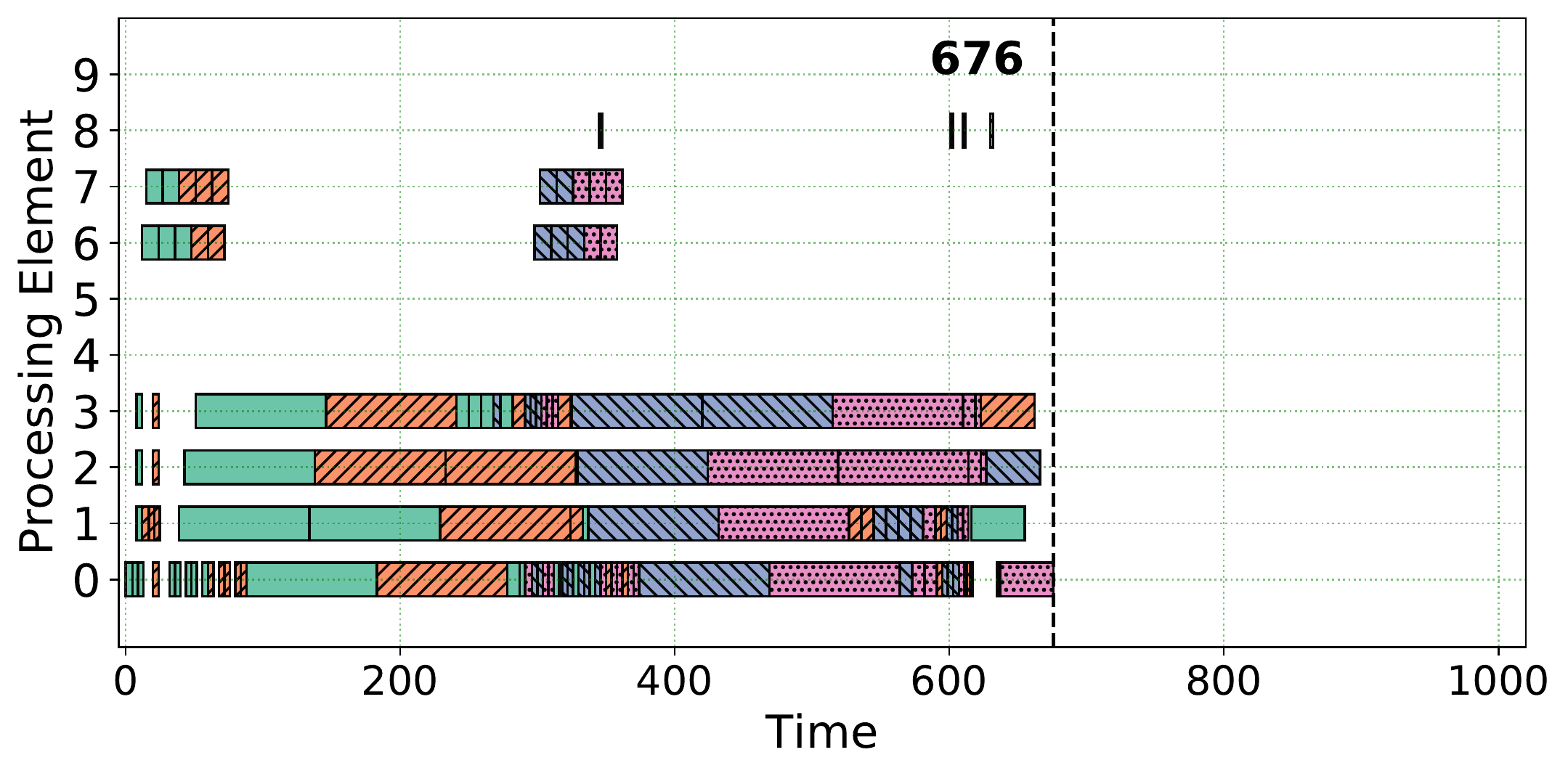}
	    \subcaption{} \label{fig:opt3}
	\end{subfigure}
	\caption{Evolution of a single workload consisting of four sequentially arriving, color-coded WiFi RX frames on SoC configuration (1) (ZCU102) from Section~\ref{sec:simsetup} across all proposed mitigation strategies. 
	(a) provides a baseline implementation, (b) builds on (a) and enables merging existing DAGs with the incoming job, (c) builds on (b) and enables fixing currently running tasks as constraints, and (d) builds on (c) and enables dynamic dependencies. Processing Elements 0-3 are ARM A53 CPUs, 4-5 are Matrix Multiply accelerators, 6-7 are FFT accelerators, and 8-9 are Viterbi decoders.}
	\label{fig:opt}
	\vspace{-6pt}
\end{figure*}

\begingroup
\subsection{Static Scheduling: Challenges \& Solutions} \label{subsec:staticvsdynamic}
\endgroup

As HEFT is a static list scheduling algorithm, we begin here by first exposing the drawbacks that come with deploying baseline HEFT~\cite{Topcuoglu02} -- referred to as HEFT\textsubscript{Base} -- in a dynamic runtime environment with interleaving applications through discussion of three key challenges.
Interleaved with those discussions, we also incrementally address each of these three challenges and demonstrate the performance benefits via execution time analysis with a workload consisting of four frames of WiFi-RX data. 
The incremental results of addressing each of these challenges is captured in Fig.~\ref{fig:opt}.

\textit{First}, in a runtime system, frames (full application DAGs) can arrive at any point in the simulation regardless of whether or not existing tasks are still being processed.
This presents a challenge to static scheduling as it disallows statically scheduling and executing each DAG serially.
Previous scheduling decisions that have yet to be fulfilled need to be considered when scheduling subsequent DAGs.
A simple approach to handle this is to consider all previous scheduling assignments that have been generated as fixed and use them as constraints on the scheduling of the incoming frame.
However, this choice can easily lead to suboptimal assignment of resources.
As there are potentially a large number of previously scheduled tasks that have not had their execution dependencies satisfied yet, they can clearly be reassigned. 

Instead, the proposed approach maintains a DAG representing the current system state, excluding tasks that are currently executing, and uses the Common Entry and Common Exit method proposed by Zhao et al.~\cite{Zhao06} to merge the newly arrived DAG with the current system state.
Ranking-based composition was also tested, but it was not found to be beneficial relative to its increased complexity.
This enables the scheduler to balance the needs of the incoming DAG with the needs of the outstanding task nodes in the system.
Without DAG merging, HEFT is only provided a local/application-level view for each DAG in the system. As HEFT is a stateless algorithm, this then means that 4 instances of a single application back-to-back all produce identical schedules and identical executions as shown in Fig.~\ref{fig:opt0}.
With DAG merging, HEFT is then provided a global view of all applications currently being processed by the system, and it can make more informed decisions based on interleaving the execution of independent applications more intelligently to better utilize the system resources.
This leads in a reduction in the achieved makespan from 995 to 791, a 20.5\% improvement compared to Fig.~\ref{fig:opt0}.

\textit{Second}, many static scheduling approaches like HEFT assume that the underlying platform they are scheduling on is idle, and as such, all time slots are available for scheduling.
In runtime systems, however, this is not necessarily the case.
Hence, we need to ensure that currently running tasks are represented as constraints on the scheduling problem so that the algorithm doesn't attempt to schedule in an occupied time slot.
As it turns out, HEFT in particular is quite amenable to the integration of running task constraints.
They can simply be excluded from the ``$rank_u$" list prioritization phase and treated as tasks that have already been assigned to resources in the main ``pop and schedule" loop from Fig.~\ref{fig:list_scheduler_process}.
With these constraints in place, the remaining standard processor assignment algorithm can be invoked.
Taken together, these first two issues solve two complementary problems with regards to representing the current system state: solving the first issue allows the scheduler to have a global view of all the unscheduled tasks in the system while solving the second issue allows the scheduler to also have a global view of all currently executing tasks in the system. 
In total, by addressing these two issues, a stateless list scheduler can be provided with all the context needed to have a holistic view of the system state when performing scheduling decisions.

Compared to Fig.~\ref{fig:opt1}, as shown in Fig.~\ref{fig:opt2}, running task constraints allow for earlier execution of the FFTs on the FFT accelerators (PEs 6 and 7) near timestep 300.
This allows the remainder of the instance 3 and instance 4 nodes to be completed earlier to reduce the overall makespan. 
As a result we observe a drop in the achieved makespan from 791 to 722, an 8.7\% reduction over Fig.~\ref{fig:opt1} and 27.4\% overall compared to Fig.~\ref{fig:opt0}.

\textit{Third}, our last key challenge in applying static scheduling policies in a runtime environment centers around ensuring, after scheduling is complete, that the generated schedule is executed faithfully by the runtime framework.
The output of a static schedule is traditionally thought of as a lookup table mapping tasks in the runtime to their associated processing elements.
As each task has its dependencies satisfied, the runtime then checks this table and dispatches based on its guidance.
However, as these schedulers are scheduling before tasks become ready, they do not know exactly when two independent tasks, for example A and B, will have their predecessors completed.
Despite this, the scheduler may want to enforce a particular execution order of those two tasks on the same PE to, i.e., prioritize the DAG's critical execution path.
Meanwhile, at runtime, variation in the completion of dependencies of these tasks could cause the runtime to dispatch differently from the requested order due to i.e. discrepancies in estimating task execution time. 
Supposing that B has its predecessors met slightly before A, the runtime could attempt to launch B before A despite the scheduler requesting that A runs first and is followed by B.
If B is a particularly long running task and A has many dependent nodes, this can cause massive divergences between the static schedule execution and observed execution.
To address this, static schedulers must produce, along with their resource allocation decisions, a set of \textit{dynamic dependencies} that the runtime can use to ensure that the desired execution order is maintained.
These dynamic dependencies are then verified in the same fashion as any other dependencies by the runtime with the exception that they may be modified after any \textit{Job Generation} event (shown in Fig.~\ref{fig:framework}), so special care must be taken to ensure that no old data from previous scheduling epochs causes cyclic dependencies and subsequently system deadlock.

We observe the result of enabling dynamic dependencies in Fig.~\ref{fig:opt3}.
Compared to Fig.~\ref{fig:opt2}, this further compresses the node executions together, filling nearly all the gaps present in the Gantt chart.
It does this by increasing the amount of application interleaving performed, with nodes from all 4 DAG instances spread across the entire execution.
Notably, this change leads to a large delay in the execution of instance 1. However, the overall execution of all four DAGs together is still able to finish sooner than it was able to previously despite this delay.
This leads to a reduction in the achieved makespan from 722 to 676, a 6.3\% reduction over previous and a 32.1\% reduction overall compared to the baseline.
We refer to this implementation as HEFT\textsubscript{Dyn}.

\subsection{Algorithmic Optimizations of HEFT\textsubscript{Dyn}}

\textit{Finally}, as an optimization on top of HEFT\textsubscript{Dyn}, we integrated it directly with the \textit{Scheduler} interface of Fig.~\ref{fig:framework}.
This is distinct from HEFT\textsubscript{Dyn} and HEFT\textsubscript{Base} as they have all been integrated through the \textit{Job Generation} interface where the full application DAG is available.
Scheduling decisions for these previous schedulers were then written to a lookup table for later reference by the \textit{Scheduler} component.
When integrated with the \textit{Scheduler} component directly, HEFT\textsubscript{Dyn} is only provided the ready queue of tasks at each scheduling epoch.
This lowers the work per scheduling decision as it introduces a number of opportunities to simplify the HEFT algorithm. 
By the nature of the ready queue, all tasks are independent, so ``$rank_u$" calculation reduces to prioritizing based on the mean computation time.
Consequently, operations that depend on graph traversal in HEFT -- like the upward ranking calculation or ready time availability during processor assignment -- can be simplified down to a single set of independent tasks.
Because this implementation only utilizes information available at runtime rather than information about the full DAG to make its scheduling decisions, we refer to this \textit{runtime} implementation as HEFT\textsubscript{RT}.

\begin{figure}[tb]
	\centering
	\begin{subfigure}[]{0.75\linewidth}
	    \centering
	    \includegraphics[width=\linewidth]{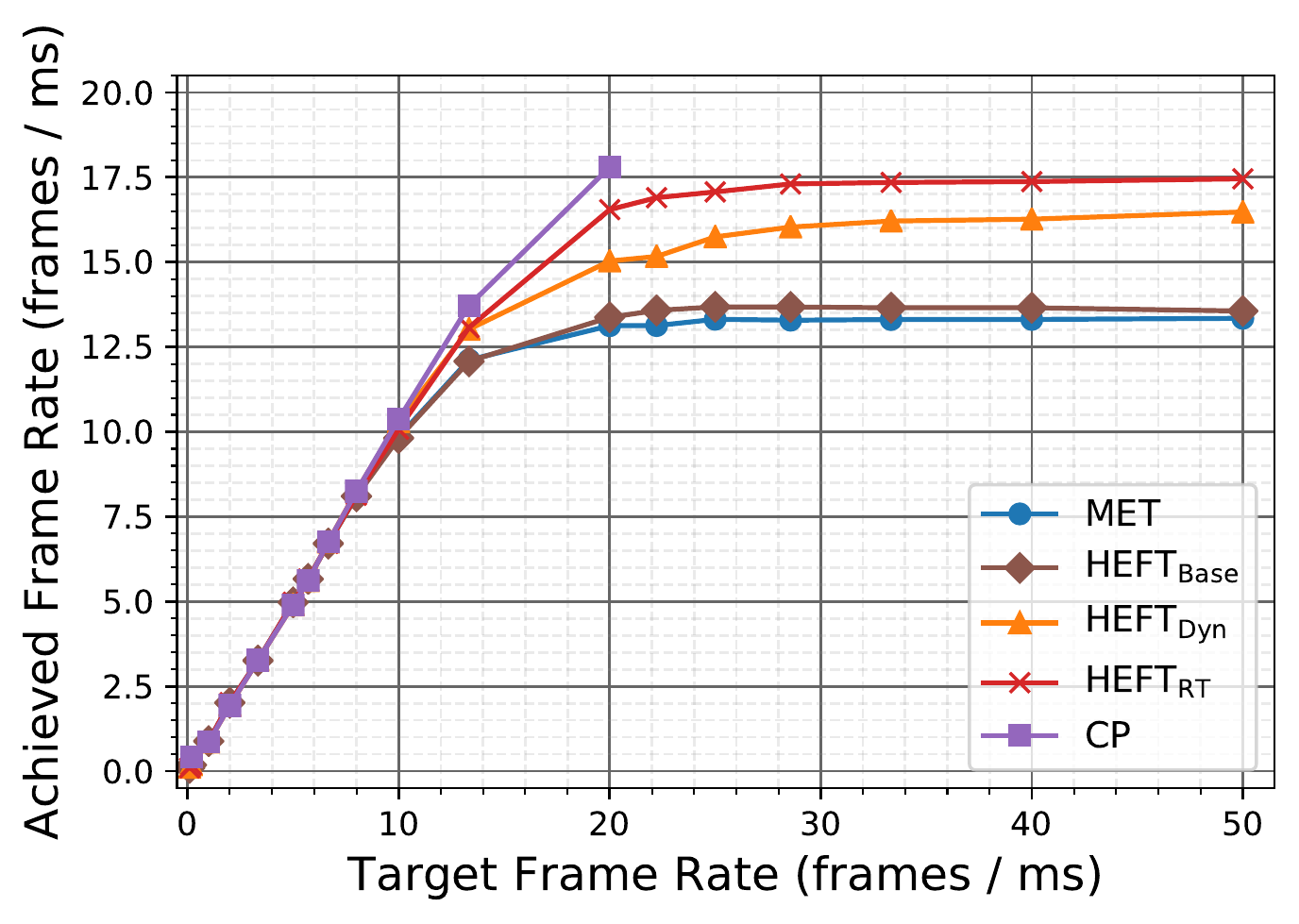}
	    \subcaption{}
	    \label{fig:baseline_saturation}
	\end{subfigure}
	\begin{subfigure}[]{0.75\linewidth}
	    \centering
	    \includegraphics[width=\linewidth]{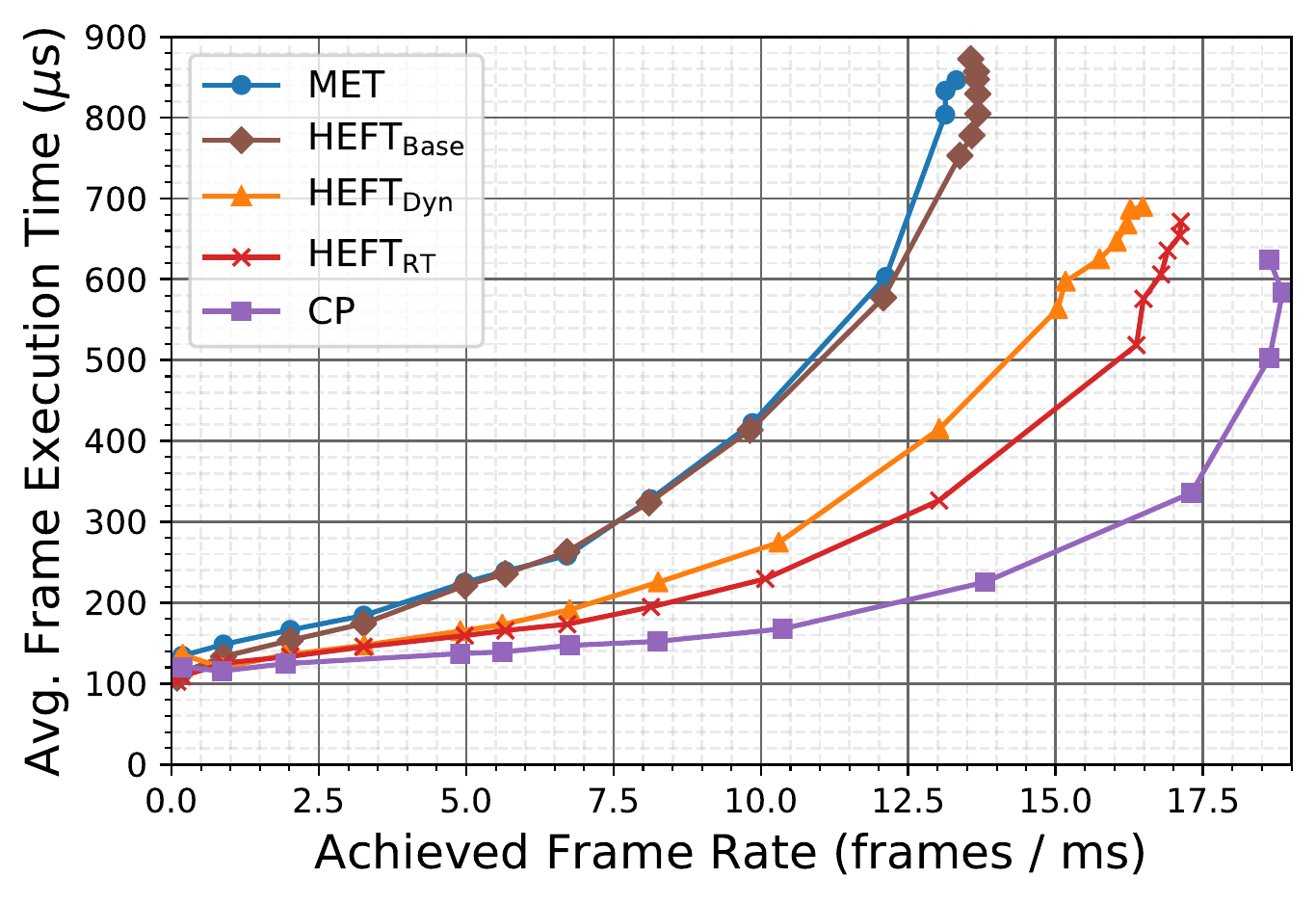}
	    \subcaption{}
	    \label{fig:baseline_exec}
	\end{subfigure}
	\caption{
	Comparison of schedulers on SoC Configuration (1) (ZCU102) from Section~\ref{sec:simsetup}. (a) Target frame rate versus achieved frame rate per scheduler -- i.e., at a target frame rate of 20 frames/ms, HEFT\textsubscript{Dyn} achieved 15 frames/ms. (b) Achieved frame rate versus average frame execution time.
	}
	\vspace{-4mm}
	\label{fig:baseline}
\end{figure}

\subsection{Scheduler Overhead and Throughput Analysis} \label{subsec:execution_time_analysis}
Next, we extend our analysis of HEFT\textsubscript{Base} against HEFT\textsubscript{Dyn} and HEFT\textsubscript{RT} using a more demanding workload scenario. 
We contextualize the results with DS3's built-in MET and CP schedulers in Fig.~\ref{fig:baseline}.
The CP scheduler sets an effective lower bound on frame execution time with the understanding that lower times are potentially possible if CP were given longer to search the solution space.
We generated a sample workload consisting of an 80\%/20\% mixture of WiFi-TX/WiFi-RX flow graphs and evaluated the average execution time for each frame in the system as a function of the frame rate.
As our SoC configuration, we modeled configuration (1) (ZCU102) from Section~\ref{sec:simsetup}.
In Fig.~\ref{fig:baseline_saturation}, we observe the ability of each scheduler to cope with the target frame rate as described in Section~\ref{sec:simsetup}.
The x-axis presents the target frame rate set in DS3 and the y-axis presents the achieved frame rate for each of the schedulers.
We see that, up to 10 frames/ms, each scheduler is able to handle the incoming workload effectively and therefore the plot of the achieved frame rate is linear.
Above 10 frames/ms, though, each scheduler begins to be unable to cope with the rate of frame arrival, and the achieved frame rate experiences a drop for each scheduler.
For instance, the CP scheduler achieves a frame rate of 17.5 frames/ms at a target frame rate of 20 frames/ms.
We see in these plots that HEFT\textsubscript{Base} performs nearly identically to the greedy MET scheduler, saturating near 13.5 frames/ms. 
After resolving each of the challenges discussed above, HEFT\textsubscript{Dyn} saturates at 16.1 frames/ms, a 22\% higher throughput than HEFT\textsubscript{Base}.
Also, we see that HEFT\textsubscript{RT} extends the saturation to 17.5 frames/ms, a 32\% higher throughput than HEFT\textsubscript{Base}.

Additionally, Fig.~\ref{fig:baseline_exec} illustrates the average frame execution time observed at each of the achieved frame rates.
In this plot, the x-axis presents the achieved frame rate (the y-axis from Fig.~\ref{fig:baseline_saturation}) against the average frame execution time for frames executed by each particular scheduler.
As expected based on the previous saturation plot, HEFT\textsubscript{Base} consistently performs worse than HEFT\textsubscript{Dyn} and HEFT\textsubscript{RT}. 
However, the performance suffers not only in level of throughput achieved but also in the quality of the schedule per application frame.
We observe that, at saturation, the average frame scheduled by HEFT\textsubscript{Base} takes nearly 750 microseconds while HEFT\textsubscript{Dyn} takes 680 and HEFT\textsubscript{RT} 640 -- 10\% and 15\% reductions respectively.
As such, we find that these algorithms improve not only application throughput, but they do so by improving the quality of the generated schedules.

Upon initial observation, it is somewhat surprising that HEFT\textsubscript{RT} outperforms HEFT\textsubscript{Dyn}, particularly at high frame rates, given that the primary design goal of HEFT\textsubscript{RT} was to optimize algorithmic efficiency.
While one might assume that HEFT\textsubscript{Dyn} would take advantage of having the full application DAG to make less greedy decisions that consider impacts to successor nodes when scheduling a particular node, HEFT by itself does not perform these kinds of calculations.
These "lookahead"-style modifications are the subject of other works~\cite{Bittencourt10,Arabnejad14}.
As such, neither HEFT nor HEFT\textsubscript{RT} take advantage of any lookahead-style considerations, and HEFT\textsubscript{RT}, by nature of being called at the moment of scheduling, is always provided the most up to date information with regards to system execution state when it performs scheduling.
Finally, by nature of each individual scheduling problem being smaller, HEFT\textsubscript{RT} schedules in a much smaller combinatorial space.

\begin{figure}[tb]
    \centering
    \includegraphics[width=\linewidth]{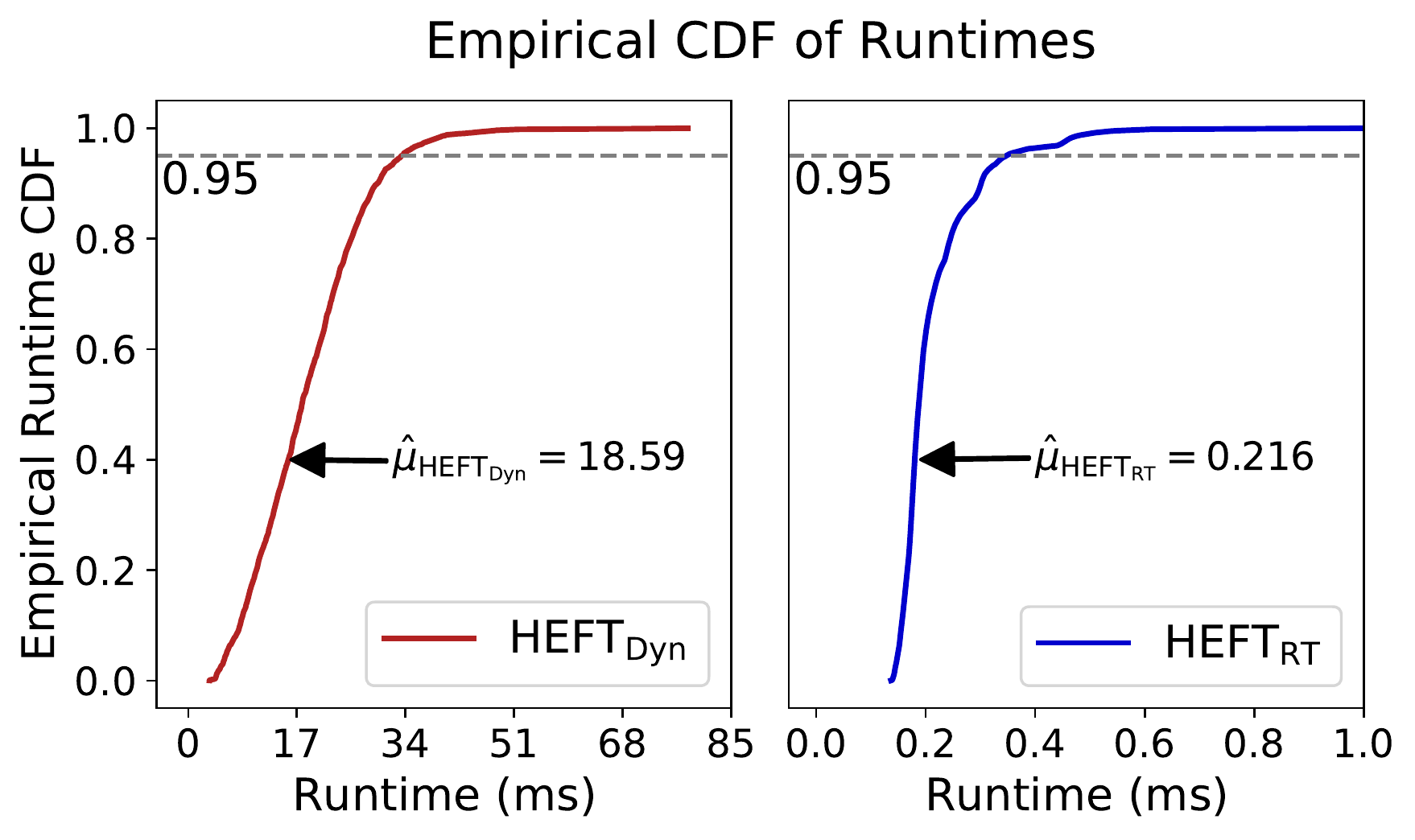}
    \caption{Empirical runtime CDFs for HEFT\textsubscript{Dyn} versus HEFT\textsubscript{RT} across an even mixture of all 6 DS3 applications with a total of 2338 executed application instances.}
    \label{fig:cdf_dyn_rt}
    \vspace{-8pt}
\end{figure}

With this analysis in mind, it is worth exploring whether HEFT\textsubscript{RT} achieved its primary design goal: optimizing algorithmic efficiency and reducing scheduling overhead.
To explore this, we profiled the execution times of each implementation inside the DS3 simulation environment.
We generated a workload consisting of an even mixture of all 6 applications available in the DS3 environment in order to ensure variety, and we swept the target frame rate from 0.1 to 10 frames/ms. 
Across this full sweep, we executed a total of 2338 instances of each of these application DAGs, profiling the scheduling calls made by HEFT\textsubscript{Dyn} and HEFT\textsubscript{RT} during execution.
As the host system can affect this performance, in this instance, DS3 was executed on Ubuntu 18.04LTS with an Intel Core i7 8700.
Fig.~\ref{fig:cdf_dyn_rt} displays the empirical cumulative distribution function (CDF) of the runtimes observed.
We observe that the average HEFT\textsubscript{Dyn} call requires 18.59 ms while the average HEFT\textsubscript{RT} call requires only 0.216 ms, and 95\% of HEFT\textsubscript{Dyn} calls fall below 33.36ms while 95\% of HEFT\textsubscript{RT} calls fall below 0.347ms.
Taken together, we find that, on average, HEFT\textsubscript{RT} observes an 86x speedup.
This does not take into account differences in call frequency, however: HEFT\textsubscript{Dyn} is called once for each application DAG while HEFT\textsubscript{RT} is called every time the ready queue requires scheduling.
We find that HEFT\textsubscript{Dyn} is called 2338 times while HEFT\textsubscript{RT} is called 27780 times. 
When this is multiplied by the average execution to estimate total time spent scheduling, we observe that HEFT\textsubscript{Dyn} spent 43.46 seconds scheduling while HEFT\textsubscript{RT} spent 6.00 seconds, still yielding a total overall speedup of 7.24x for an equivalent workload.
Together with Fig.~\ref{fig:baseline}, we see that HEFT\textsubscript{RT} provides better scheduling performance with lower scheduling overhead.
Therefore, we will use this algorithm as the foundation for development of energy aware scheduling policies.

\subsection{Energy Aware Scheduling}

To develop an energy-aware variant of HEFT, we start by redesigning the ranking method. 
Specifically, we take HEFT\textsubscript{RT} as a baseline -- given its excellent execution-time performance -- and modify the original upward rank calculation from Section~\ref{sec:background} to incorporate a $v\times q$ power cost matrix $P$ to complement HEFT's $v\times q$ computation cost matrix $W$, where $v$ is the number of tasks and $q$ is the number of PEs. 
To enable this, we utilized power consumption measurements on hardware as explained in Section~\ref{sec:simsetup}. 
If analytical power models are available, estimates from those can be tabulated into a power cost matrix similarly.
In this context, let the average power cost of a task $n_i$ be $\bar p_i = \sum\limits_{j=1}^{q}\frac{P_{i, j}}{q}$, where $P_{i,j}$ is the power consumption of task $n_i$ on PE $j$.
Using this power matrix, we define a new rank metric as:
\begin{equation} \label{eq:ranku_edp}
rank_{u}^{edp}(n_i) = \bar{w_i}^2 \bar{p_i} + \max\limits_{n_j\in succ(n_i)}(\bar{c_{ij}} + rank_{u}^{edp}(n_j))
\end{equation}
where $\bar{w_i}$ is the average computation cost for task $n_i$, $\bar{p_i}$ is the average power cost for task $n_i$, $\bar{w_i}^2 \bar{p_i}$ is the average EDP consumption of task $n_i$, $succ(n_i)$ is the set of immediate successors of task $n_i$, and $\bar{c_{ij}}$ is the average communication cost of edge $(i,j)$.
During the task prioritization phase of HEFT, this metric enables ranking of tasks based on their impact on DAG EDP.
\edit{
In practice, we apply this new upward rank formulation to HEFT\textsubscript{RT}.
It is then worth noting that, because HEFT\textsubscript{RT} only schedules one ready queue worth of tasks per execution, its DAG effectively consists of disconnected, independent tasks that have no successors.
Consequently, the $\max_{n_j\in succ(n_i)}$ term in Eq.~\ref{eq:ranku_edp} does not influence the computation's result.
}

\begin{algorithm}[t]
	\SetAlgoLined
	\SetKwInOut{Input}{Input}
	initialize task list with $rank_{u}^{edp}$-based ranking\;
	\While{there are unscheduled tasks in the task list}{
		pop task $n_{i}$ from list for scheduling\;
		$minEDP = inf$\;
		\For{each processor $p_{k}$}{
			$sched = EFT(n_{i}, p_{k})$\;
			$edp = (sched.end - sched.start)^{2} * P[n_{i}, p_{k}]$\; \label{alg1:schedStart} 
			\If{$edp < minEDP$}{
				$minEDP = edp$; $minSched = sched$\;
			}
			\ElseIf{$edp == minEDP$ \&\& $sched.end < minSched.end$}{
				$minSched = sched$\;
			}
		}
		assign $n_{i}$ according to $minSched$\;
	}
	\caption{HEFT\textsubscript{EDP}}
	\label{alg:heft_assignment_edp_v1}
\end{algorithm}

\begin{figure}[t]
	\centering
	\begin{subfigure}[]{0.75\linewidth}
	    \centering
	    \includegraphics[width=\linewidth]{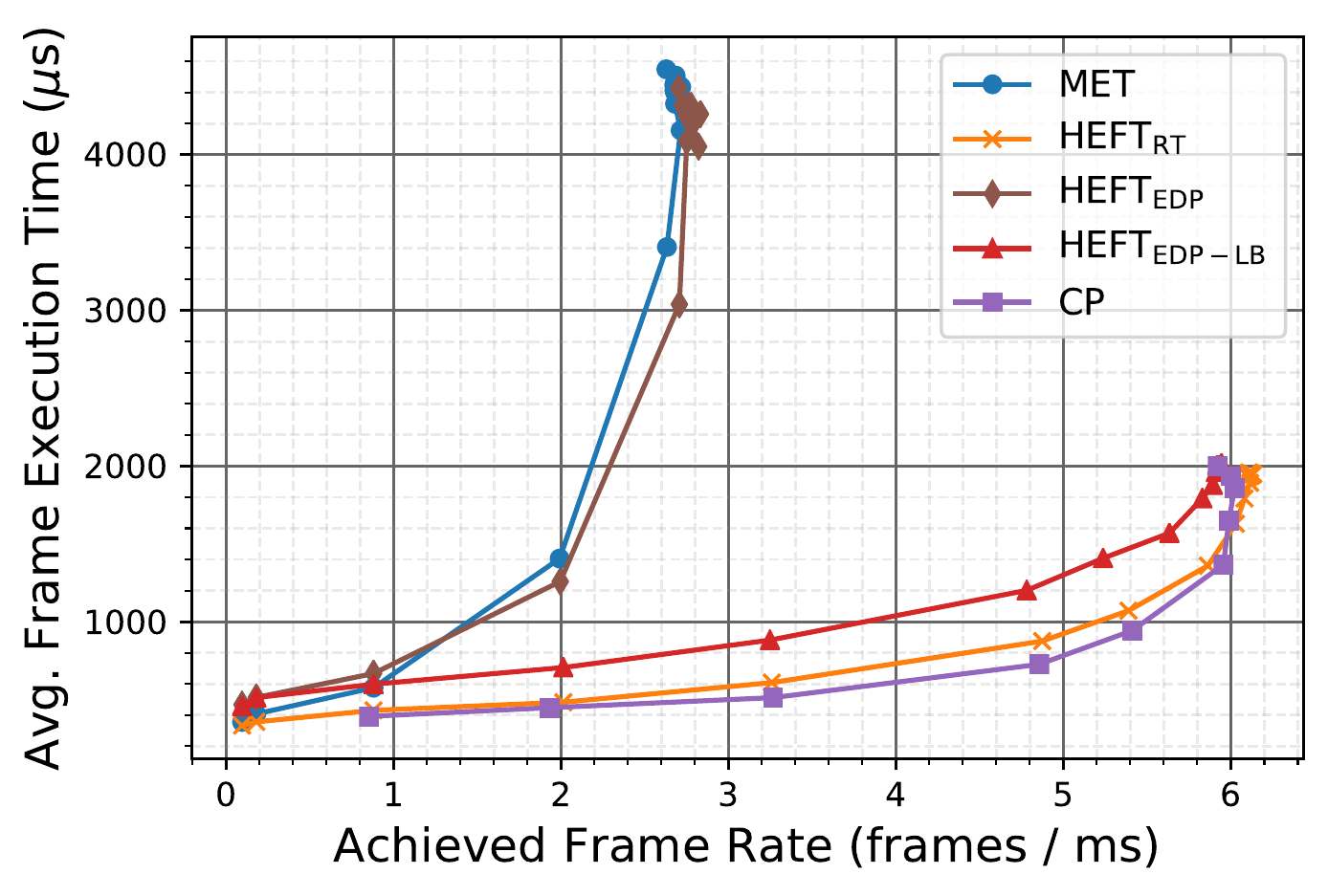}
	    \subcaption{}
	    \label{fig:energy_results_exec}
	\end{subfigure}
	\begin{subfigure}[]{0.75\linewidth}
	    \centering
	    \includegraphics[width=\linewidth]{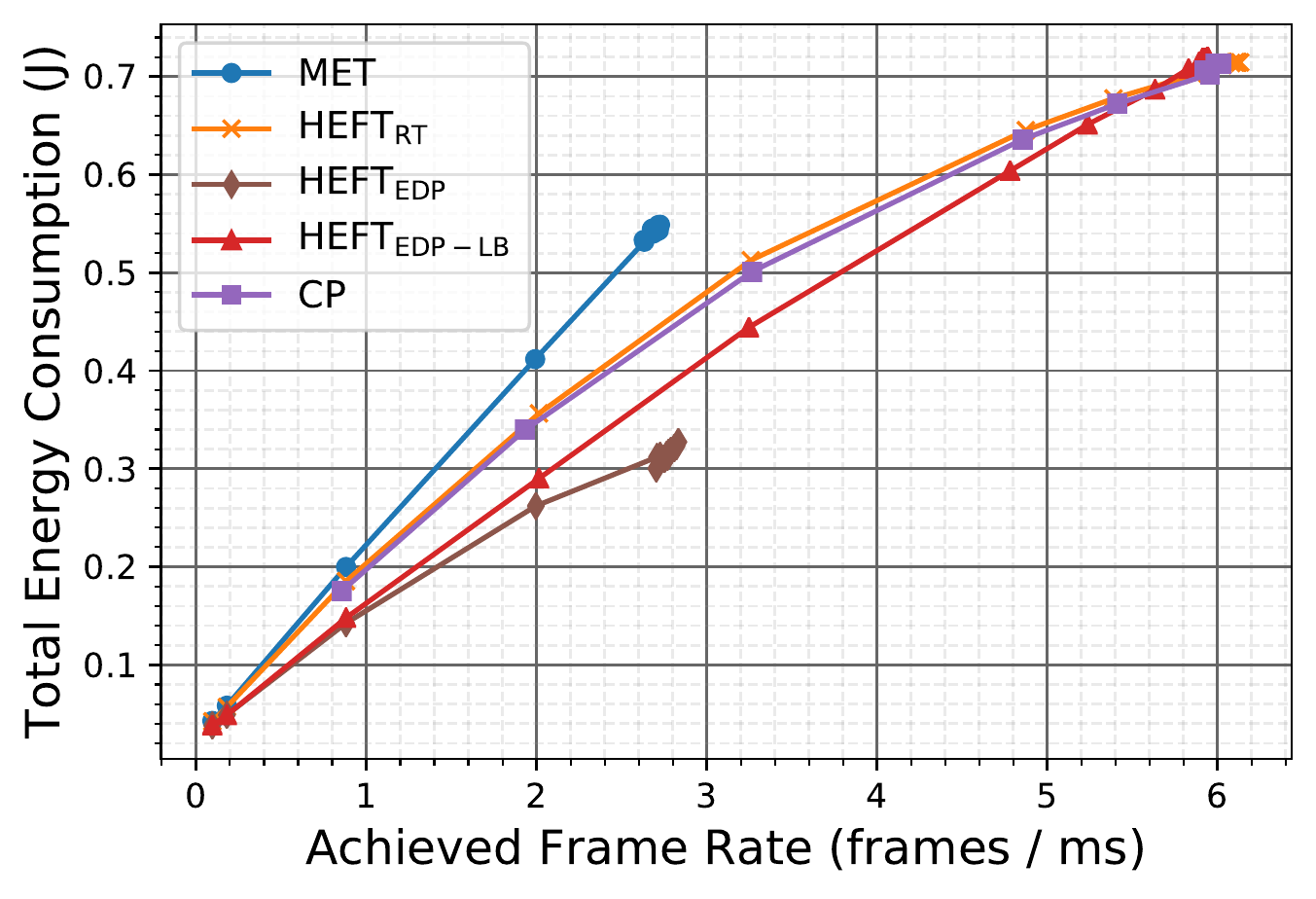}
	    \subcaption{}
	    \label{fig:energy_results_energy}
	\end{subfigure}
	\caption{(a) average frame makespan and (b) total energy versus achieved frame rate in a workload composed of 80\% WiFi TX, 20\% WiFi RX on configuration (2) (Odroid XU3) from Section~\ref{sec:simsetup}.}
	\label{fig:energy_results}
	\vspace{-6mm}
\end{figure}

With a new rank metric in place, we adjust the resource allocation phase similarly to assign tasks to PEs that minimize application EDP.
The first approach taken in this work is to perform resource assignment to the processor that minimizes task EDP, regardless of how much later an execution slot is chosen.
We refer to this implementation as HEFT\textsubscript{EDP} as shown in Algorithm~\ref{alg:heft_assignment_edp_v1}.
Based on the simulation and workload setup described in Section~\ref{sec:simsetup}, we test this algorithm and illustrate the findings in Fig.~\ref{fig:energy_results}.
In addition to average execution time, shown in Fig.~\ref{fig:energy_results_exec}, we plot total energy versus frame rate in Fig.~\ref{fig:energy_results_energy}, where the total energy is obtained by accumulating static and dynamic energy consumption at each target injection rate.
Like Fig.~\ref{fig:baseline}, the workload is a mixture of 80\% WiFi TX, 20\% WiFi RX, but we expand our evaluation to SoC configuration (2) (Odroid XU3) from Section~\ref{sec:simsetup} that is composed of 4 ARM A7 LITTLE cores and 4 ARM A15 big cores.
Many trends observed in Fig.~\ref{fig:baseline_exec} continue to hold for this configuration, with the key difference being the saturation point.
Compared to Fig.~\ref{fig:baseline_exec}, saturation occurs sooner because the SoC configuration is now composed of 4 big and 4 LITTLE CPU cores, and without accelerators, each individual frame takes longer on average.
It is worthwhile to consider this SoC configuration because, as explained in Section~\ref{sec:simsetup}, we are able to extract accurate power estimates for this SoC by using an equivalent Odroid-XU3 development board.
As such, this SoC configuration is more suitable for development of energy aware scheduling. 
Despite the drop in saturation point, HEFT\textsubscript{RT} performs quite well and is both far from MET and near CP throughout the full range of achieved frame rates.
We find that the gap between CP and HEFT\textsubscript{RT} is narrower in this instance, and this could be caused by two possibilities. 
First, it could be that there aren't any better solutions available.
Second, it could be that, on this SoC configuration, finding a schedule with constraint programming is actually more difficult. 
By their nature, accelerators are restricted in the tasks they can execute, so swapping accelerators for CPU cores actually produces more valid solutions in the combinatorial search space and increases the likelihood that CP will timeout before finding a true optimum.

Looking at HEFT\textsubscript{EDP}, we observe that while it performs similarly to MET in frame execution, it does provide the most energy efficient scheduling decisions, with a maximum energy reduction relative to HEFT\textsubscript{RT} of 58.0\% and an average reduction of 46.2\% across all frame rates.
HEFT\textsubscript{EDP} prioritizes the PE that contributes to EDP reduction most. 
Therefore, it is not sensitive to execution time-driven task to PE mapping decisions. 
Due to the increased average execution time per frame compared to HEFT\textsubscript{RT}, fewer frames get completed as the number of frames increases. 
This results in achieved throughput for HEFT\textsubscript{EDP} saturating earlier near 2.8 frames/ms compared to HEFT\textsubscript{RT} as shown in Fig.~\ref{fig:energy_results_energy}.
Next, we illustrate the cause of this limitation of HEFT\textsubscript{EDP} with an example and introduce our solution.

\begin{figure}[t]
	\centering
	\begin{subfigure}[b]{0.56\linewidth}
		\centering
		\includegraphics[width=\linewidth]{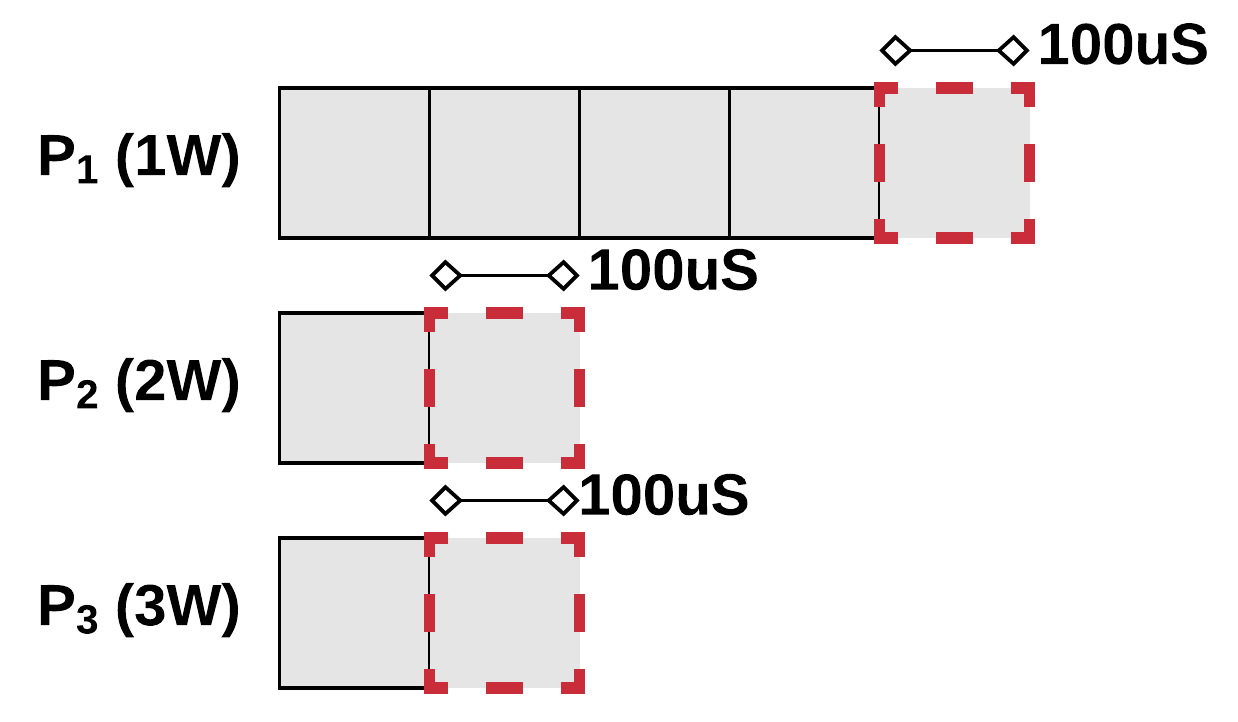}
		\subcaption{HEFT\textsubscript{EDP}} \label{fig:edp_proc_assign_a}
	\end{subfigure}
	\hfill
	\begin{subfigure}[b]{0.42\linewidth}
		\centering
		\includegraphics[width=\linewidth]{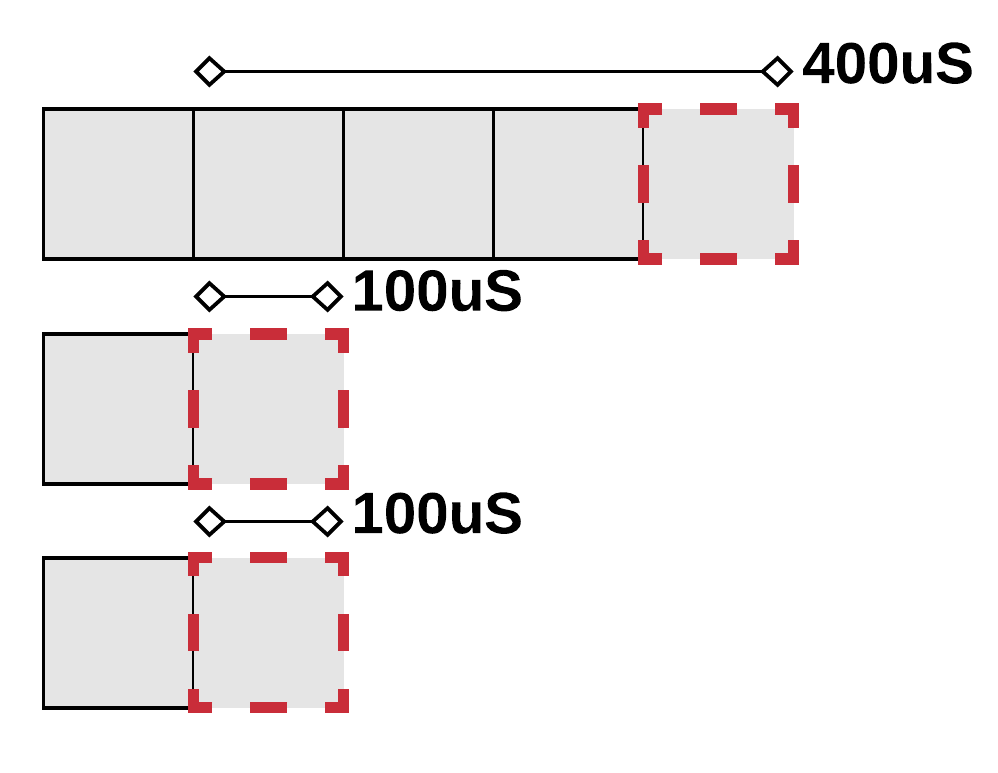}
		\subcaption{HEFT\textsubscript{EDP-LB}} \label{fig:edp_proc_assign_b}
	\end{subfigure}
	\caption{Illustration of the processor assignment heuristics for both EDP-aware methods. The red task is being considered for placement on three different processors. In the first case, P1 is chosen despite over subscription, but in the second case, P2 is chosen as a compromise between P1 and P3.}
	\vspace{-4mm}
\end{figure}

The poor scaling of HEFT\textsubscript{EDP} can be explained through Fig.~\ref{fig:edp_proc_assign_a}.
Assume there are three processing elements operating at 1 Watt, 2 Watts, and 3 Watts respectively.
Suppose the assigned workload for each PE is given by the solid grey boxes, and the goal is to decide on which PE to assign the new task highlighted in red.
On each PE, the estimated makespan is the same: $100{\mu}s$.
In standard EFT-based scheduling, P\textsubscript{2} and P\textsubscript{3} finish at an earlier time of $t=200{\mu}s$, so the heuristic would choose one of them.
HEFT\textsubscript{EDP} observes that, in a local sense, the EDP of this task can be minimized by assignment to P1 regardless of how much later the frame executes.
This has the effect of elongating the runtime for the full application's DAG.
As the heuristic disregards the absolute end time of each node and only considers the relative makespan in combination with power usage of the PE, queues for the most EDP-efficient PEs in the system grow unbounded while PEs that are less optimal remain idle, leading to an unbalanced workload distribution.

\begin{algorithm}[b]
	\SetAlgoLined
	\SetKwInOut{Input}{Input}
	initialize task list with $rank_{u}^{edp}$-based ranking\;
	\While{there are unscheduled tasks in the task list}{
		pop task $n_{i}$ from list for scheduling\;
		$minStart = minEDP = inf$\;
		\label{alg2:loop_start} \For{each processor $p_{k}$}{ 
			$sched = EFT(n_{i}, p_{k})$\;
			$schedules[p_{k}] = sched$\;
			\If{$sched.start < minStart$}{
				$minStart = schedules[p_{k}].start$\;
			}
		}\label{alg2:loop_end}
		\For{each processor $p_{k}$}{
			$sched = schedules[p_{k}]$\;
			\label{alg2:minstart_based_edp}
			$edp = (sched.end - minStart)^{2} * P[n_{i}, p_{k}]$\;
			\If{$edp < minEDP$}{
				$minEDP = edp$; $minSched = sched$\;
			}
			\ElseIf{$edp == minEDP$ \&\& $sched.end < minSched.end$}{
				$minSched = sched$\;
			}
		}
		assign $n_{i}$ according to $minSched$\;
	}
	
	\caption{HEFT\textsubscript{EDP-LB}}
	\label{alg:heft_assignment_edp_v2}
\end{algorithm}

To solve this issue by incentivizing load balancing, we propose a second heuristic, referred to as HEFT\textsubscript{EDP-LB}, as shown in Algorithm~\ref{alg:heft_assignment_edp_v2}.
This heuristic measures the delay used in EDP calculation relative to the earliest possible starting time of the new task across all processors.
This is illustrated through the loop on lines \ref{alg2:loop_start}-\ref{alg2:loop_end} in Algorithm~\ref{alg:heft_assignment_edp_v2} that define the \texttt{minStart}. Then, this \texttt{minStart} term replaces the \texttt{sched.start} value from Algorithm~\ref{alg:heft_assignment_edp_v1} on line~\ref{alg1:schedStart}.
Looking back at Fig.~\ref{fig:edp_proc_assign_b}, with this modification in place, the processor that minimizes the selection metric is P2, which strikes a balance between the short-sighted task-level minimization of P1 and the unquestionably worse decision of P3.
This helps balancing the tasks across all PEs rather than continuously assigning them to the most power efficient processor with no regard to load balancing.
Fig.~\ref{fig:energy_results_exec}  shows that HEFT\textsubscript{EDP-LB} improves the runtime significantly compared to HEFT\textsubscript{EDP}, scaling in a fashion nearly equivalent to HEFT\textsubscript{RT}.
Meanwhile, in Fig.~\ref{fig:energy_results_energy}, we see that energy is still reduced by a maximum of 20.2\% with an average reduction of 4.6\% across all injection rates relative to HEFT\textsubscript{RT}.
As the target injection rate increases, HEFT\textsubscript{EDP-LB} and HEFT\textsubscript{RT} converge to similar energy consumption values because the system has a high enough workload that it must utilize all available PEs regardless of energy impact.
Because HEFT\textsubscript{EDP-LB} only adjusts energy usage by choice of PE and not through dynamic voltage and frequency scaling (DVFS)-based measures, when all PEs are in use, the energy usage is equivalent to that of HEFT\textsubscript{RT}.
Taken together, across these results, we find that, if execution time is the largest priority, HEFT\textsubscript{RT} provides the most effective scheduling.
Meanwhile, if energy consumption is the largest priority, HEFT\textsubscript{EDP} provides the most effective scheduling, particularly at low workload rates.
Finally, if a balance of execution and energy is required, HEFT\textsubscript{EDP-LB} provides an effective method to reduce energy consumption without drastically sacrificing execution time performance.

\section{Results} \label{sec:results}

In this subsection, we conduct evaluations using DS3 across both of the hardware-validated SoC configurations discussed in Section~\ref{sec:simsetup}: the Odroid XU3-based system and the Xilinx Zynq Ultrascale+ ZCU102-based system.
For each hardware configuration, we begin by evaluating with a workload mixture consisting of an even distribution of all 6 applications available in DS3.
To provide context for the results of these experiments, we include three other schedulers: the Minimum Execution Time (MET) scheduler and Constraint Programming (CP) schedulers provided by default in DS3 as well as the well-known PEFT list scheduler~\cite{Arabnejad14}.
\edit{
MET provides a useful point of comparison as it is representative of the types of greedy heuristics commonly used for runtime environments.
Meanwhile, as discussed in Section~\ref{subsec:execution_time_analysis}, CP provides an effective lower bound on frame execution time, particularly at low injection rates where it is feaible to find optimal solutions. 
Finally, as} PEFT is a list scheduler with the same algorithmic runtime complexity as HEFT and has been shown to, broadly speaking, match or outperform it in the literature~\cite{maurya2018benchmarking}, we believed that it would be a useful point of comparison here.
PEFT was implemented in DS3 using the same approach as HEFT\textsubscript{Base} in Section~\ref{subsec:staticvsdynamic}, and as such, here we refer to it as PEFT\textsubscript{Base}.
Before running experiments, PEFT\textsubscript{Base} was validated against Tables 2 and 3 from the original work~\cite{Arabnejad14}.
For further verification, all of the implementations presented here are made available in the public release of DS3~\cite{ds3_source}.

\begin{figure}[tb]
    \centering
    \begin{subfigure}[tb]{0.75\linewidth}
        \includegraphics[width=\linewidth]{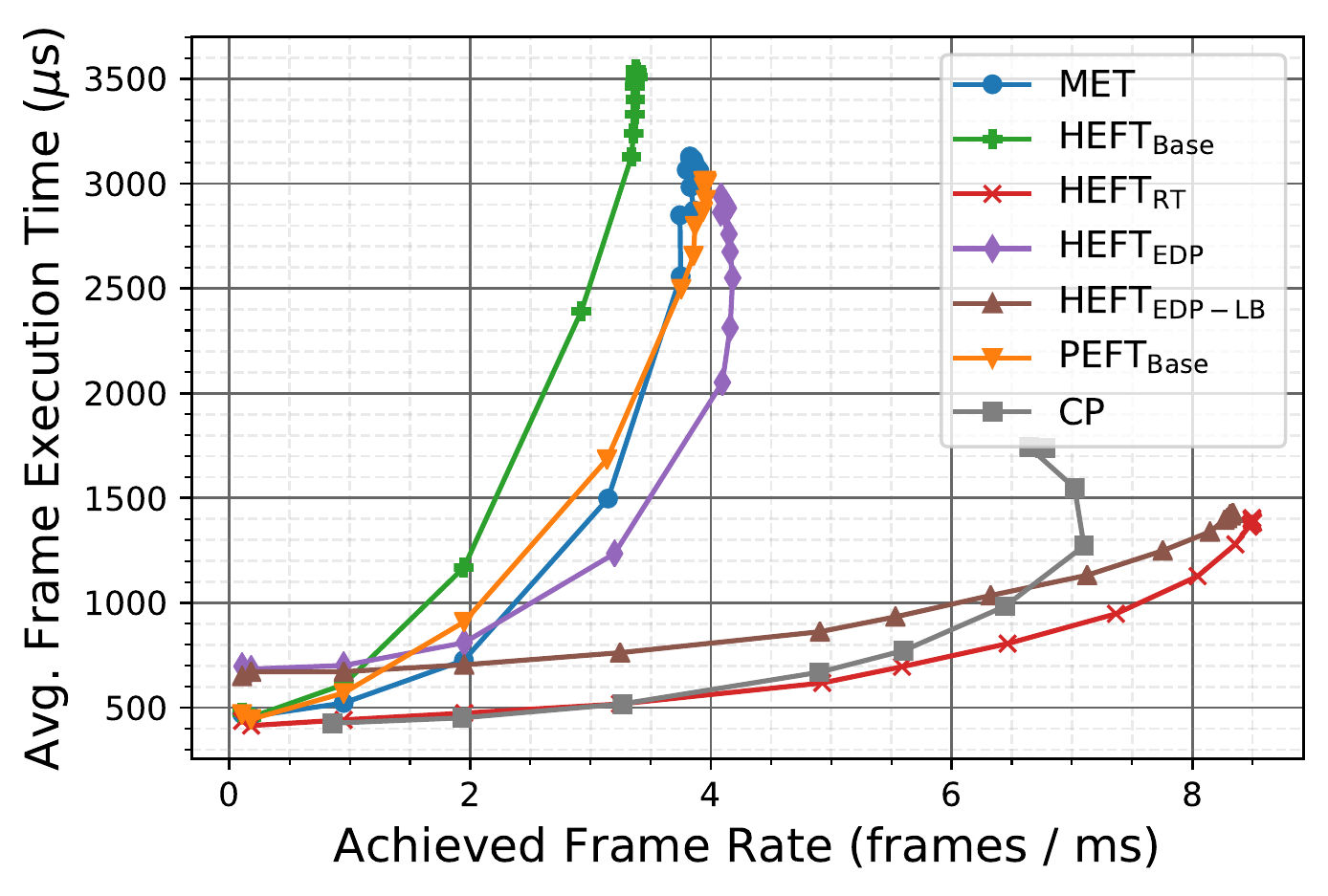}
        \subcaption{}
        \label{fig:odroid_all_6_exec}
    \end{subfigure}
    \begin{subfigure}[tb]{0.75\linewidth}
        \includegraphics[width=\linewidth]{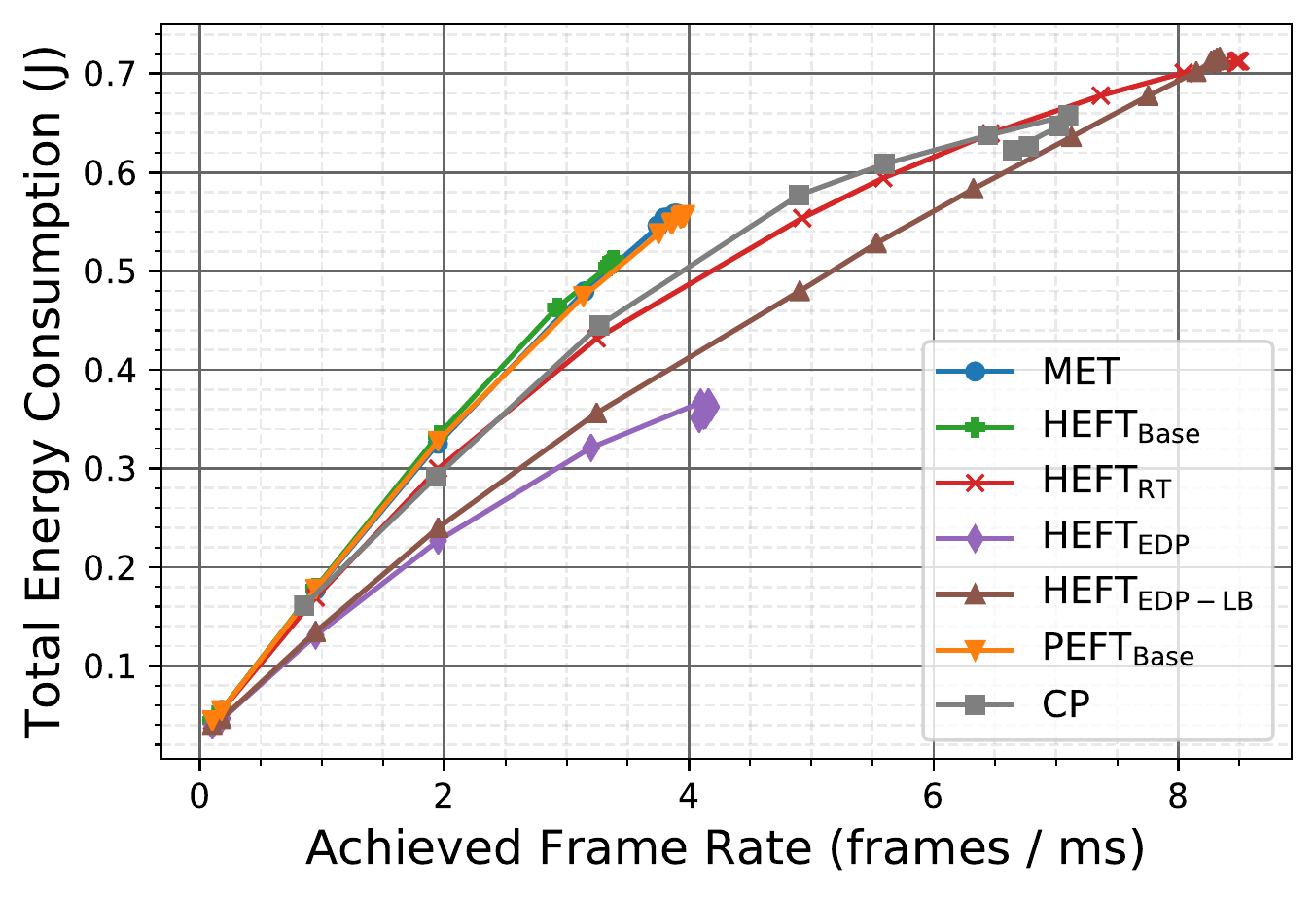}
        \subcaption{}
        \label{fig:odroid_all_6_energy}
    \end{subfigure}
    \caption{(a) average frame makespan and (b) total energy versus achieved frame rate in a workload composed of an even mixture of all six DS3 applications on configuration (2) (Odroid XU3) from Section~\ref{sec:simsetup}.}
    \label{fig:odroid_all_6_apps}
\end{figure}

Fig.~\ref{fig:odroid_all_6_apps} shows the execution time and energy results for configuration (2), the Odroid XU3-based platform.
As discussed previously, the Odroid XU3 contains a Samsung Exynos 5422 Octa big.LITTLE ARM Cortex A15.
Therefore, the primary decision for schedulers to make on this platform are whether to schedule onto the big cores or the LITTLE cores.
Comparing the trends observed here to those presented in Fig.~\ref{fig:energy_results}, we reach the same conclusions.
In execution-time analysis (Fig.~\ref{fig:odroid_all_6_exec}), HEFT\textsubscript{EDP} performs similarly to MET, HEFT\textsubscript{EDP-LB} performs slightly worse than HEFT\textsubscript{RT}, and HEFT\textsubscript{RT} outperforms all other schedulers.
Additionally, HEFT\textsubscript{Base} continues to illustrate poorest scalability, with PEFT\textsubscript{Base} slightly outperforming it as expected.
In energy analysis (Fig.~\ref{fig:odroid_all_6_energy}), we also find similar trends where HEFT\textsubscript{EDP-LB} solves the scalability issues of HEFT\textsubscript{EDP} while still managing to reduce energy consumption compared to HEFT\textsubscript{RT}. %

\begin{figure}[th]
    \centering
    \begin{subfigure}[tb]{0.75\linewidth}
        \includegraphics[width=\linewidth]{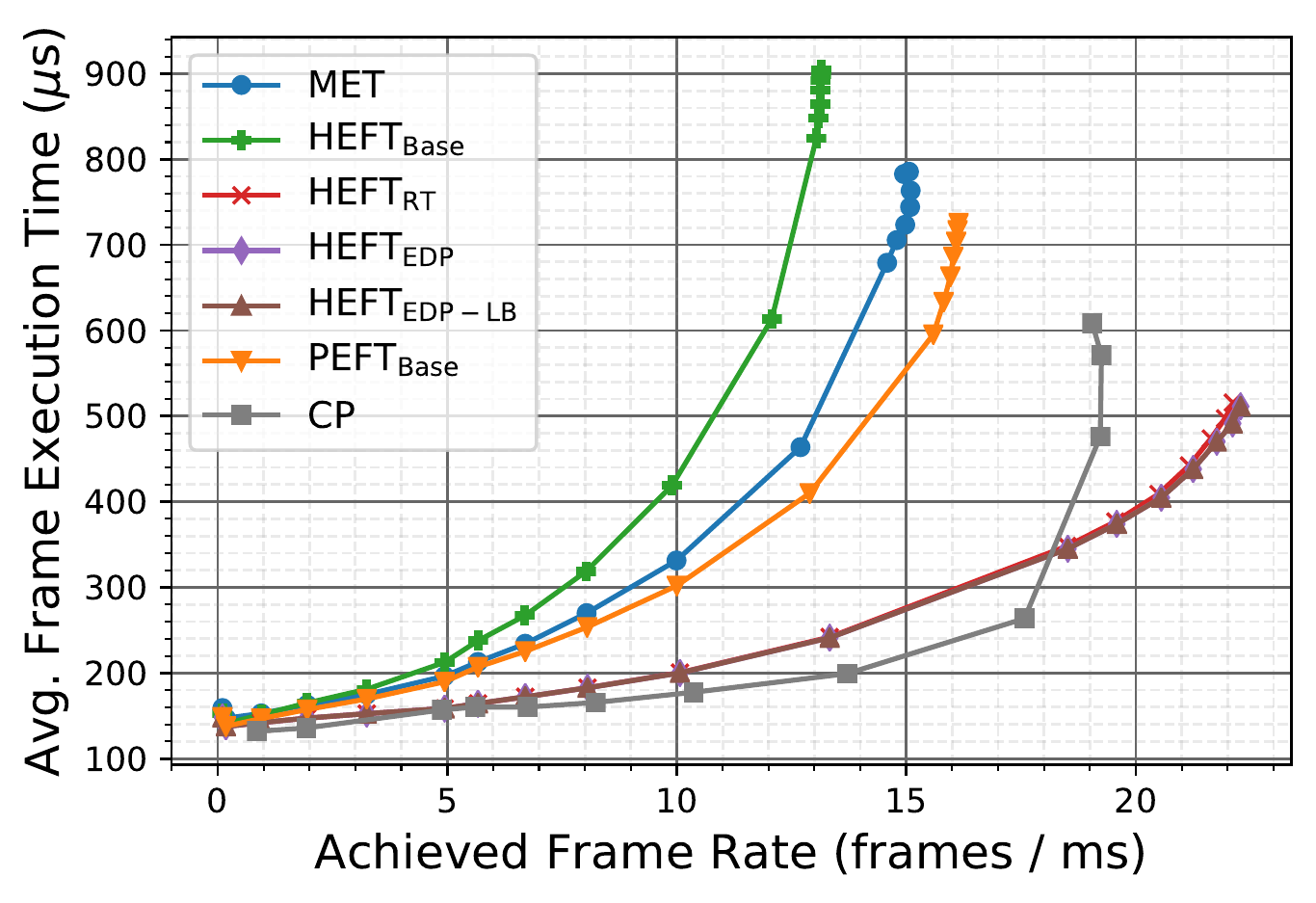}
        \subcaption{}
    \end{subfigure}
    \begin{subfigure}[tb]{0.75\linewidth}
        \includegraphics[width=\linewidth]{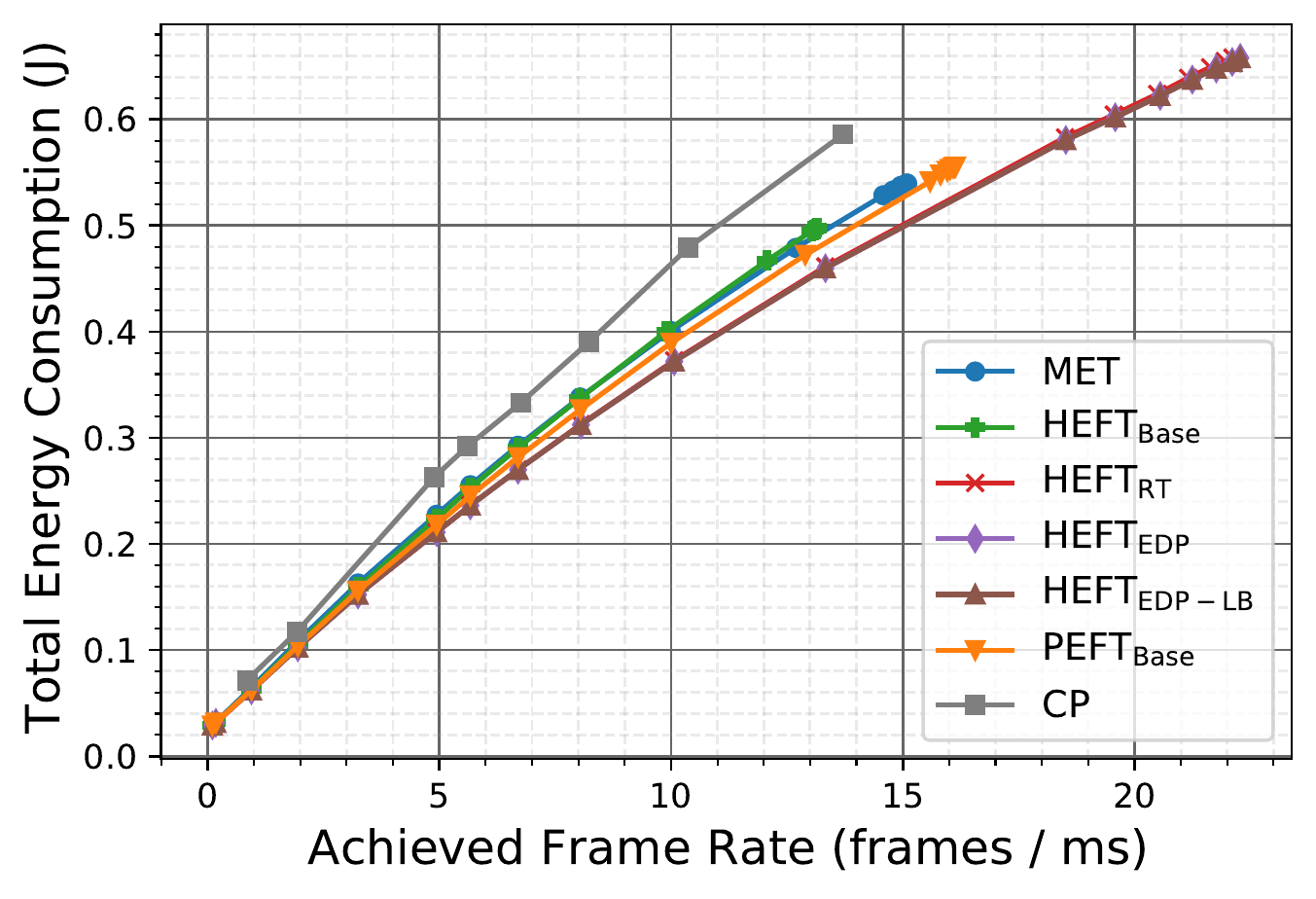}
        \subcaption{}
    \end{subfigure}
    \caption{(a) average frame makespan and (b) total energy versus achieved frame rate in a workload composed of an even mixture of all six DS3 applications on configuration (1) (ZCU102) from Section~\ref{sec:simsetup}.}
    \label{fig:zcu102_all_6_apps}
    \vspace{-6pt}
\end{figure}

Meanwhile, in Figure~\ref{fig:zcu102_all_6_apps}, we see the execution and energy results for configuration (1) (ZCU102) across all 6 applications.
In this case, we observe a notable convergence among all schedulers other than MET and HEFT\textsubscript{Base}.
This would seem to indicate that the primary issue for schedulers on this platform are simply to ensure that they are balancing jobs sufficiently across each of the processing elements.
As both MET and HEFT\textsubscript{Base} perform effectively greedy task assignment that does not consider tasks that interleave from other applications, particular accelerators in this system become overloaded and total average frame execution is elongated.
For schedulers that are able to balance across all PEs effectively, on this SoC, minimizing energy is equivalent to minimizing execution time due to the efficiencies in both execution and energy obtained from using dedicated hardware accelerators for critical kernels in these applications.

Finally, with general trends captured, we investigate application-level performance and energy efficiency gains.
Graphical plots that capture individual application execution and energy characteristics are presented for completeness in Appendix~\ref{appendix:application_plots}, Fig.~\ref{fig:odroid-single-app-results} and~\ref{fig:zcu102-single-app-results}, but the takeaways are captured in Table 2.
In this evaluation, numerical analysis is limited purely to execution time as the primary motivation for HEFT\textsubscript{RT} and its associated development in Section~\ref{subsec:staticvsdynamic} was to optimize the runtime performance of HEFT\textsubscript{Base}.
For similar reasons, while execution time characteristics of HEFT\textsubscript{EDP} and HEFT\textsubscript{EDP-LB} are captured via the same figures, their primary algorithmic motivation was to extend HEFT\textsubscript{RT} to reduce its energy consumption. As such, tabular analysis of these algorithms is limited purely to energy improvement that they provide with respect to HEFT\textsubscript{RT}.

\begingroup
\def\arraystretch{1}
\renewcommand{\tabularxcolumn}[1]{>{\centering\arraybackslash}p{#1}}
\begin{table*}[tb]
    \centering
    \begin{tabularx}{\linewidth}{|c|X|X|X|X||X|X|X|X|X|X|X|X|}
    \hline
    
    & \multicolumn{4}{c||}{Execution Improvement Over HEFT\textsubscript{Base}} &
      \multicolumn{8}{c|}{Energy Improvement Over HEFT\textsubscript{RT}} \\
    
    \hline
    
    & \multicolumn{4}{c||}{HEFT\textsubscript{RT}} &
      \multicolumn{4}{c|}{HEFT\textsubscript{EDP}} &
      \multicolumn{4}{c|}{HEFT\textsubscript{EDP-LB}} \\
    
    \hline
    
    & \multicolumn{2}{c|}{Odroid XU3} & 
      \multicolumn{2}{c||}{ZCU102} &
      \multicolumn{2}{c|}{Odroid XU3} & 
      \multicolumn{2}{c|}{ZCU102} &
      \multicolumn{2}{c|}{Odroid XU3} & 
      \multicolumn{2}{c|}{ZCU102} \\
    
    \hline
    
    Application & 
        \begin{tabular}{@{}c@{}}Avg.\\(\%)\end{tabular} & 
        \begin{tabular}{@{}c@{}}Max\\(\%)\end{tabular} &
        \begin{tabular}{@{}c@{}}Avg.\\(\%)\end{tabular} & 
        \begin{tabular}{@{}c@{}}Max\\(\%)\end{tabular} &
        \begin{tabular}{@{}c@{}}Avg.\\(\%)\end{tabular} & 
        \begin{tabular}{@{}c@{}}Max\\(\%)\end{tabular} &
        \begin{tabular}{@{}c@{}}Avg.\\(\%)\end{tabular} & 
        \begin{tabular}{@{}c@{}}Max\\(\%)\end{tabular} &
        \begin{tabular}{@{}c@{}}Avg.\\(\%)\end{tabular} & 
        \begin{tabular}{@{}c@{}}Max\\(\%)\end{tabular} &
        \begin{tabular}{@{}c@{}}Avg.\\(\%)\end{tabular} & 
        \begin{tabular}{@{}c@{}}Max\\(\%)\end{tabular} \\
    
    \hline
    
    WiFi TX &
        54.5 &
        84.1 &
        16.6 &
     	42.8 &
     	44.8 &
     	59.1 &
     	21.0 &
     	35.2 &
     	5.5 &
     	19.1 &
     	19.5 &
     	29.7 \\
    
    \hline
    
    WiFi RX &
        66.2 &
     	86.8 &
     	30.0 &
     	56.2 &
     	46.2 &
     	53.8 &
     	19.8 &
     	21.8 &
     	3.0 &
     	17.1 &
     	19.8 &
     	21.8 \\

    \hline

    \begin{tabular}{@{}c@{}} Radar\\Correlator \end{tabular} &
        68.5 &
     	90.1 &
     	47.2 &
     	78.5 &
        22.3 &
     	28.7 &
     	22.5 &
     	42.2 &
     	12.1 &
     	23.5 &
     	20.7 &
     	36.2 \\

    \hline

    \begin{tabular}{@{}c@{}} Temporal\\Mitigation \end{tabular} &
        60.0 &
     	88.8 &
     	30.1 &
     	72.9 &
     	1.5 &
     	2.8 &
     	27.3 &
     	59.3 &
     	1.1 &
     	2.5 &
     	25.2 &
     	48.7 \\

    \hline
    
    \begin{tabular}{@{}c@{}} Single\\Carrier TX \end{tabular} &
        29.9 &
     	61.6 &
     	14.1 &
     	57.0 &
     	23.2 &
     	28.7 &
     	15.1 &
     	26.2 &
     	23.1 &
     	28.7 &
     	15.1 &
     	26.2 \\

    \hline
    
    \begin{tabular}{@{}c@{}} Single\\Carrier RX \end{tabular} &
        69.8 &
     	88.4 &
     	45.3 &
     	83.7 &
     	44.5 &
     	53.0 &
     	17.7 &
     	24.5 &
     	7.1 &
     	24.9 &
     	17.7 &
     	24.5 \\

    \Xhline{3\arrayrulewidth}
    \textbf{Averages} &
        \textbf{58.2} &
        \textbf{83.3} &
        \textbf{30.6} &
        \textbf{65.2} &
        \textbf{30.4} &
        \textbf{37.7} &
        \textbf{20.6} &
        \textbf{34.9} &
        \textbf{8.7} &
        \textbf{19.3} &
        \textbf{19.7} &
        \textbf{31.2} \\
        
    \hline
    \end{tabularx}
    \caption{Improvements in average frame execution time of HEFT\textsubscript{RT} relative to HEFT\textsubscript{Base} across both SoC configurations followed by improvements in total energy consumption of HEFT\textsubscript{EDP} and HEFT\textsubscript{EDP-LB} relative to HEFT\textsubscript{RT}.}
    \label{tab:single_app_two_col}
    \vspace{-4mm}
\end{table*}
\endgroup

In Table~\ref{tab:single_app_two_col}, each row captures one of the six applications under test. 
The first four columns capture the average and maximum percentage improvements -- across both SoC configurations -- in average frame execution time for frames scheduled via HEFT\textsubscript{RT} over frames scheduled via HEFT\textsubscript{Base}.
Assuming BASE and RT are vectors containing the average frame execution time at each target frame rate, the ``Avg." entries are calculated via Eq.~\ref{eqn:table_avg_calc}.
\begin{equation}\label{eqn:table_avg_calc}
    AVG(100 * (BASE - RT) / (BASE))
\end{equation}
The maximum calculations are performed similarly, with the AVG operator replaced by MAX.
In this case, we find that HEFT\textsubscript{RT} provides an average improvement in execution time of 58.2\% across all applications on the Odroid XU3 and a 30.6\% average improvement across all applications on the ZCU102.
This difference in speedup is attributable to the accelerators present on the ZCU102 platform: while HEFT\textsubscript{Base} makes scheduling decisions that are unaware of other frames on both platforms, the shift towards accelerators on the ZCU102 platform ensures that PEs are able to become idle sooner.
As such, application DAGs are less likely to interleave in ways that degrade the performance of HEFT\textsubscript{Base}.
Looking at the last eight columns, we observe the improvements in energy consumption -- across both SoC configurations -- for frames scheduled via HEFT\textsubscript{EDP} and HEFT\textsubscript{EDP-LB} compared to those scheduled via HEFT\textsubscript{RT}.
Looking first at HEFT\textsubscript{EDP}, we find that across all applications, HEFT\textsubscript{EDP} provides an average energy savings of 30.4\% on the Odroid XU3 and 20.6\% on the ZCU102.
Meanwhile, HEFT\textsubscript{EDP-LB} provides an average energy savings of 8.7\% on the Odroid XU3 and 19.7\% on the ZCU102.
As seen in the Odroid results in Fig.~\ref{fig:odroid-single-app-results}, while HEFT\textsubscript{EDP} outperforms HEFT\textsubscript{EDP-LB} in pure energy savings, HEFT\textsubscript{EDP-LB} outperforms HEFT\textsubscript{EDP} in execution time over the same workloads.
Initially, it may seem paradoxical that HEFT\textsubscript{EDP} drops by nearly 10\% between the Odroid and the ZCU102 while HEFT\textsubscript{EDP-LB} gains nearly 10\%, but looking at the plots, we find that this, again, is attributable to the presence of accelerators. 
On the Odroid platform, the decrease in energy is accompanied by an increase in execution time, indicating that HEFT\textsubscript{EDP} continues to preferentially schedule on the LITTLE cores, while HEFT\textsubscript{EDP-LB} further utilizes the big cores and as such consumes more energy.
Meanwhile, on the ZCU102 platform, we actually observe a convergence between HEFT\textsubscript{RT}, HEFT\textsubscript{EDP}, and HEFT\textsubscript{EDP-LB}, where all three schedulers give near identical execution time performance, but HEFT\textsubscript{EDP} and HEFT\textsubscript{EDP-LB} both give nearly a 20\% reduction in energy consumption relative to HEFT\textsubscript{RT}.
As discussed previously with Fig.~\ref{fig:zcu102_all_6_apps}, this convergence occurs due to execution time minimization converging with energy minimization on this particular accelerator-rich SoC.

In summary, in this section, we present a novel and thorough analysis of HEFT\textsubscript{Base} in richly interleaving workload scenarios that, to the best of our knowledge, have not been applied previously to static schedulers like HEFT. 
Primary evaluation metrics of such schedulers typically include evaluation of single DAG instances against overly optimistic objectives such as the Schedule Length Ratio (SLR).
In contrast, the results here illustrate that, despite modern analysis continuing to show HEFT to be an effective scheduler in the case of non-interleaving DAGs~\cite{maurya2018benchmarking}, static schedulers like HEFT face a number of challenges that prevent seamless deployment in rapidly varying workload mixtures.
Finally, we find that, across all workloads tested, the conclusions of Section~\ref{sec:algorithmic_contributions} continue to hold.
For execution-focused scheduling, HEFT\textsubscript{RT} provides the most effective performance, for energy-focused scheduling -- particularly at low rates -- HEFT\textsubscript{EDP} provides the most effective performance, and for a balance of the two, HEFT\textsubscript{EDP-LB} is a good compromise.

\section{Generalizability of Proposed Optimization Techniques} \label{sec:peft_rt}
As the last case study, we apply the optimization strategies (DAG merging, running-task constraints, dynamic dependencies, and runtime-aware simplifications) on PEFT\textsubscript{Base}, and seek to evaluate the generalizability of these techniques for other list schedulers. 
We follow the methodology applied on HEFT\textsubscript{Base} to generate HEFT\textsubscript{RT} and start from the PEFT\textsubscript{Base} scheduler described in the beginning of Section~\ref{sec:results} to generate a “PEFT\textsubscript{RT}” scheduler.

\begin{figure}[tb]
    \centering
    \begin{subfigure}[tb]{0.49\linewidth}
        \includegraphics[width=\linewidth]{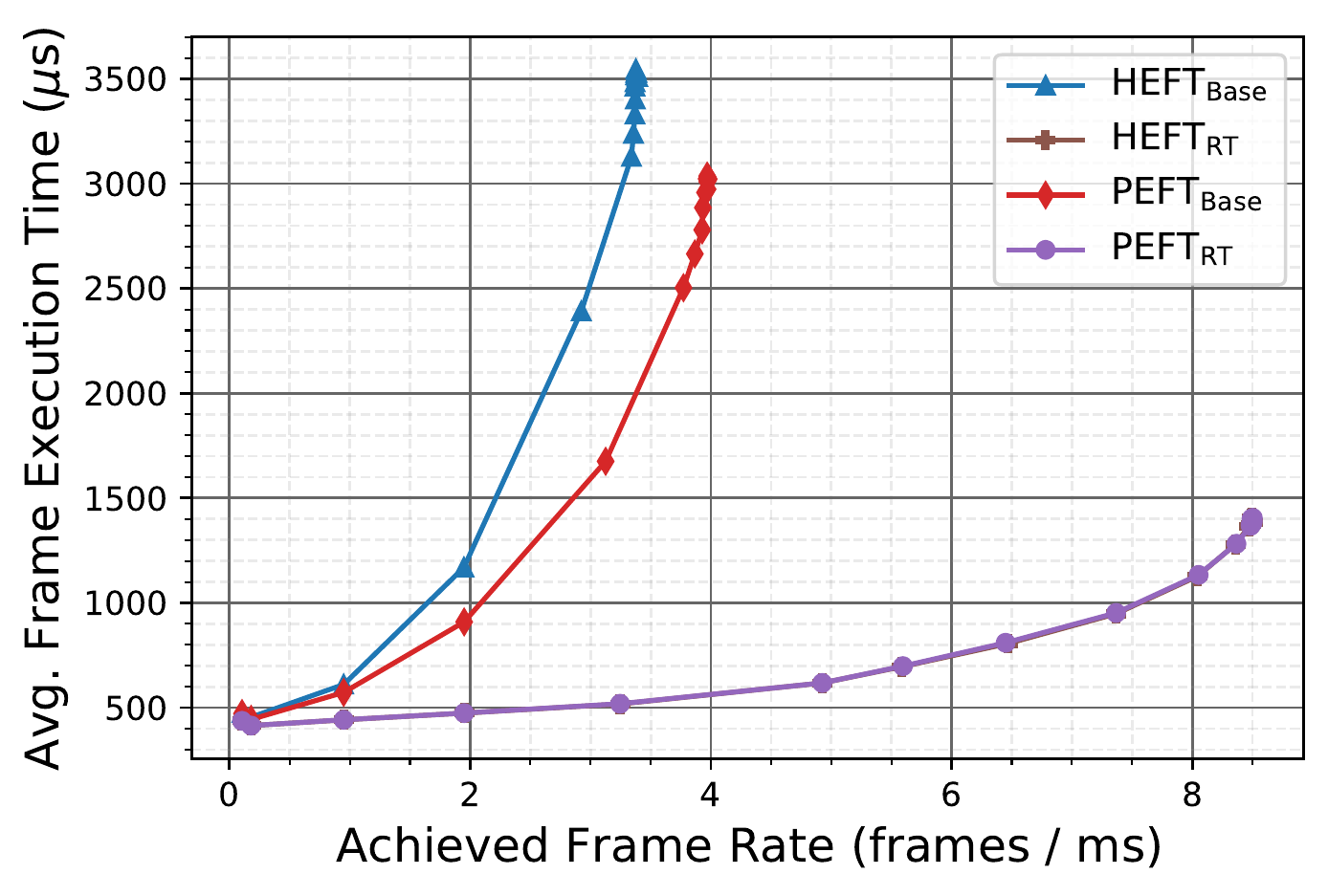}
        \subcaption{}
        \label{fig:odroid_all_6_peftrt}
    \end{subfigure}
    \begin{subfigure}[tb]{0.49\linewidth}
        \includegraphics[width=\linewidth]{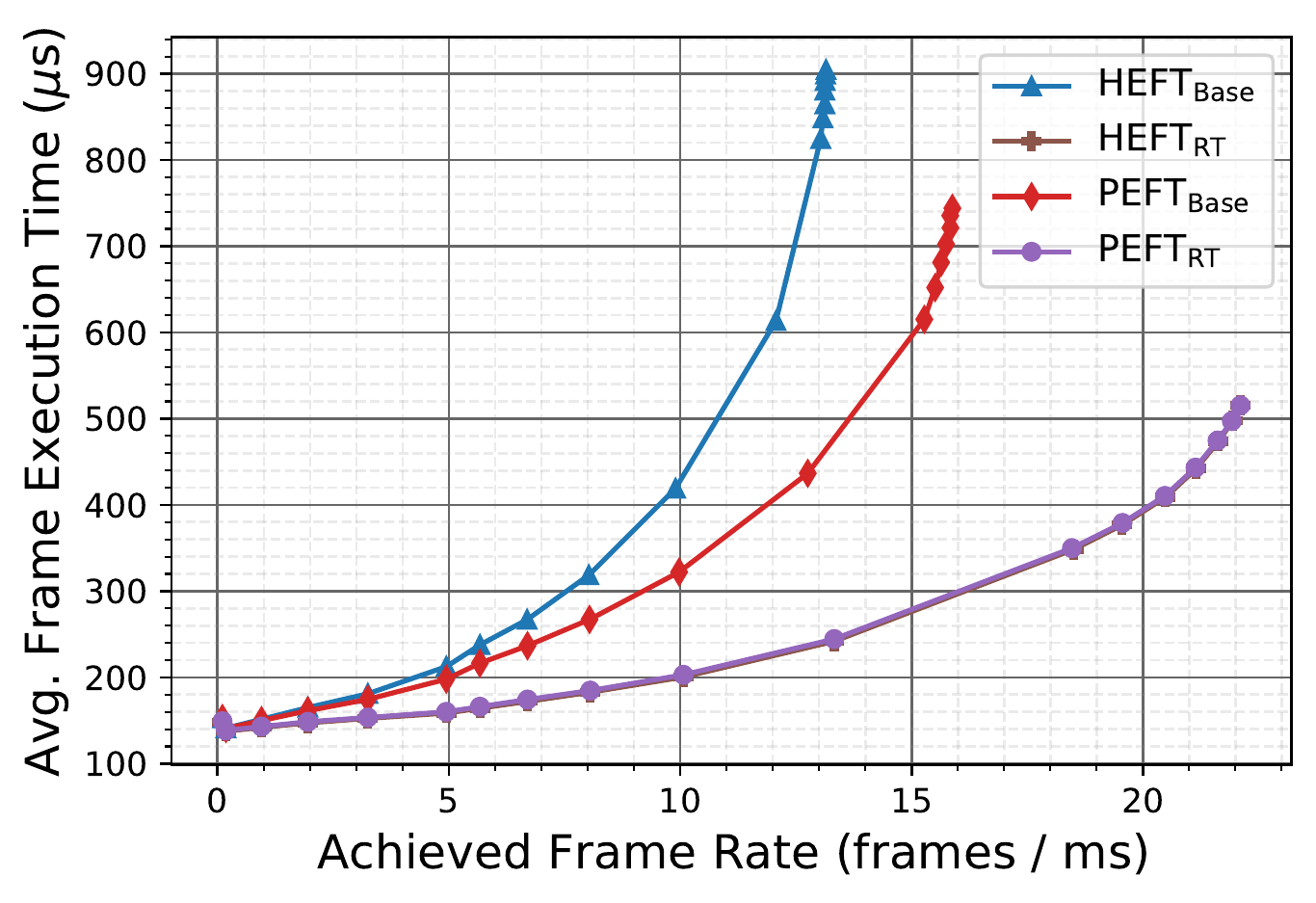}
        \subcaption{}
        \label{fig:zcu102_all_6_peftrt}
    \end{subfigure}
    \caption{average frame makespan versus achieved frame rate on (a) Odroid XU3 and (b) ZCU102 platforms with an even mixture of all six DS3 applications.}
    \vspace{-6mm}
\end{figure}

We evaluate PEFT\textsubscript{RT} on a workload consisting of an even mixture of all 6 test applications on both the Odroid XU3 and ZCU102-based platform configurations, as shown in Figures~\ref{fig:odroid_all_6_peftrt} and~\ref{fig:zcu102_all_6_peftrt}. We also include HEFT\textsubscript{Base} and HEFT\textsubscript{RT} results and observe that the relative performance gain between the baseline and run time versions of the two algorithms are consistent. 
Our optimization strategies reduce the average execution time and improve achieved frame rate for HEFT by 48.2\% and 25.0\% respectively, across all frame rates. 
Similarly, the corresponding improvements are 40.4\% and 19.2\% for PEFT.  
We also observe that PEFT\textsubscript{RT} performs almost identically to HEFT\textsubscript{RT} with the plots overlapping, which is consistent with how closely HEFT\textsubscript{RT} approached the estimated CP bounds in many workloads. 
These results provide strong evidence to suggest that the optimizations described here are generalizable to the broader class of list schedulers, and if pursued further, may help enable system designers to reach high levels of performance even in highly complex and demanding workload scenarios.

\edit{
Finally, while we do not explore it here, we believe a similar case study could be performed with regards to applying the energy optimization strategies proposed in this study to other rank-based static scheduling policies like PEFT.
Namely, in many such schedulers, the proposed techniques should be broadly applicable by substituting the appropriate computation cost terms with equivalent EDP variations, such as the transformation of $w_i$ to $w_i^2 * p_i$ in Equation~\ref{eq:ranku_edp}.
Meanwhile, the EDP-based load-balancing methodology discussed here can be broadly applied in many such schedulers as well by incorporating the ``earliest start time''-based EDP calculation methodology applied in Line~\ref{alg2:minstart_based_edp} of Algorithm~\ref{alg:heft_assignment_edp_v2}.
}
\section{Related Work} \label{sec:related_work}
While there is a large body of work related to static DAG scheduling algorithms in both makespan and energy aware contexts~\cite{baskiyar_low_2006, song_efficient_2017, zhang2017bi, Bittencourt10, Arabnejad14, zhou2017list, tariq_energy-efficient_2019, xie_energy-efficient_2017, akbar_list-based_2016, teller_scheduling_2008, teller_scheduling_2009, tang_energy-efficient_2016, lee_minimizing_2009},~\cite{li2018FluctuationAwarePredictive, wu2020MOELSMultiobjective, hu2021EnergyMinimizedScheduling, wang2021EffectiveCloud, zhu2021TaskScheduling},\edit{\cite{Wen2020IndustrialWorkflow}}, the evaluation methodologies of these works all assume that applications never interleave and primarily focus on metrics related to single-DAG static schedules like Schedule Length Ratio (SLR).
Additionally, these works evaluate their scheduling algorithms in a vacuum, independent of any particular runtime environment.
For those that are energy aware, the primary means they achieve their reduction is through DVFS, whereas the algorithms here optimize energy exclusively by choosing different processing elements.
Of the available literature, the breadth of studies evaluating runtime DAG scheduling implementations is much smaller, likely because as discussed in Section~\ref{subsec:staticvsdynamic}, there are a number of drawbacks that arise when deploying static DAG scheduling algorithms to dynamic runtime environments which this study addresses.
In the area of real time systems, Bambagini et al.\cite{bambagini_energy-aware_2016} taxonomize multicore and single-core energy aware scheduling.
For this work, the taxonomy presented for multicore energy-aware scheduling is much narrower than the corresponding alternative presented for single-core scheduling, and none of the methods presented explore the use of list schedulers.
Additionally, there are a large number of non list scheduling-based algorithms for runtime DAG scheduling, ranging from software schedulers \cite{liu_energy-efficient_2016, ali_rt-gang_2019} to hardware schedulers \cite{hounsinou_work-in-progress_2018, nasri2017offline, dellinger2011chronos, kuacharoen_configurable_2003, tang_hardware_2015, gaitan_cpu_2015, gomes_bringing_2016}.
However, as none of these works specifically explore runtime DAG scheduling with list-based algorithms, we consider the work presented here to be orthogonal to these studies.
To the best of our knowledge, \edit{there are very few publications that describe runtime performance of a HEFT-inspired list scheduler. One example of such a work} can be found in StarPU~\cite{augonnet2011starpu}, and as they discuss in a later work~\cite{thibault2018runtime}, there are actually a number of key differences between the baseline HEFT algorithm and the various implementations they provide in their \textit{dmda} family of schedulers~\cite{Augonnet10}.
Namely, the upward rank is assumed to be precomputed rather than calculated during scheduling, and these algorithms only schedule ready tasks, ignoring opportunities to implement insightful planning for future tasks that may be critical to overall makespan.
Even while this approach is, at a high level, similar to the approach taken by HEFT\textsubscript{RT}, these evaluations do not consider workloads with such a heavy degree of interleaving DAG instances as is done here.
For instance, later work~\cite{agullo_are_2016} performs a similar throughput-style analysis where performance is analyzed as a function of increasing problem size for a single application, but each instance of that application is assumed independent and executed separately from all other instances.
\edit{Meanwhile, works such as~\cite{lumpp_openvx_2021} explore the use of HEFT -- along with a proposed algorithm ``XEFT'' -- in a computer vision-focused runtime system called OpenVX.
This work similarly acknowledges that HEFT by itself produces improved, but suboptimal, execution schedules relative to greedy scheduling policies.
However, the solution proposed to improve HEFT's performance, XEFT, differs in its approach from those taken here.
XEFT instead attempts to maximize the time spent executing tasks with high levels of ``exclusive overlap'': a metric that captures whether a given set of tasks support opposing sets of resources (i.e. one task may only support CPU execution while another may only support GPU).
Because their supported resources are disjoint, they can trivially be scheduled in parallel across their respective resources.
In this work, we instead focus on optimizing HEFT across multiple DAGs in highly interleaved workload scenarios while leaving the behavior at the single-DAG level nearly unchanged.
In summary, we believe that} the experiments presented here yield valuable insights into workloads that are rarely explored in the DAG scheduling literature. 

\section{Conclusions} \label{sec:conclusions}
In summary, this work presents analysis of the well known HEFT algorithm from a new dynamic runtime perspective via HEFT\textsubscript{Dyn} and HEFT\textsubscript{RT}, and it presents novel EDP-aware modifications via HEFT\textsubscript{EDP} and HEFT\textsubscript{EDP-LB} that adapts it for use in power-constrained heterogeneous platforms.
Notably, we believe that the techniques presented here can be broadly generalized to map similar list scheduling algorithms like HEFT and PEFT for use on heterogeneous SoC platforms.
With regards to SoC design, we illustrate the benefits of pairing a suitably realistic simulation environment with effective scheduling algorithms in rapid iteration to a final SoC platform.
Future work will explore adaptively switching between HEFT\textsubscript{EDP} and HEFT\textsubscript{EDP-LB} based on system utilization.
\edit{Also, future work will explore expanding DS3 to support reference-calibrated applications that are outside the domain of communications and radar in order to evaluate our proposed algorithms in a broader set of heterogeneous workload scenarios.}
Additionally, the proposed schedulers will be extended to work with different DVFS governors present in current heterogeneous platforms.
With that, the scheduler will adapt task scheduling based on the processor's current frequency and voltage states, which affect both the system's performance and power dissipation.
\section*{Acknowledgement}\label{sec:acknowledgement}
This material is based on research sponsored by Air Force Research Laboratory (AFRL) and Defense Advanced Research Projects Agency (DARPA) under agreement number FA8650-18-2-7860. The U.S. Government is authorized to reproduce and distribute reprints for Governmental purposes notwithstanding any copyright notation thereon. The views and conclusion contained herein are those of the authors and should not be interpreted as necessarily representing the official policies or endorsements, either expressed or implied, of Air Force Research Laboratory (AFRL) and Defence Advanced Research Projects Agency (DARPA) or the U.S. Government.

\bibliographystyle{IEEEtran}
\bibliography{refs.bib}

\vskip -2\baselineskip plus -1fil

\begin{IEEEbiography}[{
\vspace*{-12mm}\includegraphics[width=0.9in,height=0.9in,clip,keepaspectratio]{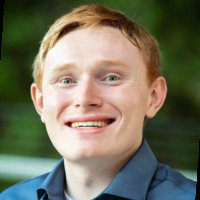}
}]{Joshua Mack} 
is a Ph.D. student in the Electrical and Computer Engineering program at the University of Arizona. His research interests include reconfigurable systems; emerging architectures; and intelligent workload partitioning across heterogeneous systems.
\end{IEEEbiography}

\vskip -4\baselineskip plus -1fil

\begin{IEEEbiography}[{
\vspace*{-8mm}\includegraphics[width=0.9in,height=0.9in,clip,keepaspectratio]{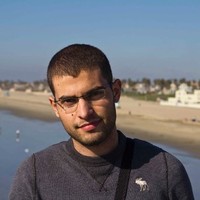}
}]{Samet E. Arda} 
received his M.S. and Ph.D. degrees in Electrical Engineering from Arizona State University in 2013 and 2016. He is currently an Assistant Research Scientist at ASU. His Ph.D. thesis focused on development of dynamic models for small modular reactors. His current research interests include system-level design and  optimization techniques for scheduling in heterogeneous SoCs.
\end{IEEEbiography}

\vskip -3\baselineskip plus -1fil

\begin{IEEEbiography}[{
\vspace*{-8mm}\includegraphics[width=0.9in,height=0.9in,clip,keepaspectratio]{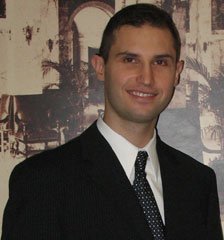}
}]{Umit Y. Ogras} 
received his Ph.D. degree in Electrical and Computer Engineering from Carnegie Mellon University, Pittsburgh, PA, in 2007. From 2008 to 2013, he worked as a research scientist at the Strategic CAD Laboratories, Intel Corporation. He is an Associate Professor at the School of Electrical, Computer and Energy Engineering, and the Associate Director of WISCA Center.
\end{IEEEbiography}

\vskip -3.5\baselineskip plus -1fil

\begin{IEEEbiography}[{
\includegraphics[width=1in,height=1.25in,clip,keepaspectratio]{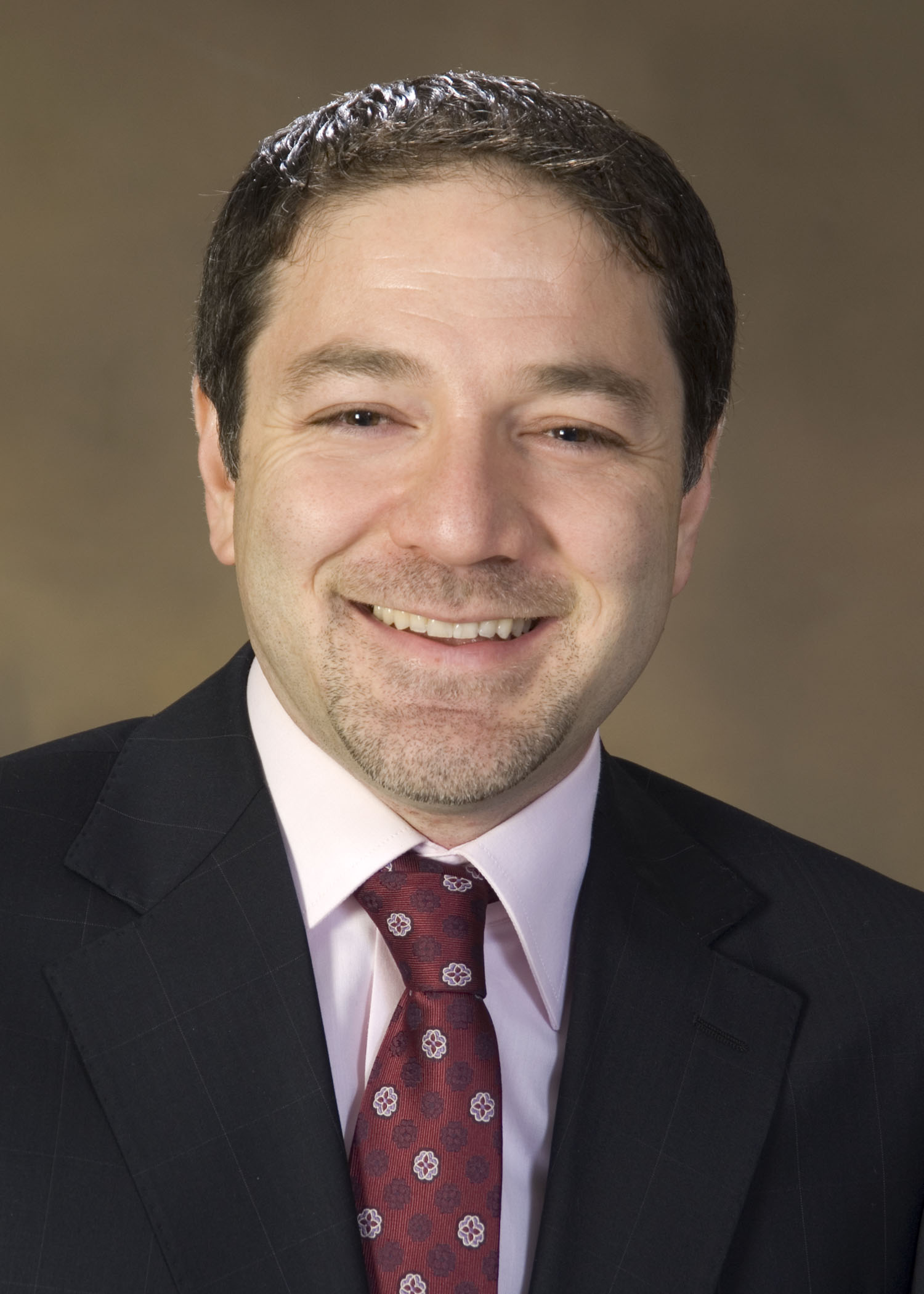}
}]{Ali Akoglu}
received his Ph.D. degree in Computer Science from the Arizona State University in 2005. He is an Associate Professor in the Department of Electrical and Computer Engineering and the BIO5 Institute at the University of Arizona. He is the site-director of the National Science Foundation Industry-University Cooperative Research Center on Cloud and Autonomic Computing. His research focus is on high performance computing and non-traditional computing architectures. 
\end{IEEEbiography}

\appendices
\renewcommand\thefigure{\thesection.\arabic{figure}}
\setcounter{figure}{0}
\renewcommand{\thetable}{\thesection.\Roman{table}}
\setcounter{table}{0}
\clearpage
\renewcommand{\thepage}{A-\arabic{page}}
\setcounter{page}{1}

\def\figwidth{0.32}

\onecolumn

\section{Supplemental Plots} \label{appendix:application_plots}

\begin{figure}[h!]
	\centering
	\begin{subfigure}[]{\figwidth\linewidth}
		\includegraphics[width=\linewidth]{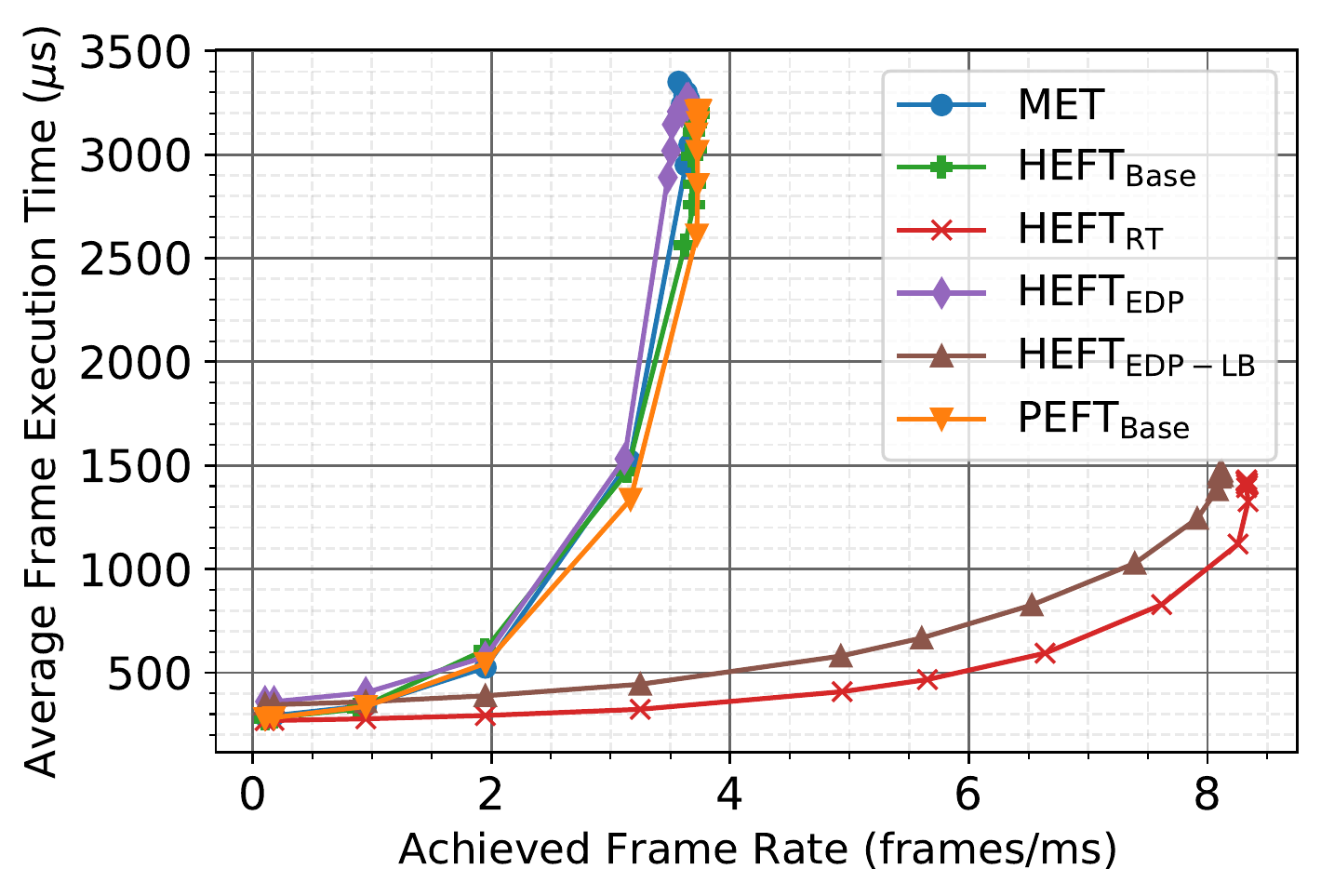}
	\end{subfigure}
	\begin{subfigure}[]{\figwidth\linewidth}
		\includegraphics[width=\linewidth]{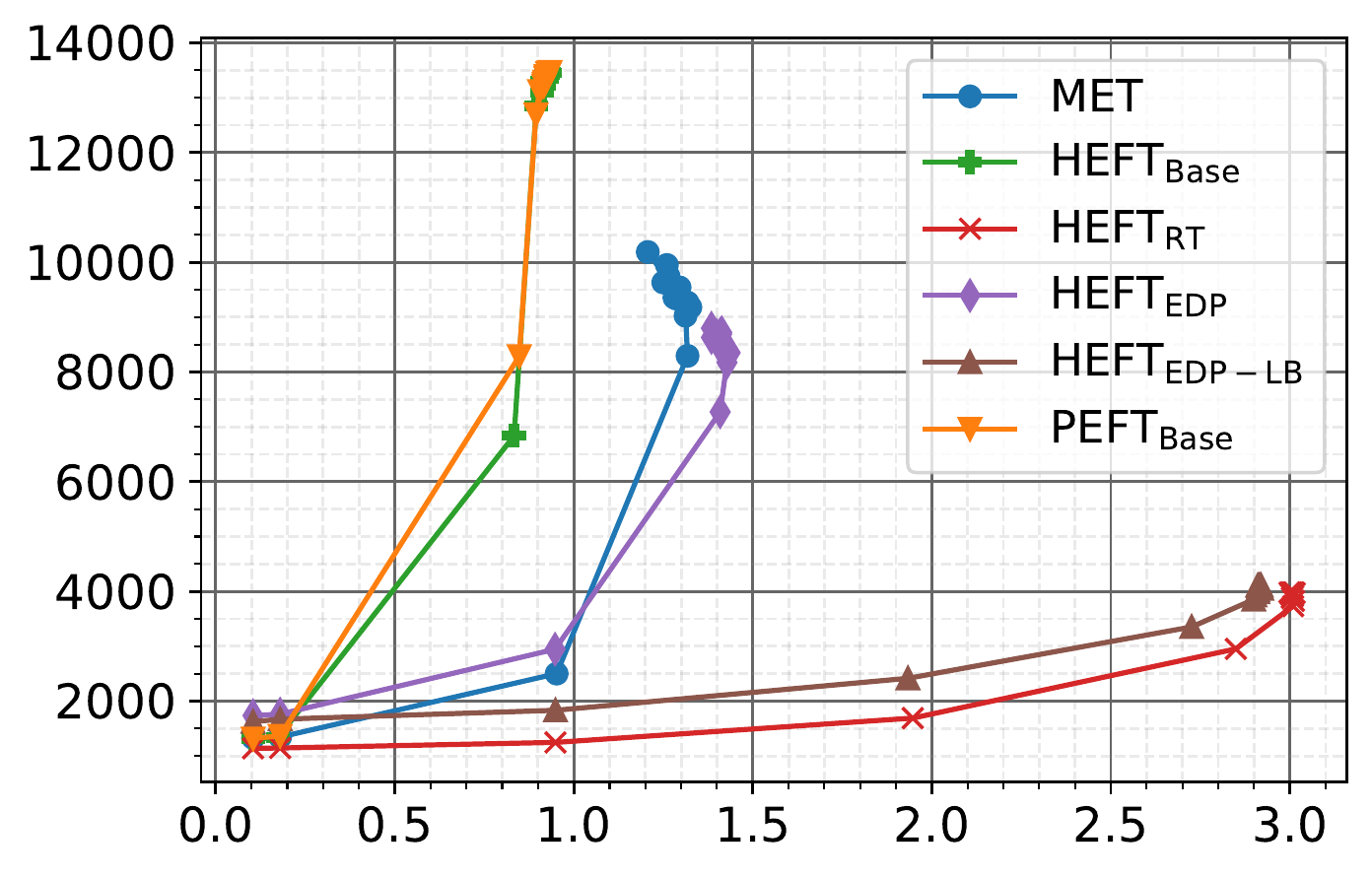}
	\end{subfigure}
	\begin{subfigure}[]{\figwidth\linewidth}
		\includegraphics[width=\linewidth]{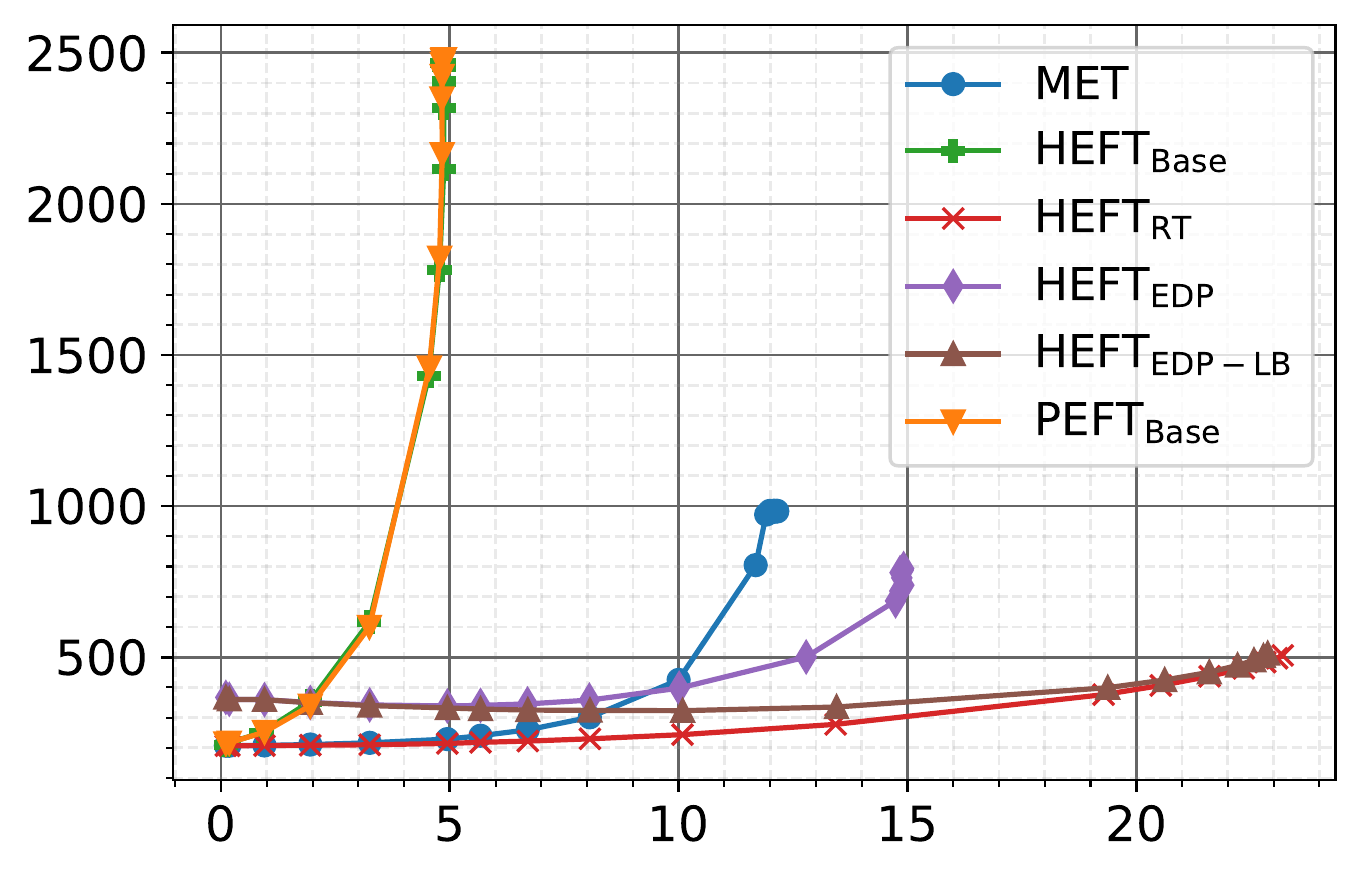}
	\end{subfigure}
	
	\begin{subfigure}[]{\figwidth\linewidth}
		\includegraphics[width=\linewidth]{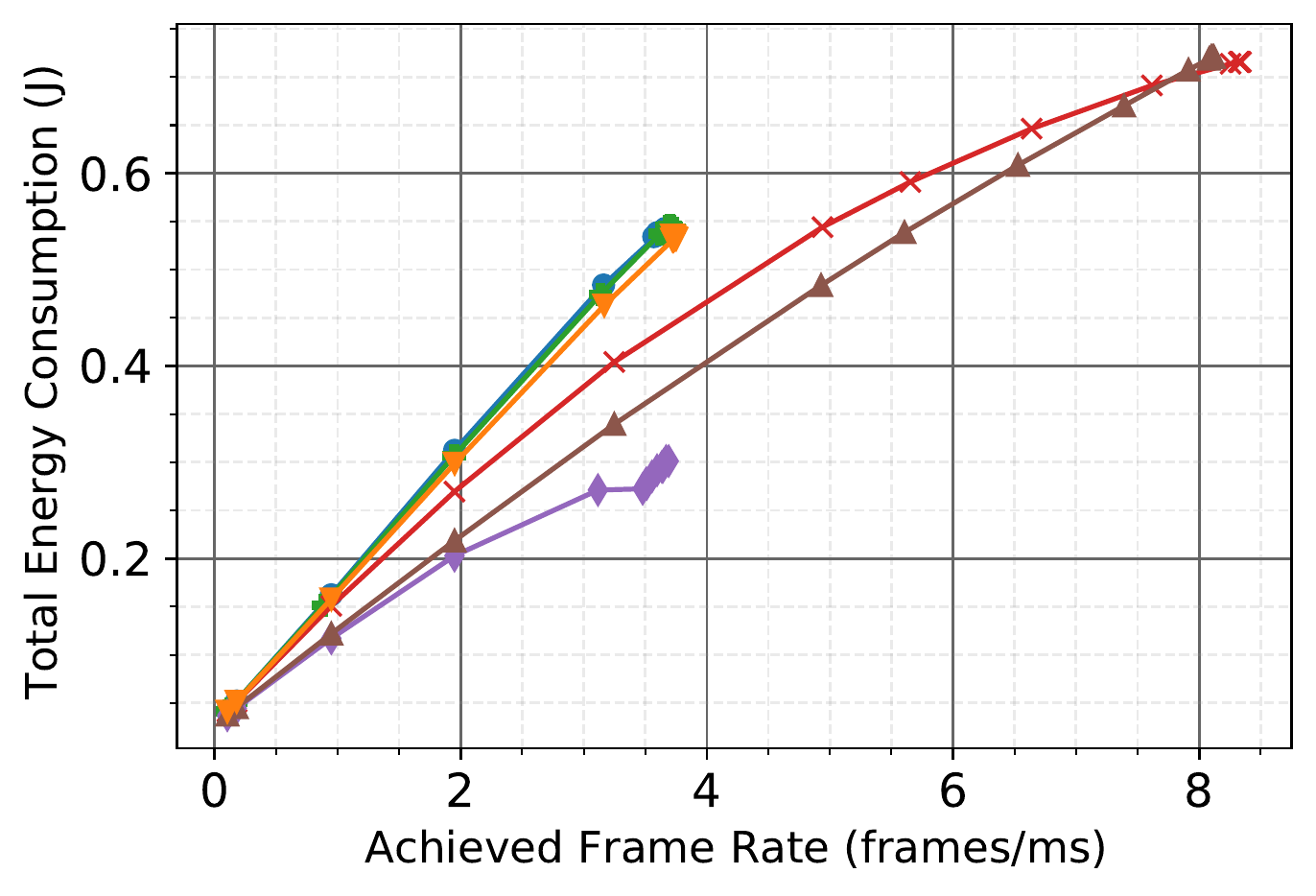}
	\end{subfigure}
	\begin{subfigure}[]{\figwidth\linewidth}
		\includegraphics[width=\linewidth]{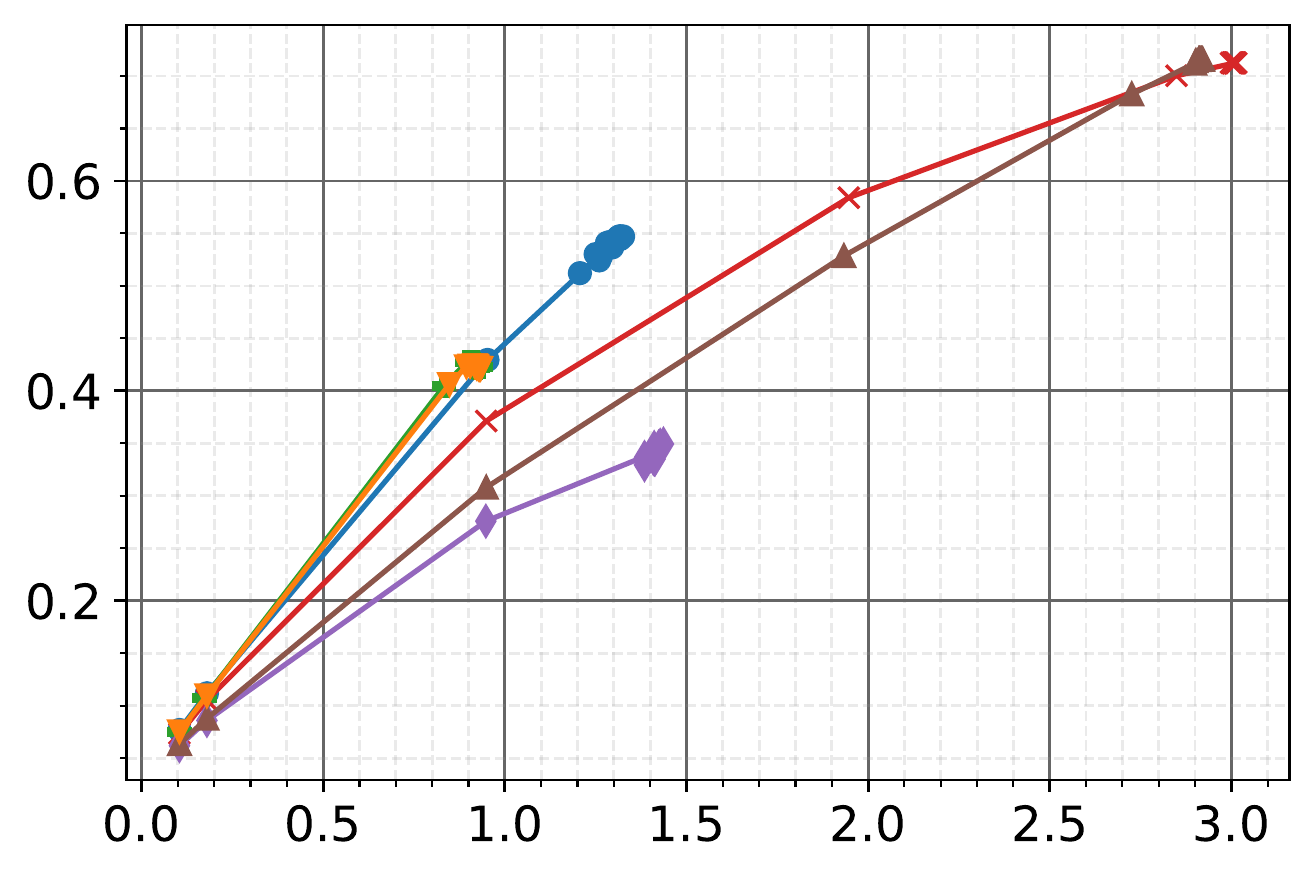}
	\end{subfigure}
	\begin{subfigure}[]{\figwidth\linewidth}
		\includegraphics[width=\linewidth]{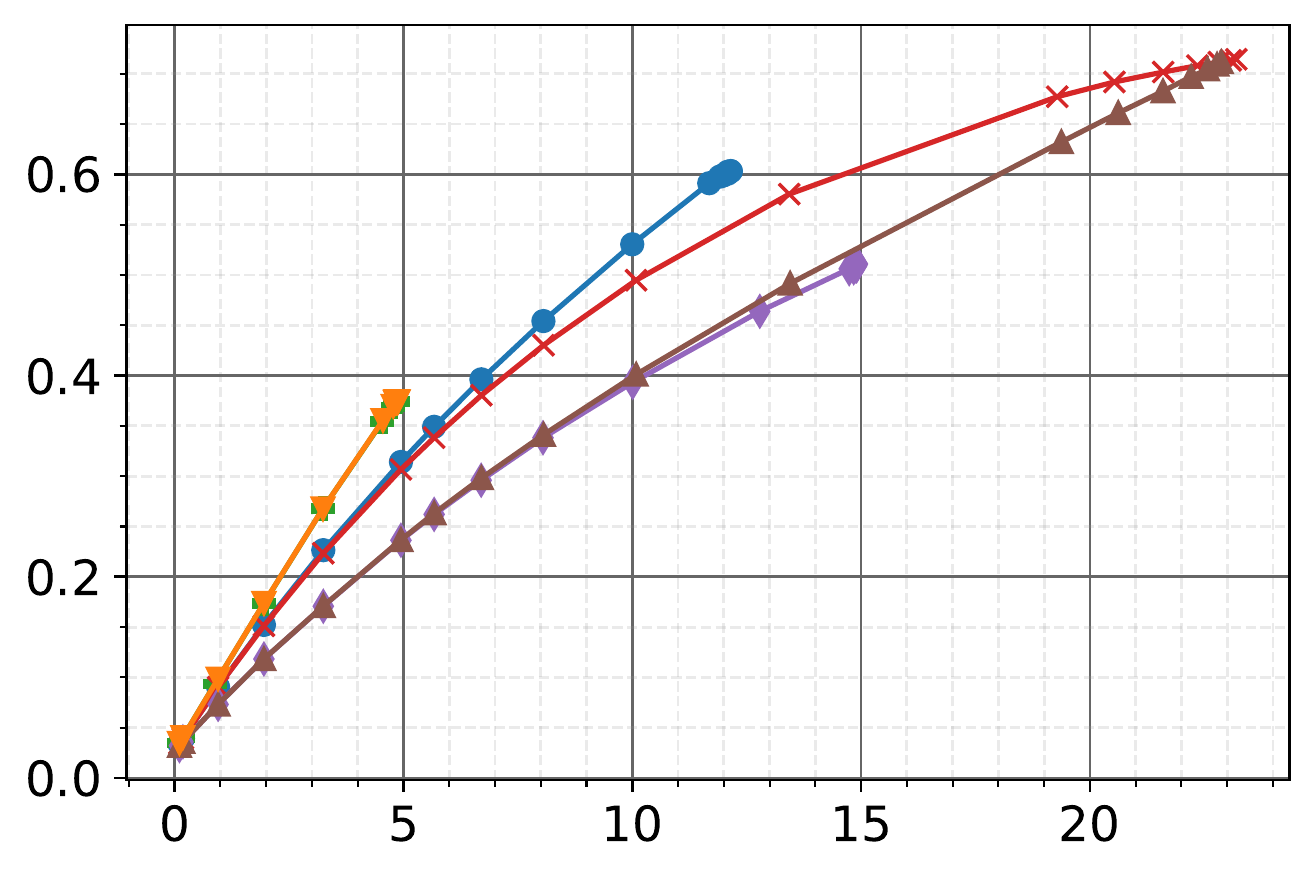}
	\end{subfigure}
	
	\begin{subfigure}[]{\figwidth\linewidth}
		\includegraphics[width=\linewidth]{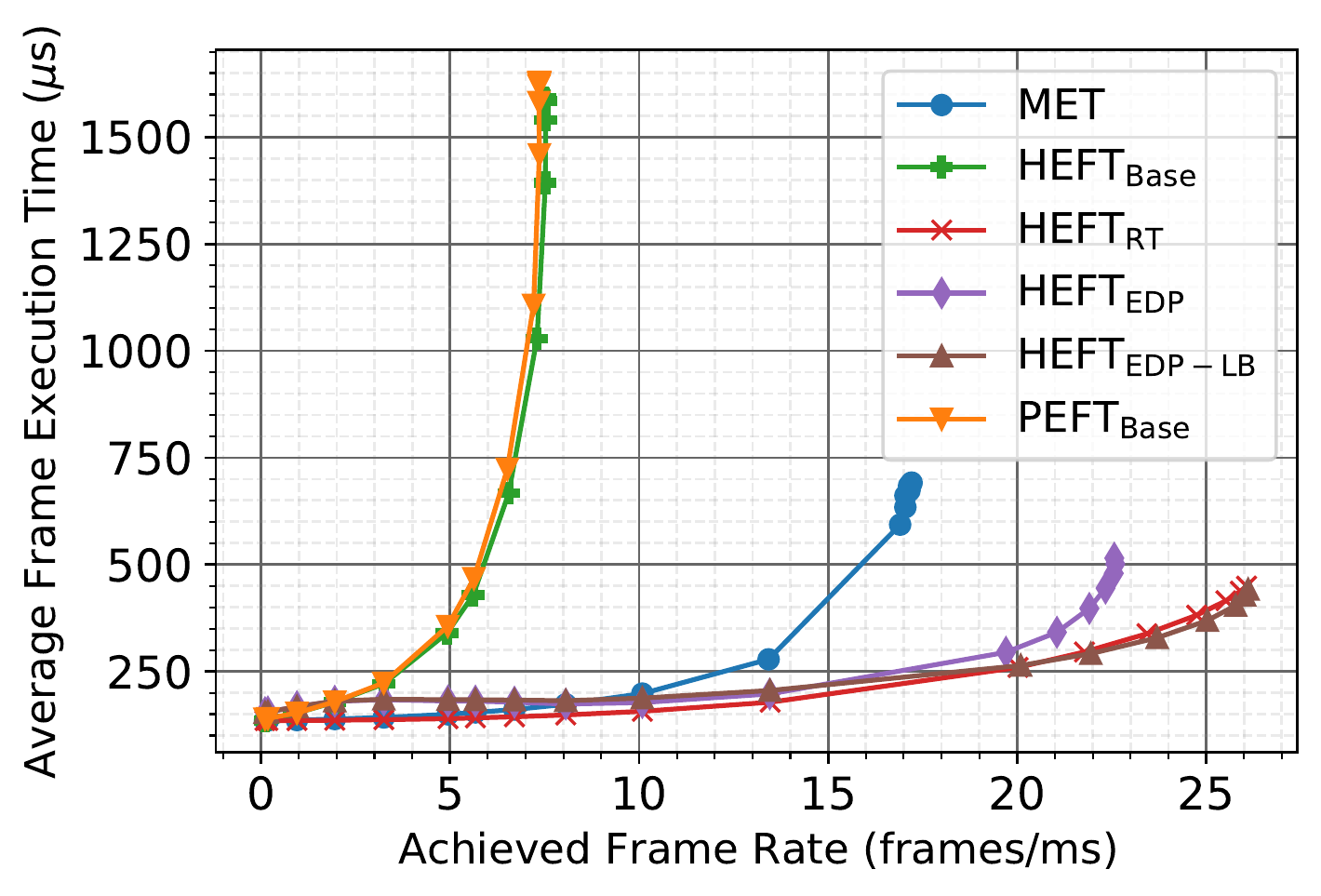}
	\end{subfigure}
	\begin{subfigure}[]{\figwidth\linewidth}
		\includegraphics[width=\linewidth]{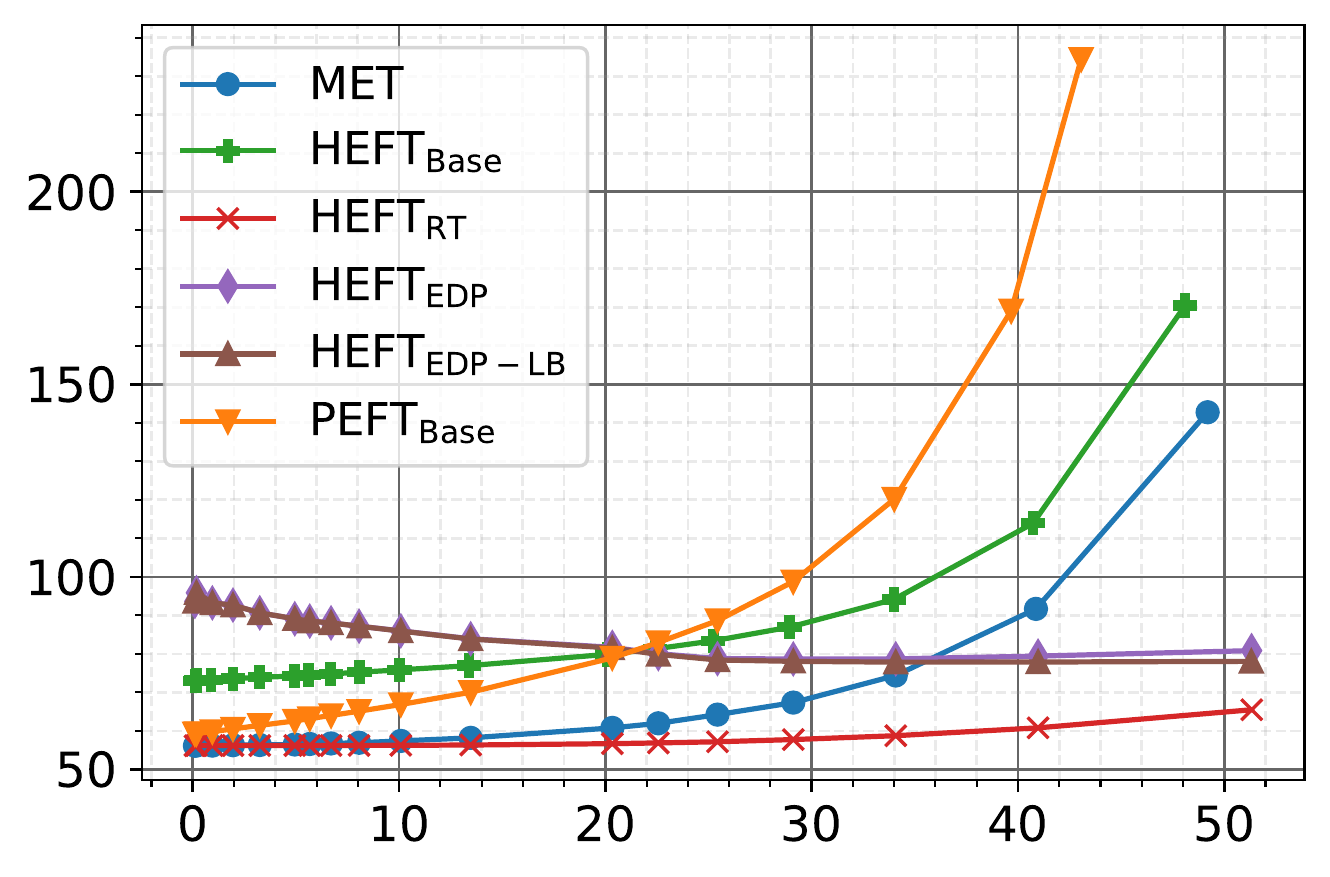}
	\end{subfigure}
	\begin{subfigure}[]{\figwidth\linewidth}
		\includegraphics[width=\linewidth]{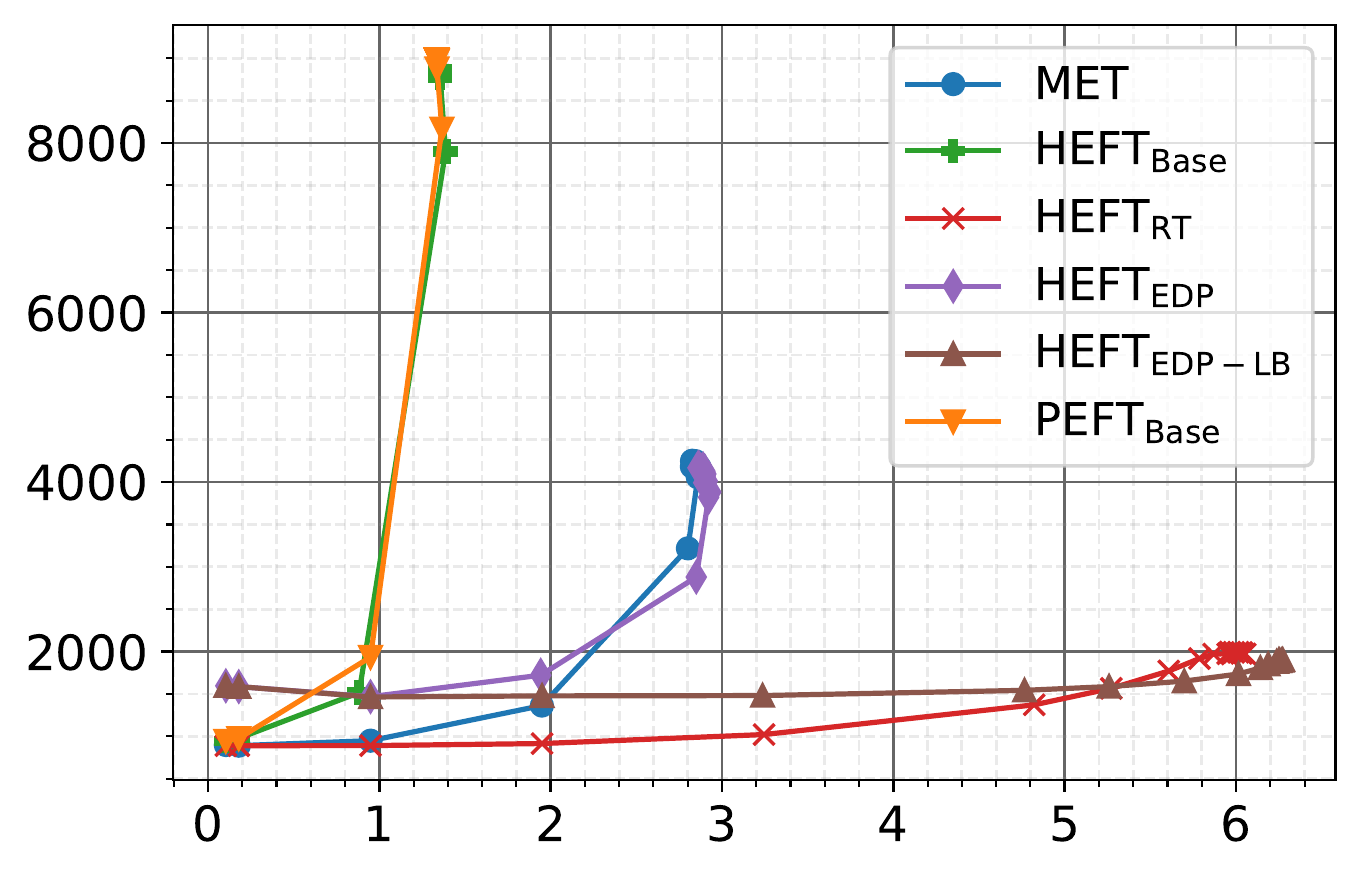}
	\end{subfigure}
	
	\begin{subfigure}[]{\figwidth\linewidth}
		\includegraphics[width=\linewidth]{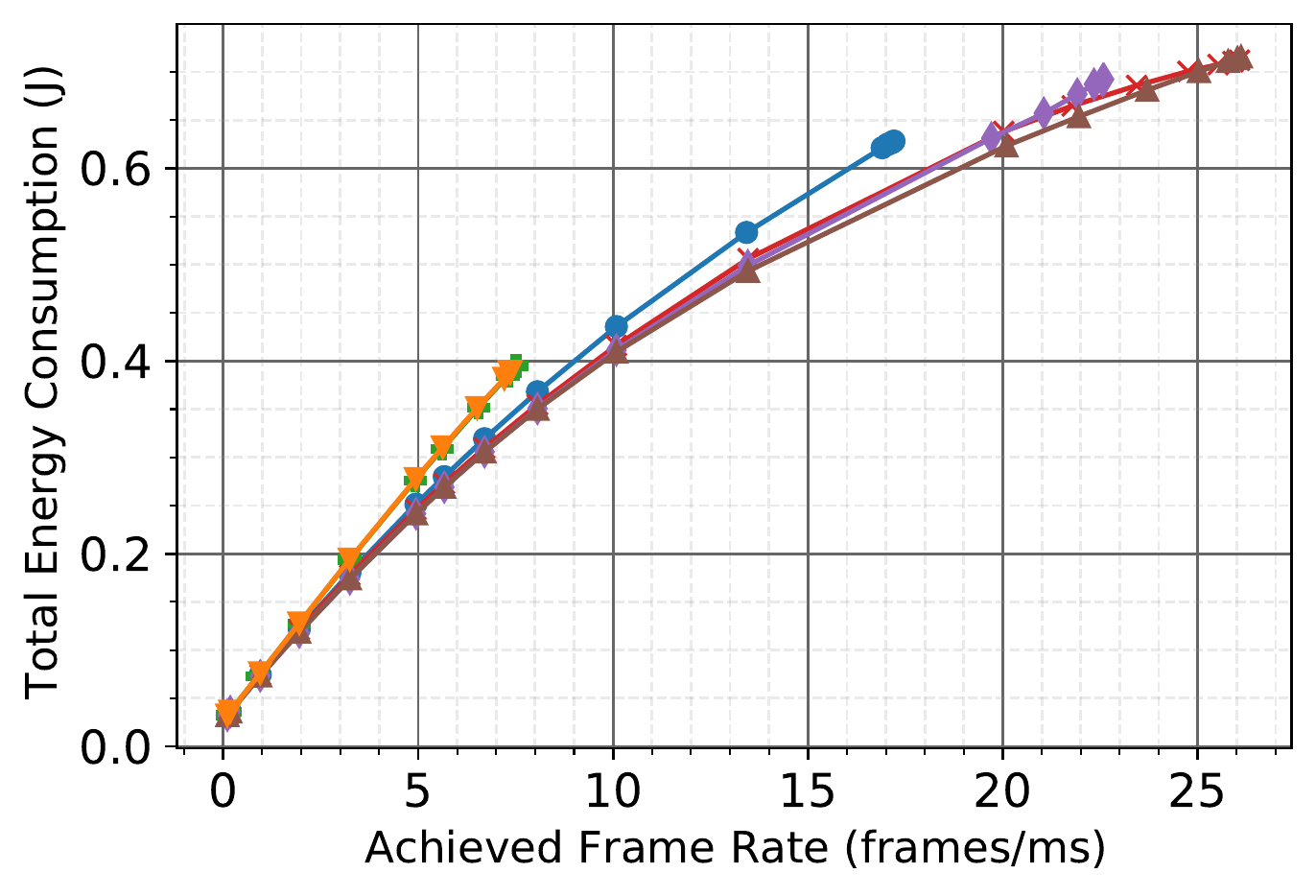}
	\end{subfigure}
	\begin{subfigure}[]{\figwidth\linewidth}
		\includegraphics[width=\linewidth]{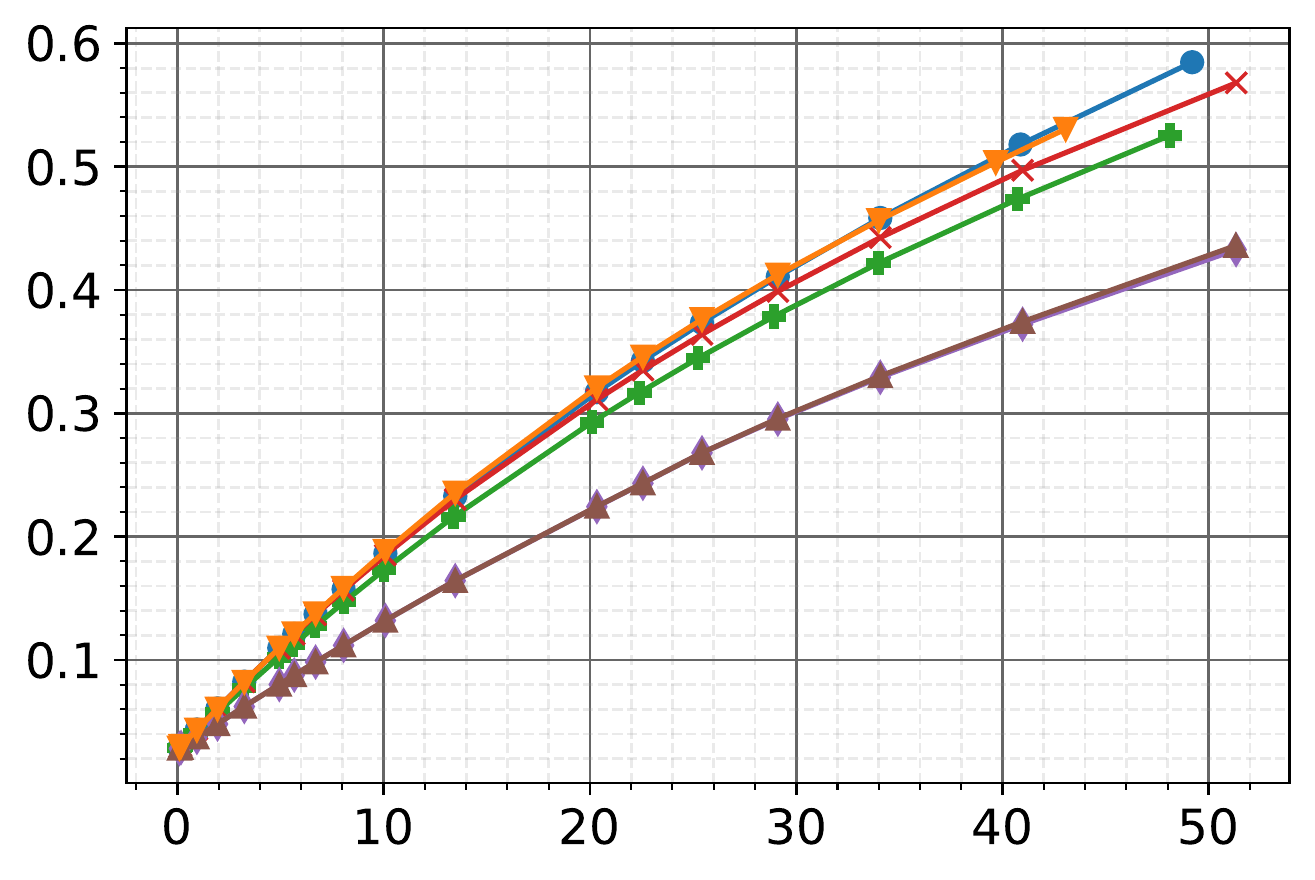}
	\end{subfigure}
	\begin{subfigure}[]{\figwidth\linewidth}
		\includegraphics[width=\linewidth]{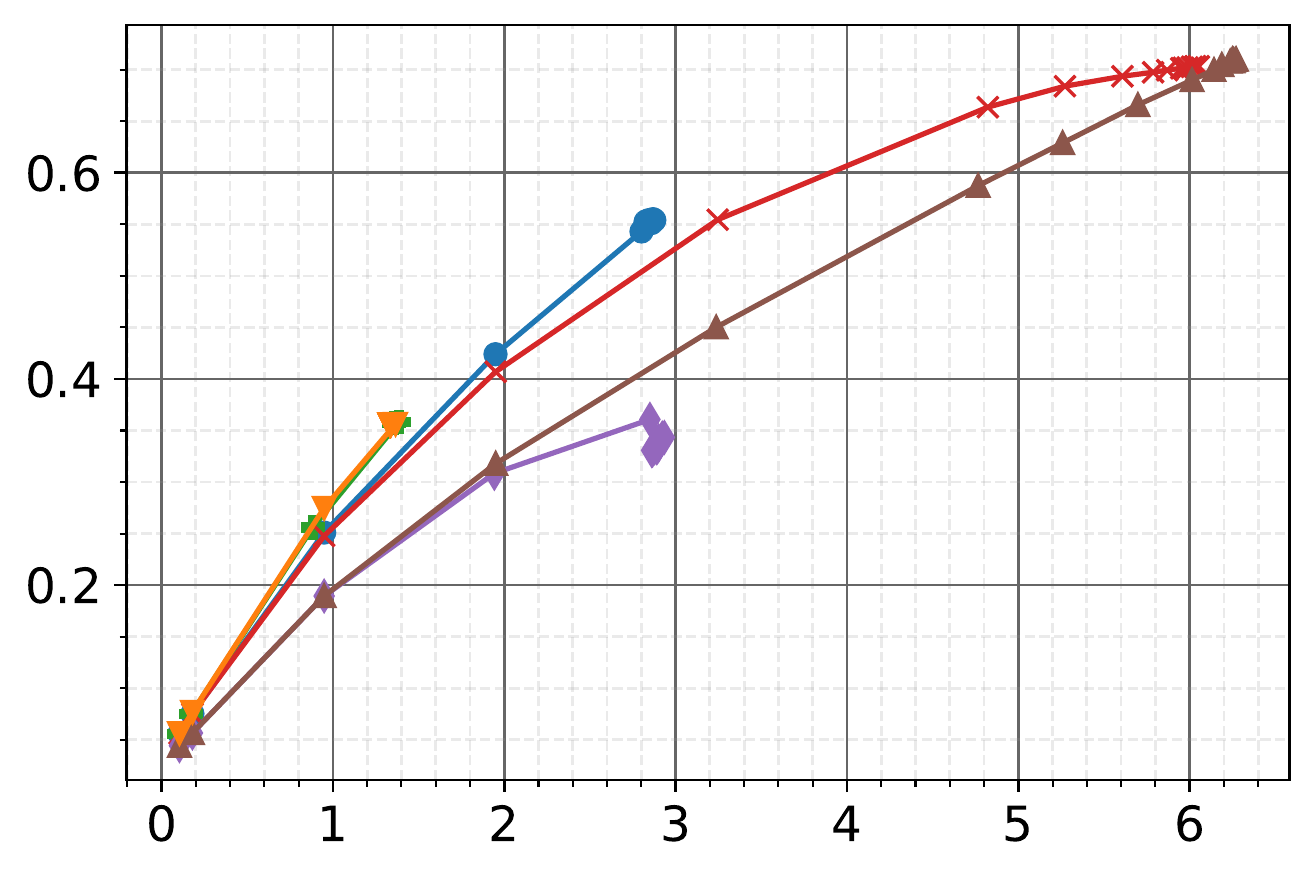}
	\end{subfigure}
	\caption{
	Odroid-XU3 application-level plots. 
	Plots are grouped into vertical pairs of Execution \& Energy. 
	Applications from left to right, top to bottom: WiFi TX, WiFi RX, Radar Correlator, Temporal Mitigation, Single Carrier TX, Single Carrier RX.
	We find that the same general trends hold as presented in Section~\ref{sec:results} across all applications individually with HEFT RT providing the best execution results, HEFT EDP providing the most energy savings, and HEFT EDP-LB falling between the two. 
	The largest outlier is the Single Carrier TX application, with all three HEFT RT, EDP, and EDP-LB schedulers performing nearly identically, due to the fact that this is the lightest workload.
	}
	\label{fig:odroid-single-app-results}
\end{figure}

\clearpage

\begin{figure}[h!]
	\centering
	\begin{subfigure}[]{\figwidth\linewidth}
		\includegraphics[width=\linewidth]{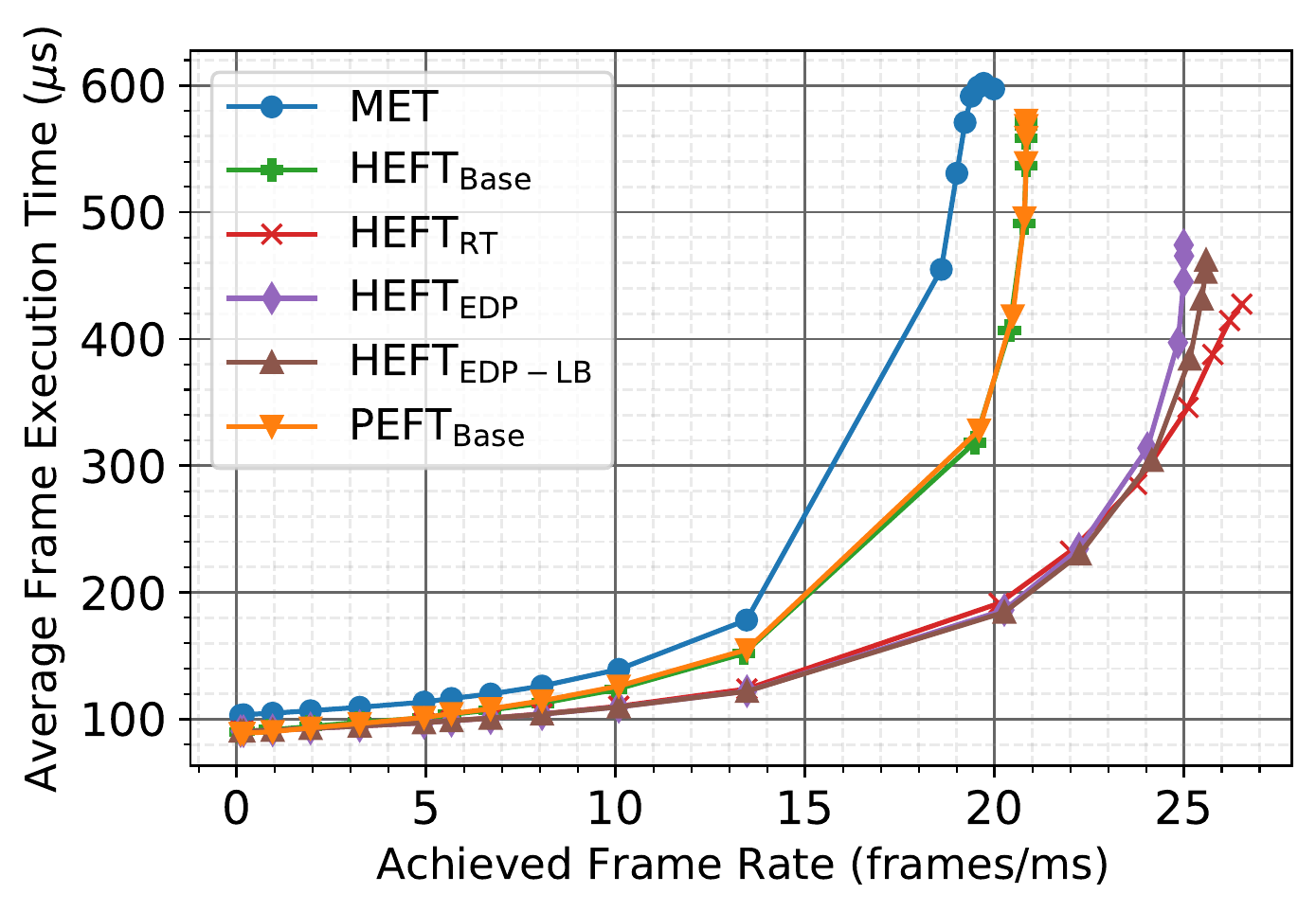}
	\end{subfigure}
	\begin{subfigure}[]{\figwidth\linewidth}
		\includegraphics[width=\linewidth]{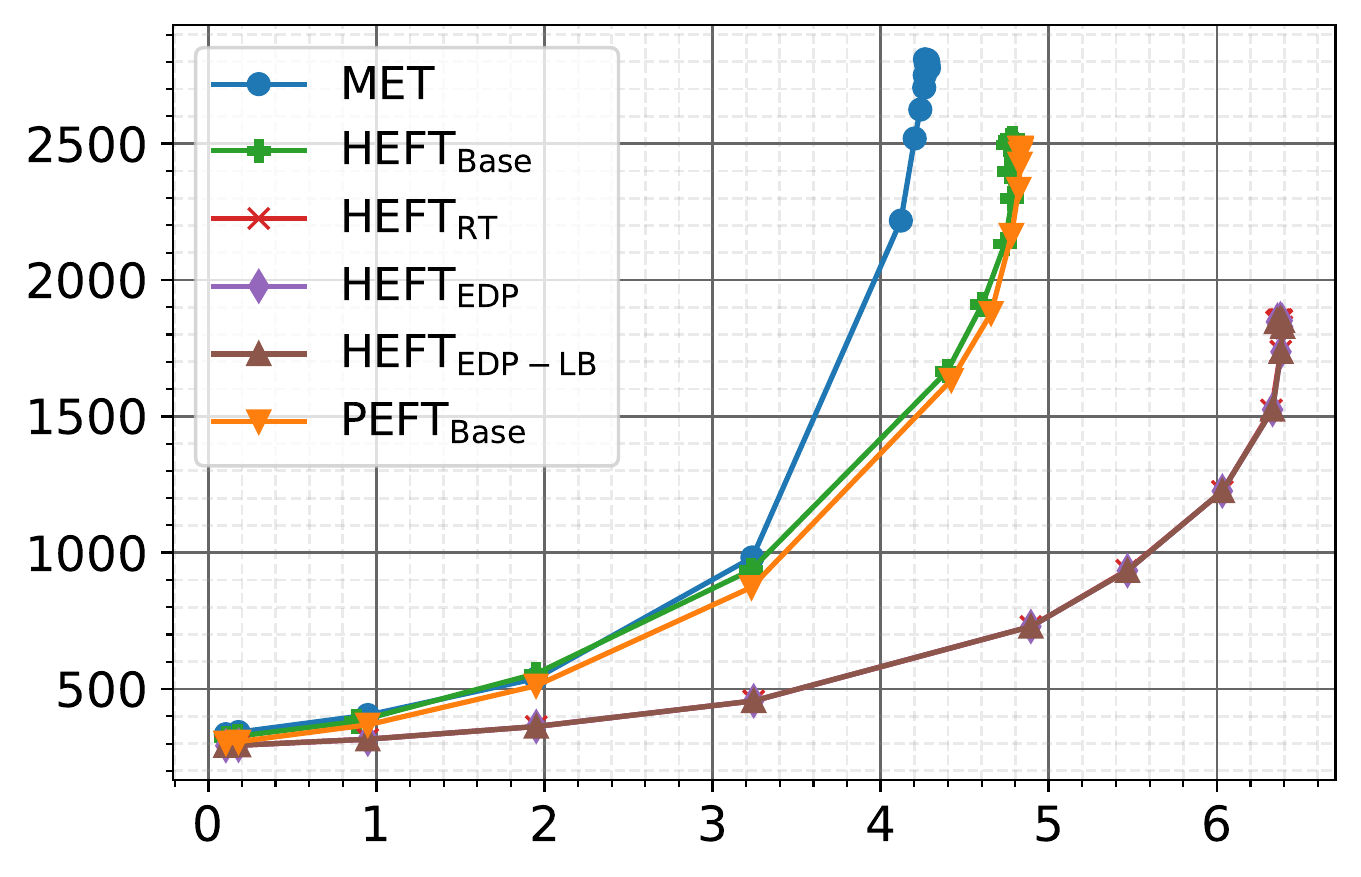}
	\end{subfigure}
	\begin{subfigure}[]{\figwidth\linewidth}
		\includegraphics[width=\linewidth]{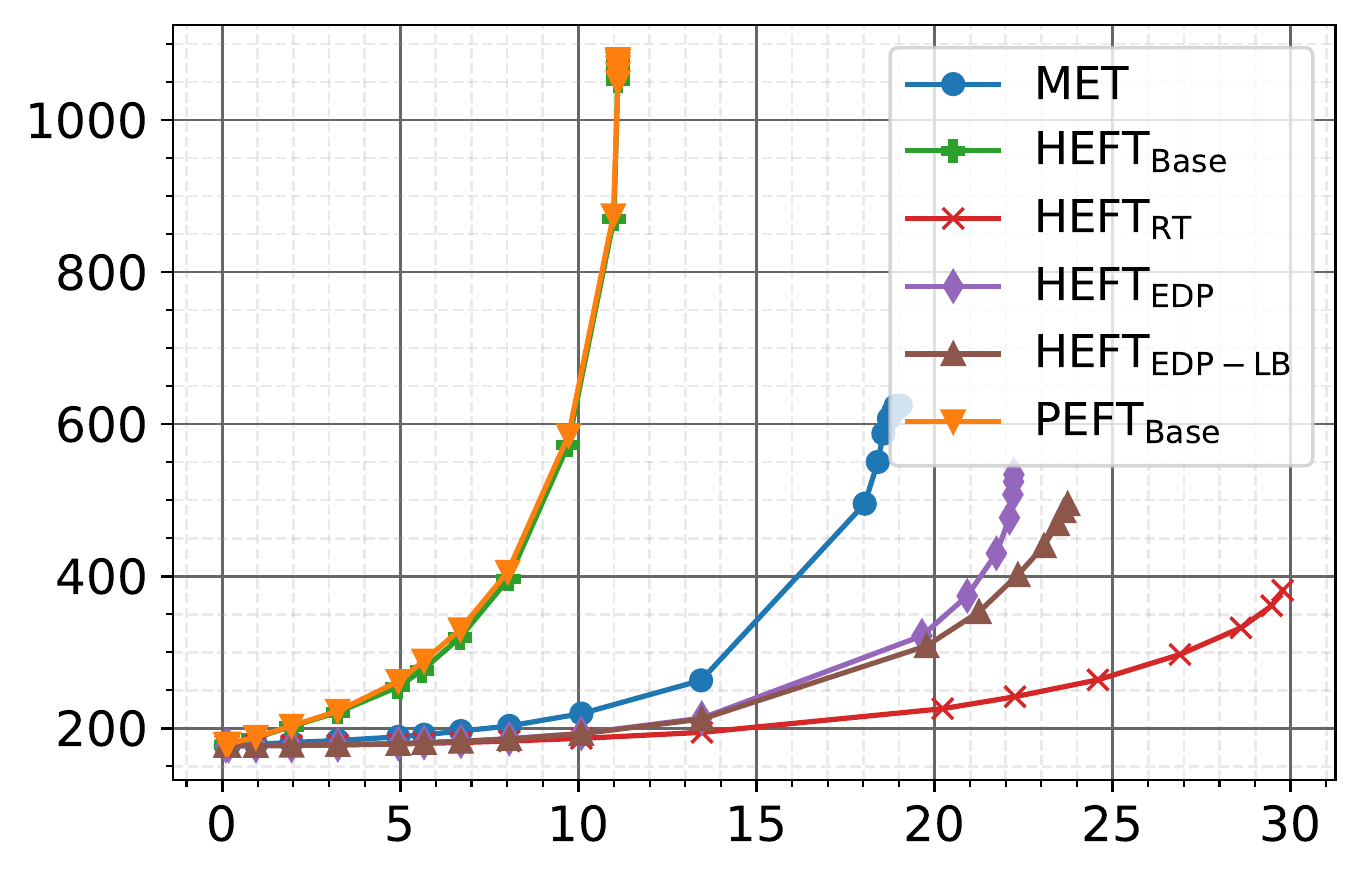}
	\end{subfigure}
	
	\begin{subfigure}[]{\figwidth\linewidth}
		\includegraphics[width=\linewidth]{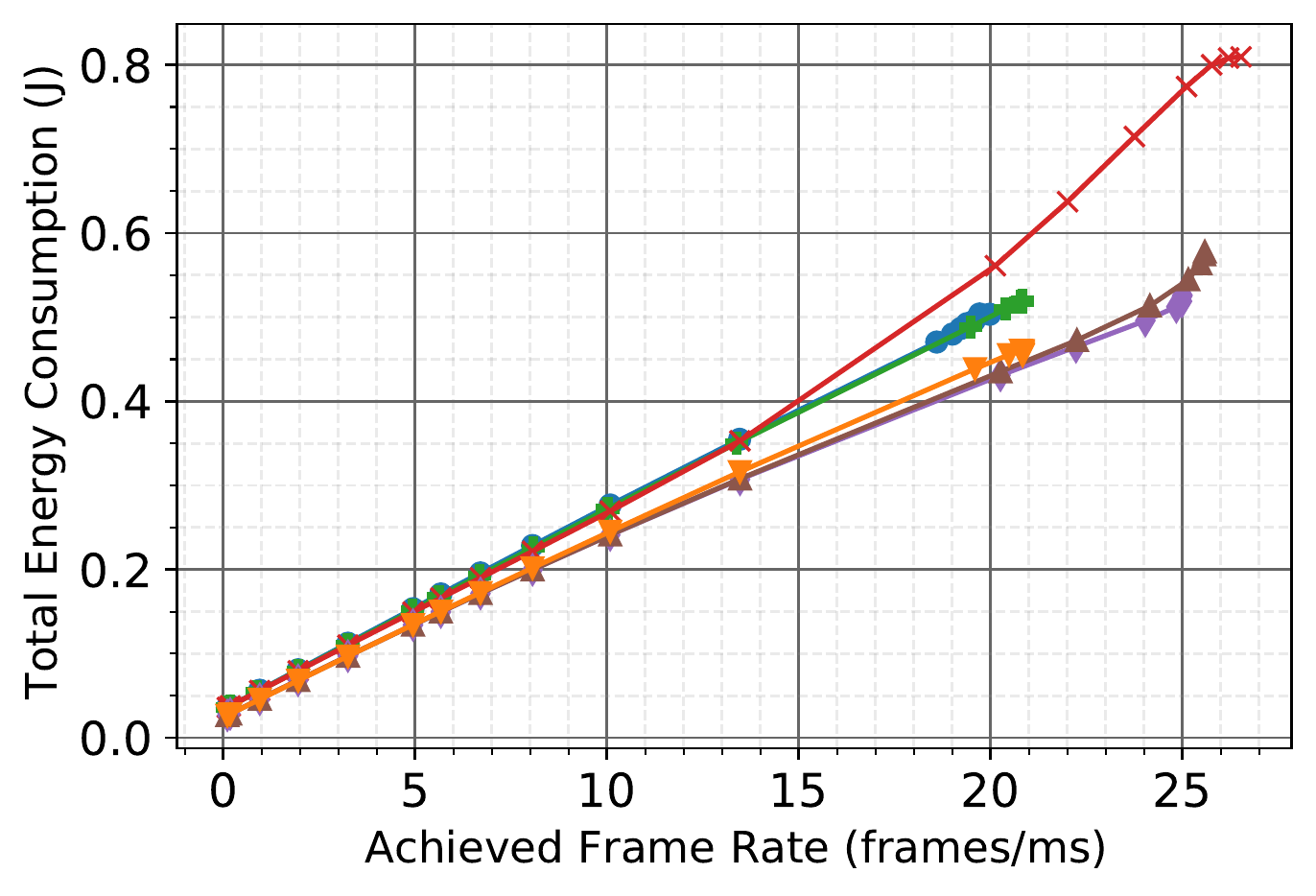}
	\end{subfigure}
	\begin{subfigure}[]{\figwidth\linewidth}
		\includegraphics[width=\linewidth]{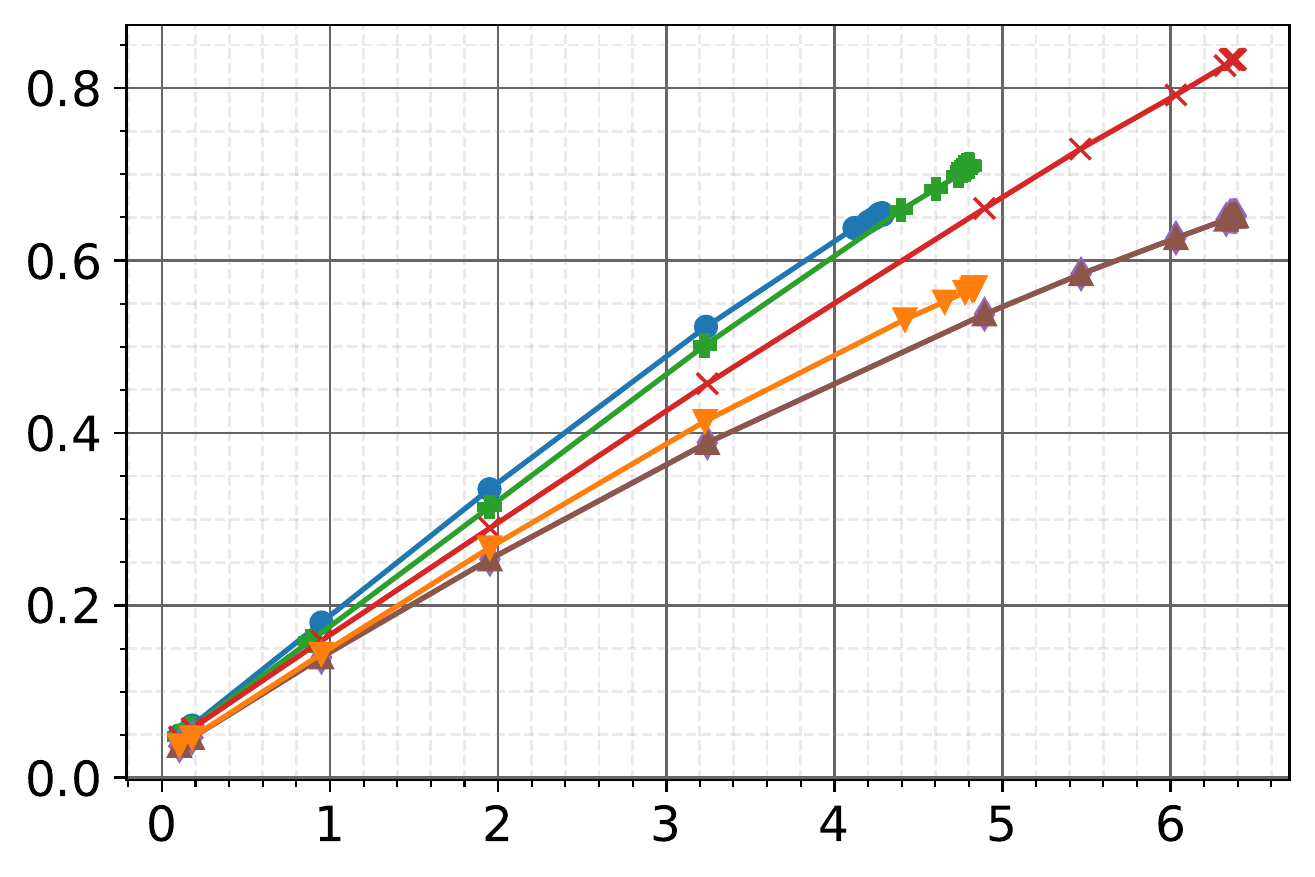}
	\end{subfigure}
	\begin{subfigure}[]{\figwidth\linewidth}
		\includegraphics[width=\linewidth]{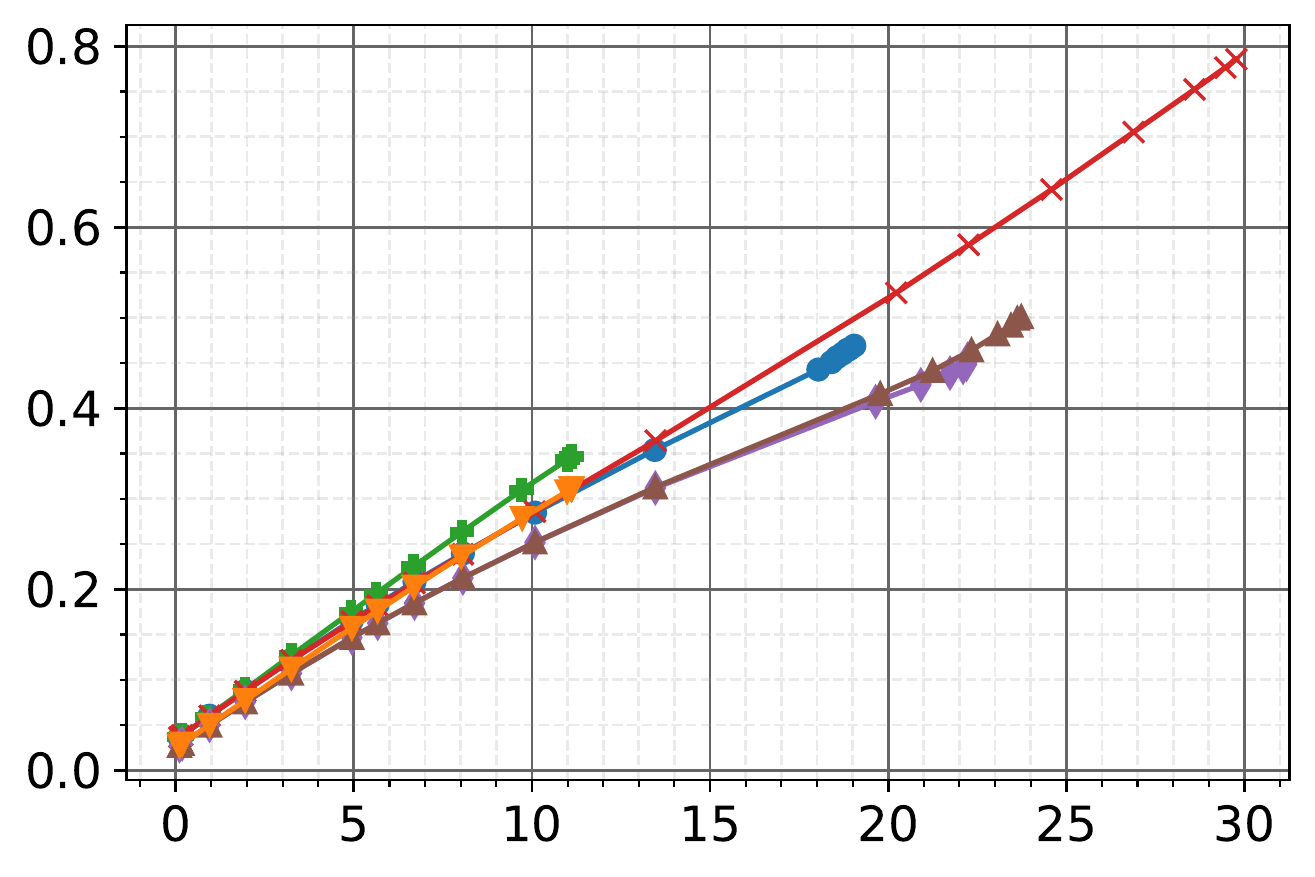}
	\end{subfigure}
	
	\begin{subfigure}[]{\figwidth\linewidth}
		\includegraphics[width=\linewidth]{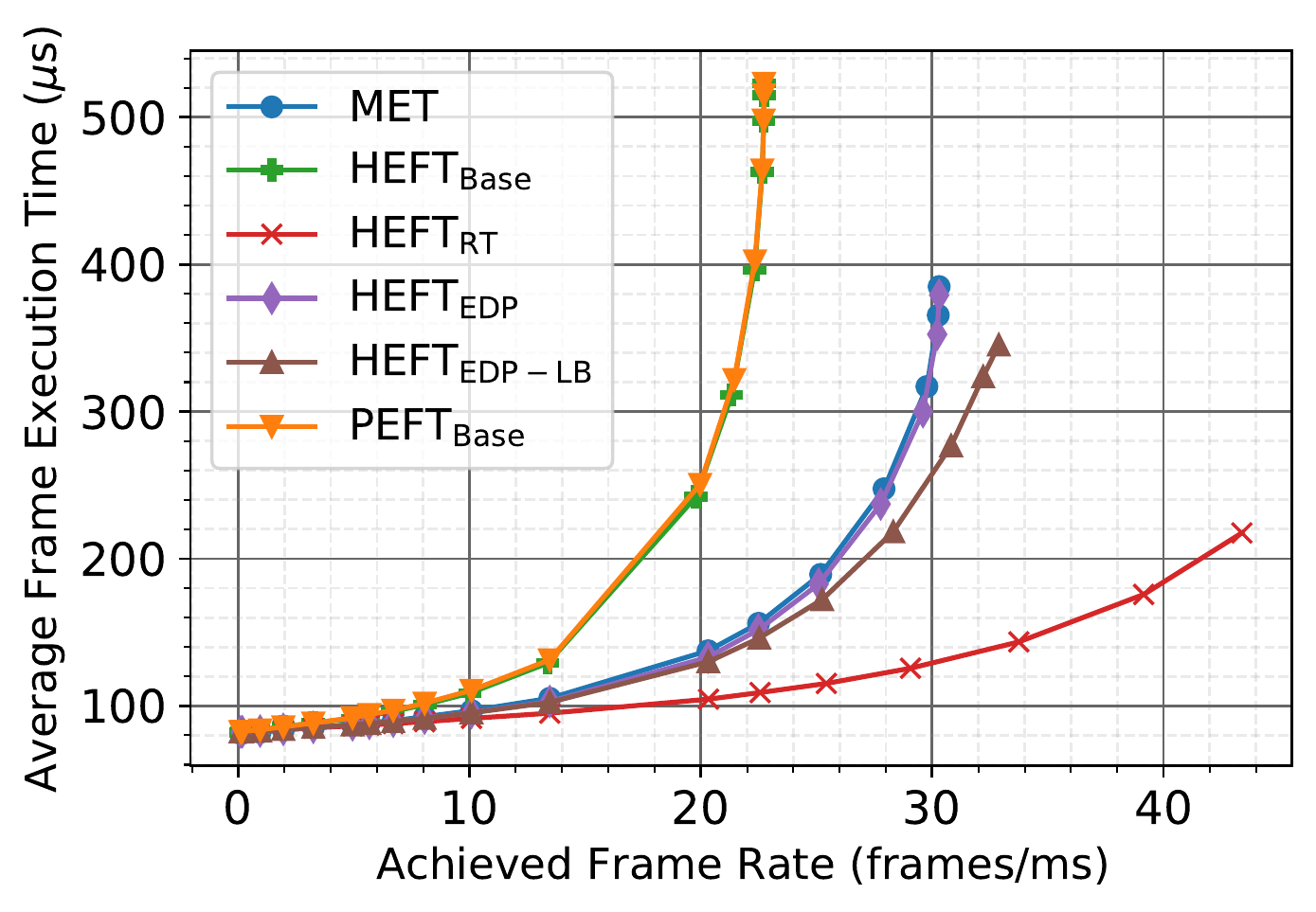}
	\end{subfigure}
	\begin{subfigure}[]{\figwidth\linewidth}
		\includegraphics[width=\linewidth]{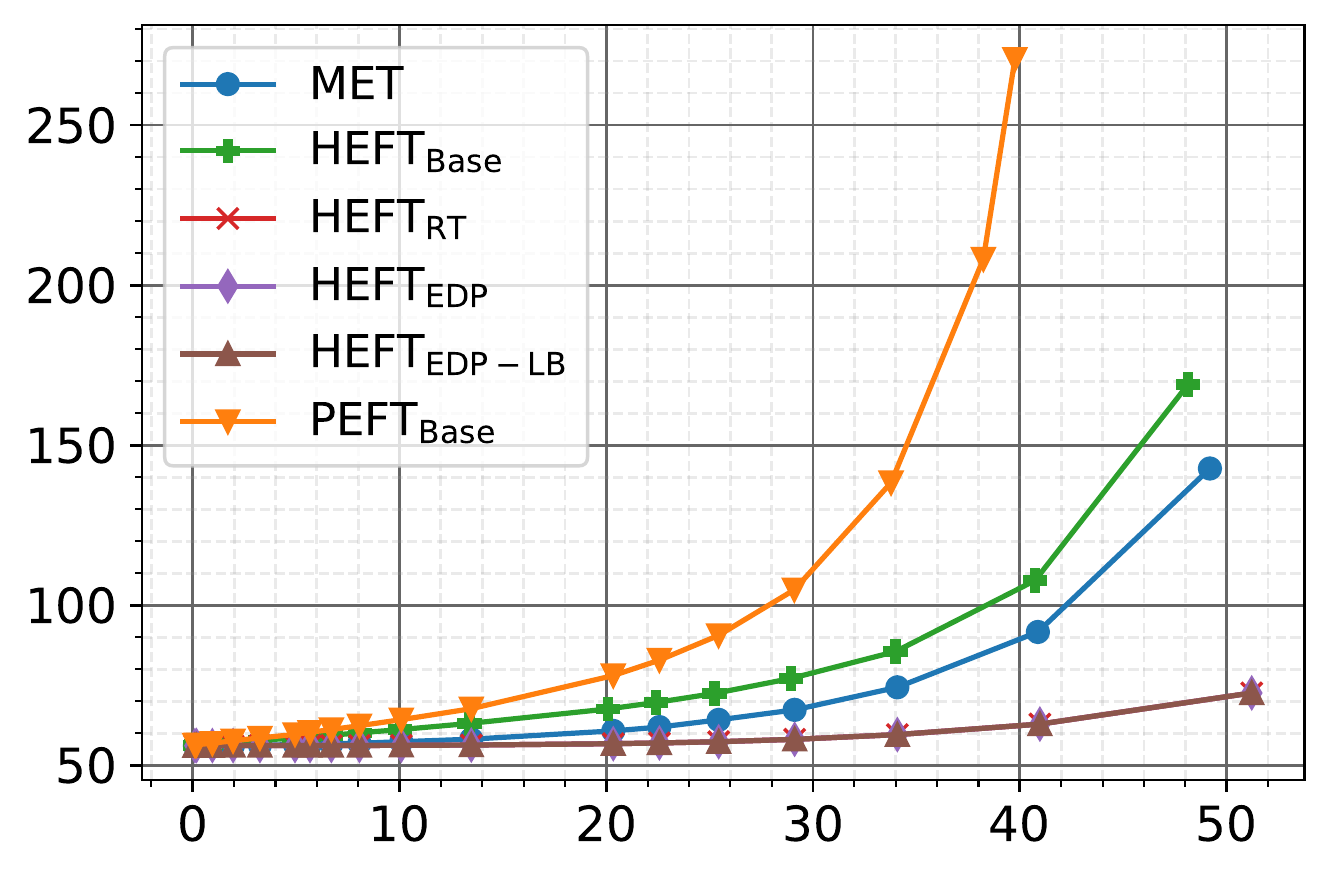}
	\end{subfigure}
	\begin{subfigure}[]{\figwidth\linewidth}
		\includegraphics[width=\linewidth]{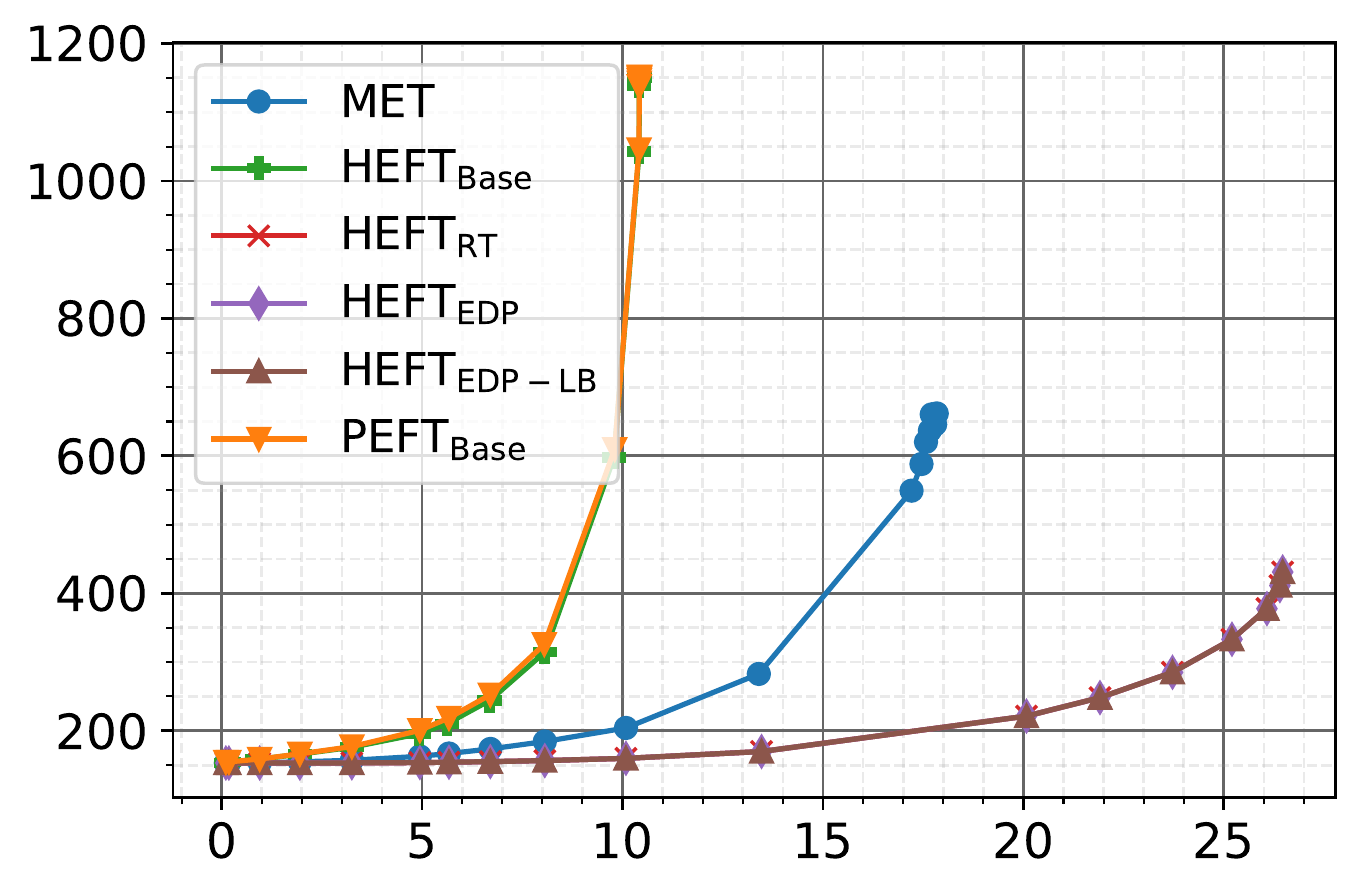}
	\end{subfigure}
	
	\begin{subfigure}[]{\figwidth\linewidth}
		\includegraphics[width=\linewidth]{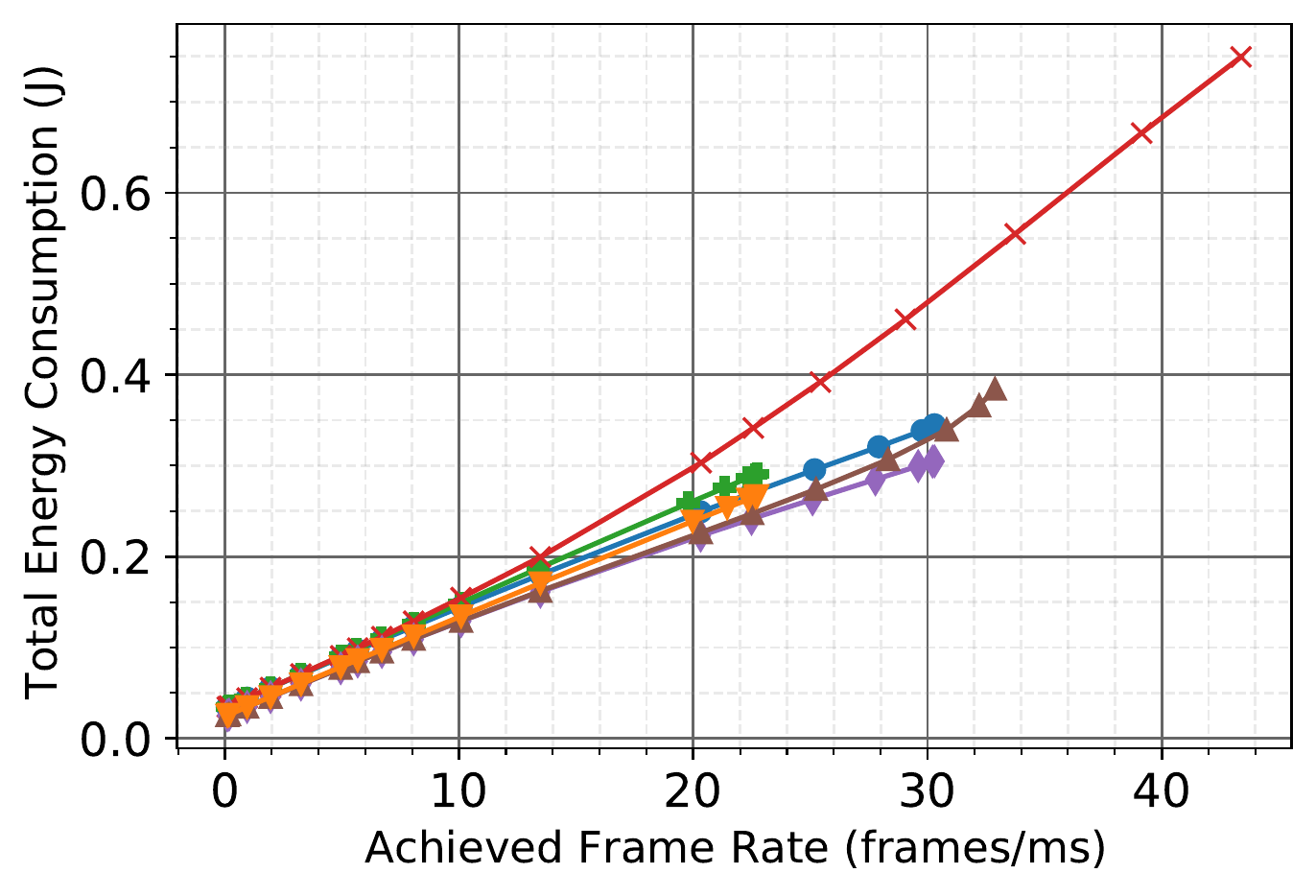}
	\end{subfigure}
	\begin{subfigure}[]{\figwidth\linewidth}
		\includegraphics[width=\linewidth]{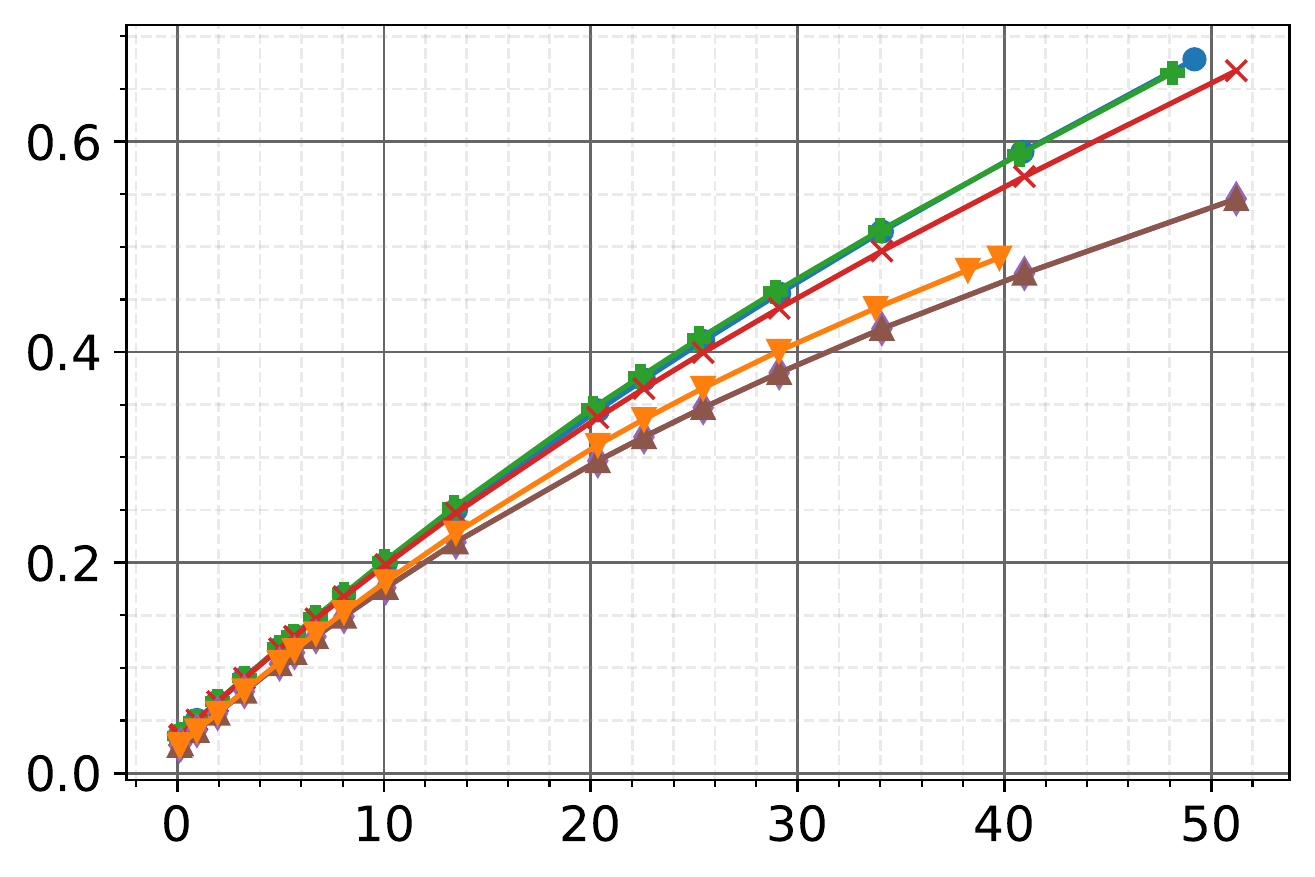}
	\end{subfigure}
	\begin{subfigure}[]{\figwidth\linewidth}
		\includegraphics[width=\linewidth]{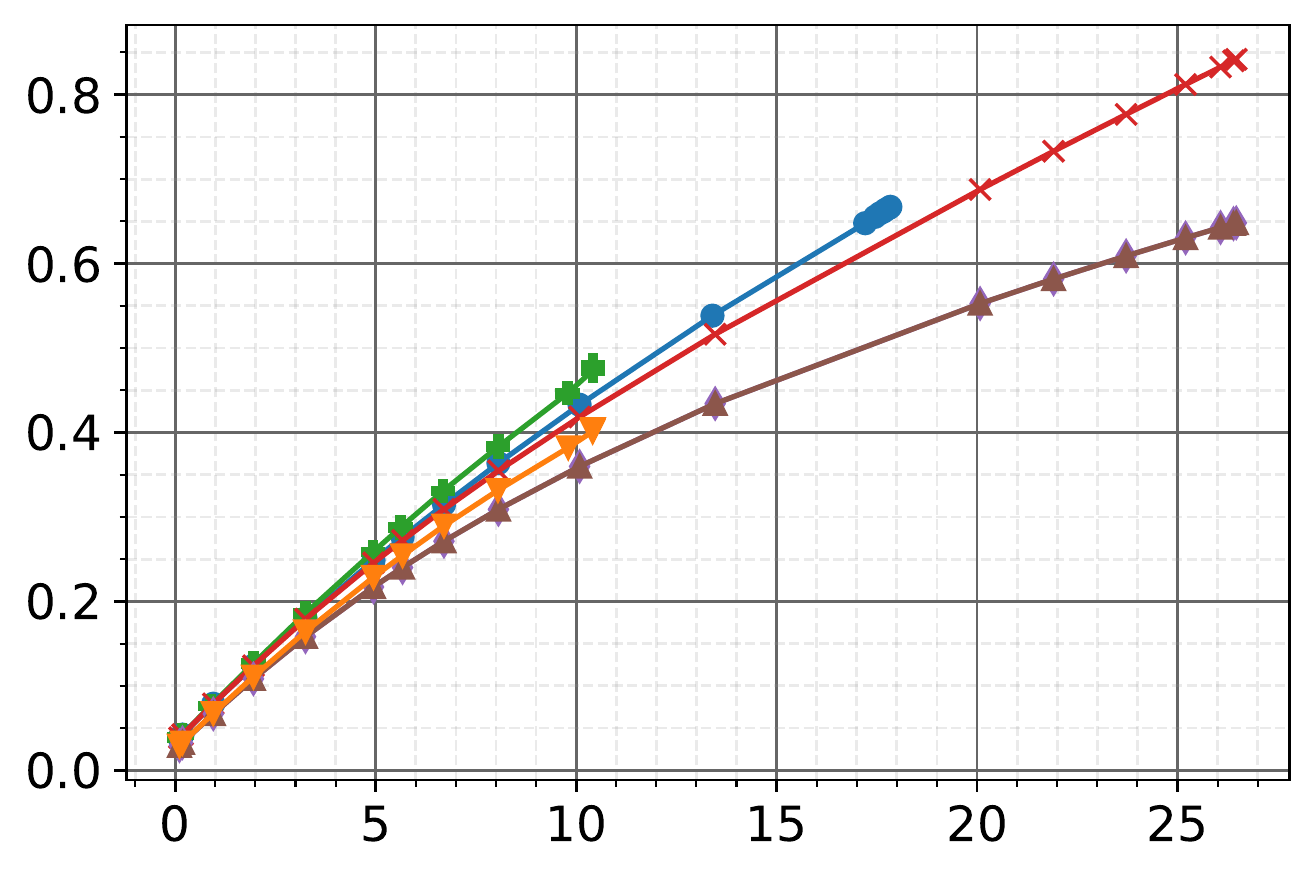}
	\end{subfigure}
	\caption{
	ZCU102 application-level plots. 
	Plots are grouped into vertical pairs of Execution \& Energy. 
	Applications from left to right, top to bottom: WiFi TX, WiFi RX, Radar Correlator, Temporal Mitigation, Single Carrier TX, Single Carrier RX.
	We find that the same general trends hold as presented in Section~\ref{sec:results} across all applications individually with HEFT RT providing the best execution results, HEFT EDP providing the most energy savings, and HEFT EDP-LB falling between the two. 
	The largest outlier is the Single Carrier TX application, with all three HEFT RT, EDP, and EDP-LB schedulers performing nearly identically, due to the fact that this is the lightest workload.
	}
	\label{fig:zcu102-single-app-results}
\end{figure}

\end{document}